\documentclass[a4paper,fleqn,usenatbib]{mnras}

\usepackage{amssymb}
\usepackage{amsmath}
\usepackage{graphicx}
\usepackage{natbib}
\usepackage{hyperref}
\usepackage{pdflscape}
\usepackage{longtable}
\usepackage{multicol}
\usepackage{multirow}
\usepackage{url}
\usepackage{graphicx}
\usepackage{color}
\usepackage{txfonts}
\usepackage{rotating}
\usepackage{color} 
\usepackage{soul} % para tachar palabras

\newcommand{\Ha}{H$\alpha$} 
\newcommand{\Hb}{H$\beta$}

\DeclareRobustCommand{\ION}[2]{%
\relax\ifmmode
\ifx\testbx\f@series
{\mathbf{#1\,\mathsc{#2}}}\else
{\mathrm{#1\,\mathsc{#2}}}\fi
\else\textup{#1\,{\mdseries\textsc{#2}}}%
\fi}

\title[The archaeological cosmic star-formation density]{SDSS IV MaNGA - An archaeological view of the Cosmic Star Formation History}

\author[S.F.S\'anchez et al.]{S.F. S\'anchez$^{1}$, 
V. Avila-Reese$^{1}$,
A. Rodr\'iguez-Puebla$^{1}$,
H. Ibarra-Medel$^{1}$,
R. Calette$^{1}$,
\newauthor 
M. Bershady$^2$,
H. Hern\'andez-Toledo$^1$
K. Pan$^3$,
D. Bizyaev$^4$ and
IA-UNAM-MaNGA Team
\\
$^{1}$Instituto de Astronom\'ia, Universidad Nacional Aut\'onoma de  M\'exico, A.~P. 70-264, C.P. 04510, M\'exico, D.F., Mexico\\
$^2$Department of Astronomy, University of Wisconsin Madison,475 N. Charter Street, Madison, WI, 53706, USA\\
$^3$Apache Point Observatory and New Mexico State University, P.O. Box 59, Sunspot, NM, 88349-0059, USA\\
$^4$Sternberg Astronomical Institute, Moscow State University, Moscow, Russia\\
}
\date{Accepted XXX. Received YYY; in original form ZZZ}
\pubyear{2018}

\begin{document}
\label{firstpage}
\pagerange{\pageref{firstpage}--\pageref{lastpage}}
\maketitle
%\author{IA-UNAM-MaNGA Team}

\begin{abstract}
We present the results of the archaeological analysis of the stellar populations of a sample of $\sim$4,000 galaxies observed by the SDSS-IV-MaNGA survey using Pipe3D. Based on this analysis we extract a sample of $\sim$150,000 SFRs and stellar masses that mimic a single cosmological survey covering the redshift range between $z\sim$0 to $z\sim$7. We confirm that the Star-Forming Main Sequence holds as a tight relation in this range of redshifts, evolving in both the zero-point and slope. This evolution is different for local star-forming (SFGs) and retired (RGs) galaxies, with the latter presenting a stronger evolution in the zero-point and a weaker evolution in the slope. The fraction of RGs decreases rapidly with $z$, particularly for RGs at $z\sim0$. We detect RGs well above $z>1$, although not all of them are progenitors of local RGs. Finally, adopting the required corrections to make the {\it survey} complete in mass in a limited volume, we recover the cosmic star-formation rate (SFR), stellar mass density, and average specific SFR histories of the Universe in this wide range of look-back times. Our derivations agree with those reported by various cosmological surveys. We demonstrate that the progenitors of local RGs were more actively forming stars in the past, contributing to most of the cosmic SFR density at $z>0.5$, and to most of the cosmic stellar mass density at any redshift. They suffer a general quenching in the SFR at z$\sim$0.35. Below this redshift the progenitors of local SFGs dominate the SFR density of the Universe.
\end{abstract}


\section{Introduction} 
\label{sec:intro}

Galaxies in the Local Universe are the consequence of their cosmological evolution, storing in their morphologies, dynamics, stellar populations and gas properties the fossil records of the processes that shaped them. There are two main observational approaches to study this evolution: (i) the analysis of the so-called cosmological surveys, i.e.,  observations of statistically significant and representative samples of galaxies that allow us to characterize the properties (stellar mass, star-formation rate, metallicity, etc.) of the bulk population at different redshift \citep[e.g.,][]{perezgonzalez08},  and (ii) the analysis of the fossil records of this evolution in the properties of galaxies observed in the Local Universe, i.e., the so-called archaeological method \citep[e.g.,][]{Thomas+2005}.

The first approach is by far the more frequently adopted one, being used by many different studies. It has allowed us to understand how galaxies evolve within the color-magnitude diagram \citep[e.g.,][]{bell06}, transforming the bulk population between star-forming to more quiescent/retired galaxies \citep[e.g.,][]{wolf05}; how disk galaxies grow in size, following an inside-out pattern \citep[e.g.,][]{bard05,vanderWel+2014}; how star formation happens more rapidly in more massive galaxies than in less massive ones \citep[e.g.][]{perezgonzalez08}, what is known as the downsizing paradigm \citep[see e.g.,][]{Cowie+1996,Fontanot+2009}; what is the general shape of the star-formation rate (SFR) density in the universe \citep[e.g.][]{Madau14,Driver17}; and how the relation between SFR and integrated stellar mass, mainly comprised in the so-called Star-forming Main Sequence (SFMS), evolves accross cosmic times \citep[see e.g.,][and references therein]{Speagle14,Rodriguez-Puebla+2017}. This approach assumes that galaxies at different redshifts are representative of the same population that evolves with time. By its nature, the approach traces the evolution of galaxies in a statistical way, being unable to trace the evolution of individual galaxies or to connect the exact same population of galaxies at different redshifts without making assumptions of how that population may evolve. On the other hand it has the major advantage that it directly probes different epochs.
%On the other hand it presents the major advantage that it is technically more simple and observationally less demanding (in general).%For example, when comparing more massive galaxies at different redshifts it is impossible to know if those galaxies are indeed 
%To understand this evolution there are two main

The second approach has been adopted by a smaller number of studies. This approach %tries 
looks to infer the evolution of individual galaxies by exploring the signatures that this evolution has produced in observational properties of galaxies. In principle it is possible to recover the star-formation and chemical-enrichment histories (SFHs and ChEHs, respectively) of galaxies, and to explore their dynamical evolution, by analyzing the spectroscopic properties, morphologies, and kinematics of both their stellar populations and ionized gas. For doing so, it is possible to adopt a particular shape for the SFHs and ChEHs \citep[e.g.,][]{Gallazzi+2005,Thomas+2010,Bitsakis+2016,zibetti17}, or to infer them in a non parametric way \citep[e.g.,][]{panter07,vale09,perez13,ibarra16,rgb17}. This approach is technically complex, prone to large uncertainties based on the adopted procedure \citep[e.g.,][]{Pipe3D_I}, the selected templates or stellar libraries \citep[e.g.,][]{rosa14}, or even the details on the errors and the assumptions within the analysis \citep[][]{cid-fernandes13,cid-fernandes14}. On the other hand, with this approach it is possible to trace the evolution of individual galaxies without any further assumption and no statistical matching between galaxy sub-samples. Adopting this procedure it has been possible to confirm  downsizing in galaxies \citep[e.g.,][]{Thomas+2005,Thomas+2010}; to demonstrate that this downsizing happens on local scales \citep[e.g.,][]{perez13,zibetti17}; that galaxies form inside-out \citep[e.g.][]{perez13,ibarra16}; that the SFHs of galaxies are different for different morphological types \citep[e.g.,][]{rgb17}; and even to reproduce the cosmic evolution of the SFR density in the Universe \citep[e.g.,][]{panter07,lopfer18} or the global metal enrichment \citep[e.g.,][]{asari07}.

As indicated above, one of the main discoveries of large galaxy surveys was the relation between the SFR and the integrated stellar mass that most galaxies follow, i.e., the SFMS \citep[e.g.][]{Brinchmann04,Salim07,Noeske+2007,Renzini15,Sparre15}. At any redshift, star-forming galaxies (SFGs) present a tight ($\sim$0.2-0.3 dex dispersion) linear correlation between both parameters in the logarithm with a sub-unity slope at low redshifts \citep[e.g.][]{Renzini15,mariana16}. The slope evolves modestly with redshift, being almost constant over cosmological time, and possibly approaching unity at higher redshifts \citep[e.g.,][]{Speagle14}. However, this evolution, although mild, could reflect a change in the overall star formation in galaxies. A unity slope  found at high redshift is consistent with an exponential SFH \citep[a shape frequently used to model the SFH of galaxies, e.g.,][]{lopfer18}, with a time delay %$\tau$ 
that regulates the downsizing. However, a sub-unity slope, typically found in low-redshift studies, implies that the SFH is shallower than an exponential one. On the other hand, the zero-point of the SFMS presents a clear shift towards larger values in the past \citep[e.g.][]{Speagle14}, following the cosmological evolution of the SFR density in the Universe \citep[e.g.][]{Katsianis15,Rodriguez-Puebla+2017}. This trend may be truncated at very high redshift if the SFR density (per comoving volume) of the universe, $\Psi_{\rm SFR}$, presents a decline as shown in the so-called Madau plot \citep[e.g.][]{Madau14,Driver17}.%, although it has not been reported so far.

Besides the (star-forming) galaxies along the SFMS, there are galaxies below this sequence, that is, with lower values of SFR for their masses. These galaxies are in a passive/quiescent mode of star formation, and if this mode prevails over time, then we can refer to these galaxies as retired (RGs, hereafter). Analysis based on cosmological surveys show that the fraction of quiescent galaxies increases with cosmic time, with RGs being extremely rare at redshfits higher than $\sim 2$ \citep[e.g., ][]{Muzzin+2013,Tomczak+2014,Martis+2016,pand17}.

More recently, \citet{zibetti17} and \citet{lopfer18}\footnote{\url{http://www.iaa.es/sites/default/files/thesis/tesis_rafael_lopez_fdo.pdf}} adopted an archaeological approach to estimate the cosmic evolution of $\psi_{\rm SFR}$, following the pioneering results by \citet{panter03} and \citet{heavens04}. In both cases they adopted a set of  SFH models and compared the observed stellar indices together with multiwavelength photometric data to derive the properties of the stellar populations for galaxies extracted from the CALIFA survey \citep{sanchez12b}. In the case of \citet{zibetti17}, the derived cosmic SFH, $\Psi_{SFR}(z)$, presents a similar shape as the one presented by \citet{Madau14}, with a decline at low- (z$<$0.5) and high-redshifts (z$>$4), but with a  much broader and shallower peak, centered at lower redshifts (z$\sim$1-2). The agreement between cosmological surveys and the results of \citet{lopfer18} is better, although the peak in the SFR density is slightly broader and shallower than previously reported results.

Following these pioneering studies, we have adopted the archaeological approach for the current study. We explore the evolution of the SFR-M$_*$ diagram of both SFGs and RGs accross cosmic times based on the analysis of the SFHs derived for $\sim$4,000 galaxies observed by the SDSS-IV MaNGA \citep{manga} survey { and adequately corrected for volume completeness. Furthermore, we calculate the global SFR and mass density evolution. We address the question of whether the predictions, based on the archaeological approach, are consistent or not with the results from cosmological surveys}. 

The flow of this article is as follows: In Sec. \ref{sec:sample} we describe the sample and data explored in this study. A summary of the analysis performed on the data is described in Sec. \ref{sec:ana}, with details on the derivation of the SFR included in Sec. \ref{sec:SFR}. The main results of the current study are presented in Sec. \ref{sec:res}, including the description of the local SFMS (Sec. \ref{sec:SFMS_0}), and its evolution accross cosmic times (Sec. \ref{sec:SFMS_t}). The quantification of this evolution for the different analyzed subsamples is explored in Sec. \ref{sec:SFMS_q}. Our estimation of the cosmic SFR density history is included in Sec. \ref{sec:madau}. The distribution of the stellar mass density of the universe at different redshifts is described in Sec. \ref{sec:m_den_t}, and the average specific SFR in Sec. \ref{sec:sSFR}. The discussion on the results is presented in Sec. \ref{sec:disc}, including a summary of the main caveats on those results in Sec. \ref{sec:caveat}, with details on the effects of mergers included in Sec. \ref{sec:merging}. The differences between the SFHs of star-forming and retired galaxies is described in Sec. \ref{sec:dSF}. The evolution of the SFMS is discussed in Sec. \ref{sec:dSFMS} with details on the turn-over at high mass discussed in Sec. \ref{sec:dturn}. Finally, the implications of the fraction of retired galaxies found along cosmological times is discussed in Sec. \ref{sec:dRGs}. The conclusions of our results are presented in Sec. \ref{sec:con}.

In this article we assume the standard $\Lambda$ Cold Dark Matter cosmology with the parameters: H$_0$=71 km/s/Mpc, $\Omega_M$=0.27, $\Omega_\Lambda$=0.73.

\section{Sample and data}
\label{sec:sample}

We use the observed sample of the Mapping Nearby Galaxies at APO \citep[MaNGA;][]{manga} survey collected through June 2017, comprising a total of 4202 galaxies. MaNGA is part of the 4th generation of the Sloan Digital Sky Survey \citep[SDSS-IV,][]{Blanton+2017}. The goal of the ongoing MaNGA survey is to observe approximately 10,000 local galaxies; a detailed description of the selection parameters can be found in \citet{manga}, including the main properties of the sample, while a general description of the Survey Design is found in \citet{renbin16b}. The sample was extracted from the NASA-Sloan atlas (NSA, Blanton M. \url{http://www.nsatlas.org}). Therefore, all the parameters derived for those galaxies are available (such as effective radius, Sersic indices, multi-band photometry, etc.). The MaNGA survey is taking place at the 2.5 meter Apache Point Observatory \citep{2006AJ....131.2332G}. Observations are carried out using a set of 17 different fiber-bundles science integral-field units  \citep[IFU; ][]{2015AJ....149...77D}. These IFUs feed two dual channel spectrographs \citep{2013AJ....146...32S}. Details of the survey spectrophotometric calibration can be found in \citet{2016AJ....151....8Y}. Observations were performed following the strategy described in \citet{law15}, and reduced by a dedicated pipeline described in \citet{2016AJ....152...83L}. These reduced datacubes are internally provided to the collaboration labeled as version 2.2.0 of the dataset. This sample includes more than 4200 galaxies at redshift 0.03$<z<$0.2, covering a wide range of galaxy parameters (e.g, stellar mass, SFR and morphology), providing a panoramic view of the properties of the population in the Local Universe. For examples of the distribution of galaxies in terms of their redshifts, colors, absolute magnitude and scale-lengths, and a comparison with other on-going or recent IFU surveys, see \citet{sanchez17a}.

The MaNGA sample comprises four different subsamples of galaxies, as described by \citet{wake17}: (i) the primary sample, design to cover at least 1.5 r$_e$ within the FoV of the different fiber bundles; (ii) the secondary sample, designed to cover at least 2.5 r$_e$; (iii) the color enhanced sample, designed to increase the galaxies within the so-called green-valley; and (iv) a set of different subsamples of ancillary or complementary objects included to make use of fiber bundles unable to be allocated by the previous three categories. According to \citet{wake17} it is feasible to perform a volume correction for the two first subsamples (that comprise nearly 90\%\ of the objects) based on the classical V$_{max}$ procedure \citep{schmidt68}. {  At the start of this work, no volume correction was available for the currently adopted dataset, either in the public domain or distributed within the MaNGA collaboration. Therefore, we calculated our own volume corrections following the prescriptions described in Appendix \ref{sec:Vmax}. We have performed a set of cross-checks of our volume corrections for the subsample of galaxies for which there is now a publically available volume correction computed as described by \citet{wake17}, finding no major differences for galaxies  above M$_*\sim $10$^9$M$\odot$. However, below this stellar mass, our volume corrections seem to provide better corrections when comparing the derived 
luminosity and mass functions with those determined from volume-complete samples. We will discuss our approach in a forthcoming article (Calette et al., in prep.).}

\section{Analysis}
\label{sec:ana}

We analyze the datacubes using the {\sc Pipe3D} pipeline \citep{Pipe3D_II}, which is designed to fit the continuum with stellar population models and to measure the nebular emission lines of IFS data. This pipeline is based on the {\sc FIT3D } fitting package \citep{Pipe3D_I}. The current implementation of {\sc Pipe3D} adopts the GSD156 library of simple stellar populations \citep[SSPs][]{cid-fernandes13} that comprises 156 templates covering 39 stellar ages (from 1Myr to 14.1Gyr), and 4 metallicities (Z/Z$\odot$=0.2, 0.4, 1, and 1.5) \footnote{Details of the actual ages are given in \url{https://data.sdss.org/datamodel/files/MANGA_PIPE3D/MANGADRP_VER/PIPE3D_VER/PLATE/manga.Pipe3D.cube.html}}. These templates have been extensively used within the CALIFA collaboration \citep[e.g.][]{perez13,rosa14}, and for other surveys \citep[e.g.][]{ibarra16,laura17}. Details of the fitting procedure, dust attenuation curve, and uncertainties on the processing of the stellar populations are given in \citet{Pipe3D_I,Pipe3D_II}. 

{ Prior to any analysis a spatial binning is performed in order to
  increase the S/N without altering substantially the original shape
  of the galaxy. For doing so, two criteria are adopted to guide the
  binning process: (i) a desired S/N for the binned spectra, and (ii)
  a maximum difference in the flux intensity between adjacent
  spaxels. The first criterion selects a S/N per \AA\ of 50, that
  corresponds to the limit above which the recovery of the stellar
  population properties have an uncertainties of $\sim$10-15\%
  \citep{Pipe3D_I}. The second criterion selects a maximum difference
  in the flux intensity of a 15\%. This corresponds to the typical
  flux variation along an exponential disk of the average size of our
  galaxies in a range of 1-2 kpc, and shorter scale-lengths for more
  early-type galaxies.

  The application of the two criteria and the spatial binning proceeds
  in the following way: First, the S/N per \AA\ is derived at each
  spaxel (spatial pixel) by constructing a narrow-band image centered
  in $\sim$5000\AA\ and comparing the mean flux intensity per
  \AA\ (signal) with the root-square of the variance within the
  considered wavelength range (noise). Each spaxel within the datacube
  with a S/N above the desired goal (S/N$>$50) is considered as an
  independent tessella. Thus, for those spaxels \citep[roughly 10-20\%
    of the total ones][]{ibarra16}, no bining is performed. Those spaxels with a S/N below the desired goal are ordered by
  their flux intensities. Then the non-binned spaxel
  with the highest flux intensity is binned with any adjacent
  one if (i) the adjacent one does not already belong to a previously
  defined tessella and (ii) the difference in the flux intensity
  between them is lower than the considered limit. The S/N within the new defined tessella 
is then re-evaluated by comparing
  the average flux intensity of the spaxels that comprise the bin with
  the propagated noise. This process takes into account the
  co-variance between adjacent spaxels. If the S/N of the binned data
  is larger than the foreseen goal, then the agregation of spaxels to
  this tessella stops, and the process starts with a new
  spaxel (following the defined flux intensity order). If the S/N of
  the tessella is still lower than the goal, then the agregation
  process is repeated by selecting non-binned adjacent spaxel within
  the flux intensity limit (using the mean flux intensity within the
  tessella as the new comparison value). If no spaxel is found
  fulfilling this criterion the agregation process stops for this
  tessella, and a new tessella is created starting from the
  non-binned spaxel with the highest flux intensity. The procedure is
  described and discussed in detail in \citet{Pipe3D_II},
  \citet{ibarra16} and \citet{casa17}.}

{ As a result of this binning process the original spaxels, with a
  size of 0.5$\arcsec$$\times$0.5$\arcsec$
  \citep[e.g.][]{2016AJ....152...83L}, are agreegated in tessellas of
  variable size. } The typical size of the tessellas range between 2-5
spaxels in most of the cases, with a few larger ones in the outer
regions of the galaxies \citep[e.g., Fig. 3 and 4 of ][]{ibarra16}.
Contrary to other binning schemes, the original shape of the galaxy is
better preserved by the adopted procedure, not mixing adjacent regions
corresponding to clear different structures (e.g.,
arm/inter-arms). The disadvantage is that it does not provide with an
homogenous S/N distribution across the entire FoV and the S/N limit is
not reached in all the final bins/voxels.  { This S/N limit of 50
  was selected based on the extensive simulations described in
  \citet{Pipe3D_I} in order to recover reliably the SFHs and stellar
  properties in general.  For lower S/N those properties are recovered
  in a less precise but still accurate way.  The tessellas with
  lower S/N are found mostly in the outer regions, where
  there are still a large number of individual bins. Therefore, averaging the stellar
  properties (including the SFHs) either radially or integrated across
  the entiry FoV provide uncertainties similar to the ones from individual but
  larger S/N bins.}  This was
already shown in \citet{ibarra16}, and it is discussed in Appendix
\ref{sec:bin}. { The adopted procedure provides a more accurate SFH than
  what would be derived from  coadding all the spectra within the FoV into a single one and analysing it, according
  to recent results (Ibarra-Medel et al. submitted).
}

Once performed the spatial binning/segmentation, the
spectra from spaxels in each tessella are co-added prior to any
further analysis. Then, a stellar population fit of the co-added
spectra within each spatial bin is computed. The fitting procedure
involves two steps: first, the stellar velocity and velocity
dispersion are derived together with the average dust attenuation
affecting the stellar populations (A$_{V,ssp}$). Second, a multi-SSP
linear fitting is performed, using the library described before and
adopting the kinematics and dust attenuation derived in the first
step. This second step is repeated including perturbations of the
original spectrum within its errors; this Monte-Carlo procedure
provides the best coefficients of the linear fitting and their errors,
which are propagated for any further parameters derived for the
stellar populations. { At the end of this analysis we have a model of
  the stellar populations for each tessella.}

{ Finally, } we estimate the stellar-population model for each spaxel by re-scaling the best fitted model within each spatial bin (tessella) to the continuum flux intensity in the corresponding spaxel, following \citet{cid-fernandes13} and \citet{Pipe3D_I} { , a standard procedure in this kind of analysis}. This model is used to derive the average stellar properties at each position, including the actual stellar mass density, light- and mass-weighted average stellar age and metallicity, and the average dust attenuation. In addition, the same parameters as a function of look-back times are derived, which comprise in essence the star formation and chemical enrichment histories of the galaxy at different locations. In this analysis we followed \citet{Pipe3D_II}, but also \citet{cid-fernandes13}, \citet{rosa16a}  \citet{rosa17} and \citet{rgb17}. In a similar way as described in \citet{mariana16} and \citet{ibarra16} it is possible to co-add, average or azimuthal average those parameters to estimate their actual (and/or time evolving) integrated, characteristics or radial distributions.  

The stellar-population model spectra are then subtracted from the original cube to create a gas-pure cube comprising only the ionized gas emission lines (and the noise and residual of the stellar population modeling). Individual emission line fluxes were then measured spaxel by spaxel fitting both a single Gaussian function for each emission line and spectrum, and also making a weighted moment analysis, as described in \citet{Pipe3D_II}. For this particular dataset, we make use of the flux intensities and equivalent widths of \Ha\ and \Hb \citep[although a total of 52 emission lines are analyzed,][]{Pipe3D_II}. All intensities were corrected for dust attenuation. For doing so, the spaxel-to-spaxel \Ha/\Hb\, ratio is used. Assuming a canonical value of 2.86 for this ratio \citep{osterbrock89}, and adopting a \citet{cardelli89} extinction law and a R$_{\rm V}$=3.1 (i.e., a Milky-Way-like extinction law), the spatial dust attenuation in the V-band ($A_{V,gas}$) is derived. Finally, using the same extinction law and  derived attenuation, the correction for each emission line at each location within the FoV was applied. 

After a detailed quality control analysis we restricted the sample to 4101 galaxies, excluding blank fields pointings, very low signal-to-noise targets, galaxies with bright foreground field stars and galaxies at the very edge of the FoV of a MaNGA IFU.

%%%%%%%%%%%%%%%%%%%%%%%%%%%%%%%%%%%%%%%%%%%%%%%%%%%
% SFMS
\begin{figure*}
 \minipage{0.99\textwidth}
    \includegraphics[width=9.5cm]{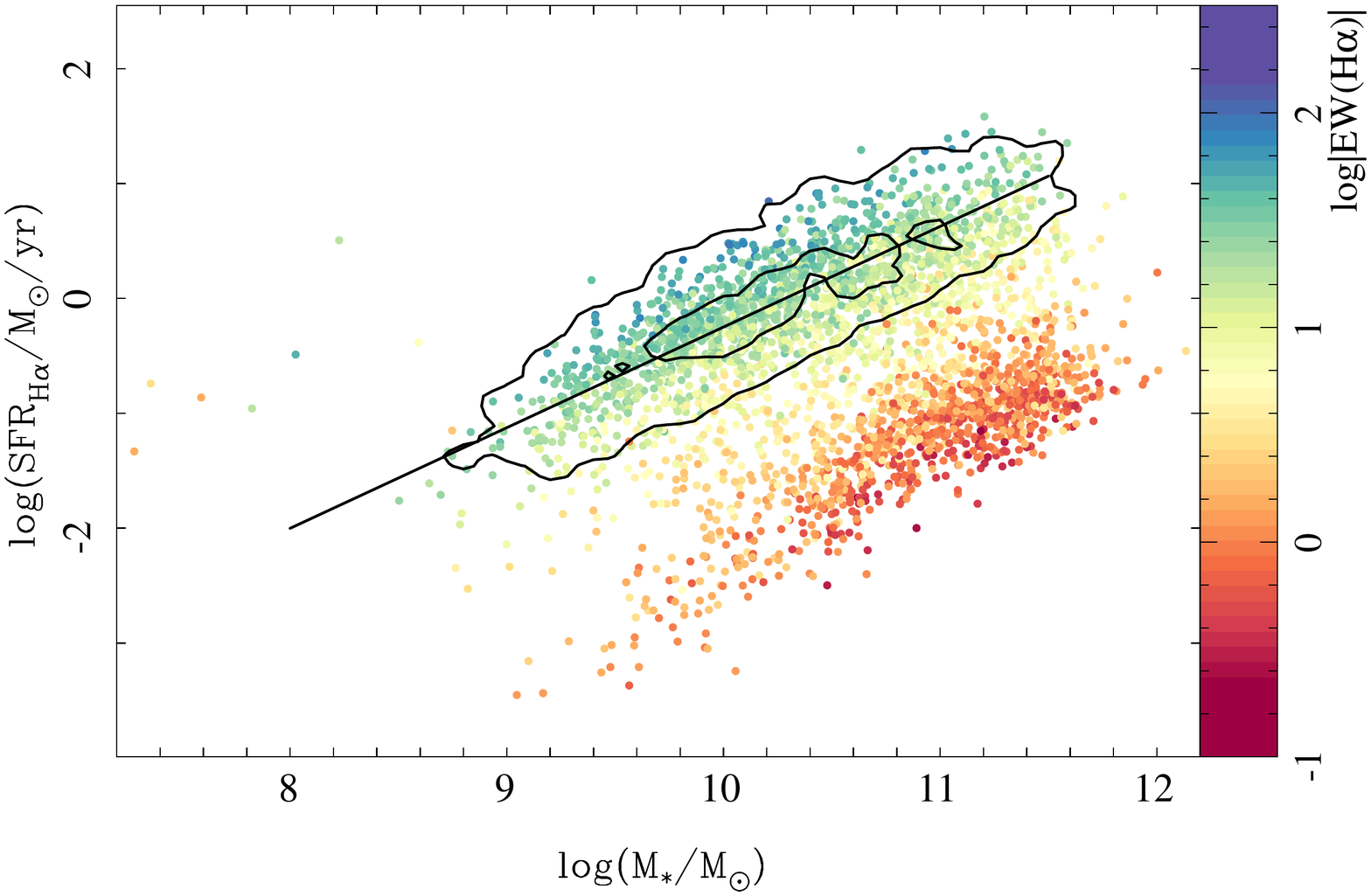}\includegraphics[width=9.5cm]{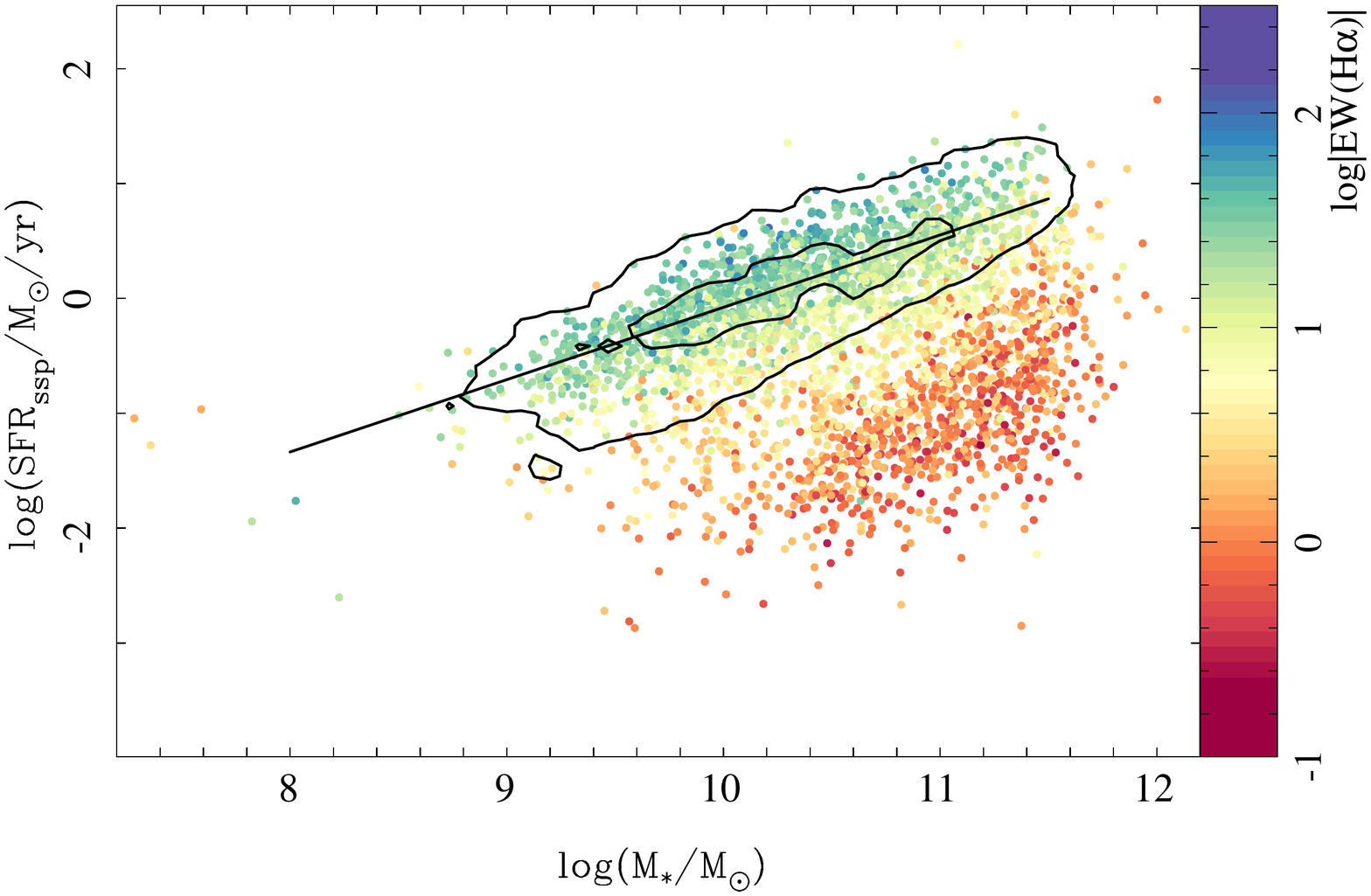}
 \endminipage
\caption{{\it left panel:} Distribution of the SFR derived from the dust corrected H$\alpha$ luminosity versus the integrated stellar mass derived from analysis of the stellar populations for each galaxy in the sample. {\it Right panel:} Similar distribution for the SFR derived based on the stellar-mass accumulated in the last 100Myr, based on the stellar population analysis. Each galaxy in each panel is shown as a solid circle colored based on the EW of H$\alpha$ averaged across the FoV of each datacube. Contours indicate the density for the current SFGs at the redshift where the galaxies were observed, selected as indicated in the text. Outer and inner contours encircle 95\% and 25\% of the galaxies. Solid black line shows the result from a linear fitting for the SFMS derived at each redshift range.}
  \label{fig:SFMS}
\end{figure*}
%%%%%%%%%%%%%%%%%%%%%%%%%%%%%%%%%%%%%%%%%%%%%%%%%%%

\subsection{Star Formation Rate and stellar mass}  
\label{sec:SFR}

The SFR was derived using two different procedures: (i) based on the \Ha\ luminosity, and (ii) based on the stellar population synthesis analysis. In the first case we use the \Ha\ intensities for all the spaxels with detected ionized gas. The intensities are transformed to luminosities (using the adopted cosmology) and corrected for dust attenuation as indicated before. { Then,} we apply the \citet{kennicutt98} calibration to obtain the spatially-resolved distribution of the SFR surface density. A Salpeter Initial Mass Function (IMF) was adopted \citep{salpeter55}, the same assumed for the SSP library. We use all the { original } spaxels irrespective of the origin of the ionization. By doing so, we take into account the PSF wings in the star-forming regions that may present equivalent widths below the cut applied in \citet{sanchez17a} and \citet{mariana16} (as we will explain in the following sections). On the other hand, we are including in our SF measurement regions that are clearly not ionized by young stars. For SFGs, that contribution is rather low, due to the strong difference in equivalent widths, as already noticed by \citet{catalan15}, and therefore the SFR is only marginally affected. However, for the RGs, the ionization comes from other sources, including AGN ionization, post-AGB stars, or rejuvenation in the outer regions \citep[e.g][]{sarzi10,papa13,singh13,Gomes16a,Gomes16b,Belfiore17a}. Therefore, the \Ha-based SFR for RGs should be considered as an upper limit, as recently demonstrated by \citet{bit18}. Hereafter we will refer to this star-formation rate as SFR$_{\rm H\alpha}$. We discuss in Appendix \ref{sec:comp_SFR_spax} in detail why adopting either integrated SFRs or selecting only those regions that we are totally sure are ionized by local star-formation (at the scale of the kpc) do not alter significantly the current analysis.
  
%In addition to this estimation 
In the second case we derive the current SFR using the decomposition of the stellar populations in a multi-SSP analysis described above. For each galaxy at each spaxel, it is possible to assign the fraction of light that corresponds to a certain age (by co-adding the fractions of light from different metallicities at the considered age). Taking into account the Mass-to-Light ratio of each particular SSP and the luminosity in each spaxel it is straight-forward to determine the mass that corresponds to stars of a certain age \citep[as described in ][]{sanchez17a} and at a certain location. Then, by co-adding these masses within the FoV of the datacubes we obtain the integrated mass of stars of a certain age, M$_{*,age}$. Once we derive this distribution of masses it is possible to integrate from the earliest times up to a certain look back time and obtain the cumulative mass of the galaxy versus cosmic time \citep[as described in][]{ibarra16}. To do so it is necessary to consider the redshift of the object in order to derive the correct look back time that is matched against each age within the SSP library. The derivative of this cumulative mass function is, by construction, the SFR at each redshift (or look back time), thus, the SFH of the galaxy \citep[e.g.][]{rosa17,rgb17}. { In particular, the SFR calculated at the observed redshift is the current one, and we refer to it as SFR$_{ssp,0}$.}
%%Creo que es mejor cerrar primero la discusion sobre las dos formas de calcular la SFR local y luego continuar ya con las SFHs que se infieren con el SPS, lo cual lo pondria definitivamente como otra subseccion. Por ende, he traido aqui este parrafo que estaba casi al final:

{ The two estimations of the SFR at $z\sim$0, SFR$_{\rm H\alpha}$ and SFR$_{ssp,0}$, do not follow a one-to-one correspondence, although they present a tight correlation, with an offset of $\sim$0.11 dex and a dispersion of $\sim$0.32 dex. These differences are expected due to the different nature of their derivations that are based on different assumptions, as we discuss in detail in Appendix \ref{sec:comp_SFR}. We maintain that using either derivation of the SFR at $z\sim$0 provides similar qualitative results, and that they can be transformed from one to the other adopting a linear relation with a slope near unity. This relation is given in Appendix \ref{sec:comp_SFR}. }

%% Sugiero que esto sea otra subseccion, en cual caso a la anterior habría que llamarla "CURRENT SFR and stellar Mass": 
%\subsection{Star Formation Rates at each look-back time}  
%\label{sec:SFRH}
Contrary to those archaeological methods based on parametric SFHs, our method cannot provide the SFR at any time (SFR$_t$), since it samples the SFH in a discrete way for each galaxy, limited by the ages included within the SSP library \citep[e.g.][]{rgb17,lopfer18}. Our currently adopted SSP library comprise 39 ages. Therefore, we can derive 38 SFRs sampled in the time steps between each two consecutive ages. Due to the range in redshift of the MaNGA sample each galaxy samples the SFH at slightly different cosmic times, whichis particularly important at low redshifts. { Some examples of the individual SFHs derived from our inversion methods were presented in previous publications \citep[][Fig. 2, 3 and 4]{ibarra16}. We present just the median SFHs in different stellar mass bins in Figure \ref{fig:SFH} for reference. To derive these SFHs it was necessary to apply an interpolation to the individual and temporally discrete values and re-sample them to a common look-back-time, since, as indicated before, each galaxy samples cosmic time in a different way due to its redshift. Once interpolated, we obtain the median of the SFRs along cosmic time for all galaxies within a considered mass bin. The time range sampled by galaxies of different stellar mass is different due to the strong correlation between this parameter and the redshift in the MaNGA sample \citep[e.g.][]{manga}. It is beyond the scope of this article to discuss the details of these typical SFHs, and indeed this is not required for the analysis we present here. As a brief summary, it can be said that more massive galaxies have a stronger SFR at earlier times, while less massive galaxies have peak SFRs at lower redshift. This is in agreement with current knowledge of the SFHs in galaxies, known as the downsizing scenario \citep[e.g.][]{perezgonzalez08,Thomas+2010}, and already seen in both the pioneering studies using the fossil record method \citep{panter03,panter07}, and in more recent analysis \citep[e.g.][]{perez13,ibarra16,rgb17,lopfer18}. We will discuss the shape of the SFHs elsewhere (Ibarra-Medel et al., in prep.).}

{The main result of our analysis is that for each galaxy we estimate its stellar mass and SFR at 38 look-back-times that corresponds to 38 different redshifts for each galaxy.} All together our procedure generates a total of 155,838 individual pairs of stellar masses and SFRs, { that are the result of combining the 38 estimates for each of the 4101 analyzed galaxies. This final {\it sample} covers a wide redshift range between $z\sim$0.005 and $z>8$}. { Although, due the limitations of our adopted procedure, our exploration is reliable only up to z$\sim$3, as we will see later}. We correct stellar masses for the mass-loss at the observed time, adopting the prescriptions by \citet{Bruzual:2003aa} that depend on the age and metallicity of the population at each redshift.

The SFR derived in this way at the redshift of the object would be the current SFR, that we will label as SFR$_{ssp}$. This procedure was used recently by \citet{rosa16a} in their exploration of the radial structure of SFR in galaxies. They derive the SFR$_{ssp}$ by integrating the stellar mass formed in the last 32Myr and dividing by this time scale. This procedure is in essence the same as the one used in the derivation of any SFR calibrator, such as the ones presented by \citet{kennicutt98}, although in these theoretical calibrators a certain SFH is assumed a priori. Following \citet{Speagle14} we adopte a time range of 100 Myr in our derivation of the SSP$_{ssp}$, although assuming any range between 10 and 100 Myr would not make any significant difference. { A direct comparison between the SFR$_{ssp}$ derived using 10Myr, 32Myr and 100Myr leads to a systematic offset toward larger SFRs as the time range increases, with  increments of 0.06$\pm$0.20 dex between 10Myr and 32Myr and 0.15$\pm$0.18 dex between 32Myr and 100Myr. Aside from this offset, there is a clear one-to-one trend; a linear regression between SFRs using any two time ranges yields slopes ranging between 0.82 and 0.97, being always compatible with one. Therefore, as claimed before, adopting any of these time ranges would not change significantly the results.}

Finally, we { should stress} that the integrated mass, SFR, and their evolution obtained from IFS spatially-resolved observations are more accurate than those obtained from a single integrated spectrum (the case of single-aperture observations) as shown in Ibarra-Medel et al. (2018; in prep.), who applied both analysis to simulated galaxies.  

\section{Results}
\label{sec:res}

\subsection{The local Star-forming Main Sequence}
\label{sec:SFMS_0}

Despite the differences between the H$\alpha$- and SSP-based SFRs, both present similar general trends when compared with other parameters of the galaxies. In particular, they show a similar trend with the stellar mass. Figure \ref{fig:SFMS} shows the distribution of the SFRs derived using both estimations versus stellar mass, color codded by the EW(H$\alpha$) averaged across the entire FoV of the datacubes. Two clear trends are seen for galaxies with EW(H$\alpha$)$>$3\AA\ and galaxies with EW(H$\alpha$)$<$3\AA, for both derivations of the SFR, as already noticed by previous authors \citep[e.g.,][]{mariana16}: (i) the Star-forming Main Sequence, SFMS, which shows a linear correlation between SFR and M$_{*}$ in logarithmic scales, with a slope slightly lower than one, %\citep[$\sim$0.8; e.g.,][]{mariana16}, 
and (ii) the sequence of passive or retired galaxies, RGs, which shows a linear correlation only for the SFR$_{\rm H\alpha}$ estimation, and a cloud for the SFR$_{ssp}$. 

As already discussed by previous authors \citep[e.g.][]{sanchez17b}, the nature of those two trends is intrinsically different. The former correlation indicates that when galaxies are actively forming stars, the integrated SFR follows a power of the look-back time (not an exponential profile as generally assumed), as discussed by \citet{Speagle14}. On the other hand, the later correlation, evident only for the H$\alpha$ derivation of the SFR, does not reflect precisely a connection between SFR and M$_{*}$, since actually the dominant ionizing source for galaxies in the RG sequence is not compatible with SF: \citet{mariana16} already have shown that their ionization is located in the so-called LINER-like (or LIER) area of the BPT diagram, being most probably dominated by some source of ionization produced by old-stars \citep[e.g., post-AGBs;][]{keel83,binn94,sta08,binn09,sarzi10,cid11,papa13,singh13,Gomes16a,Gomes16b,Belfiore17a}. Indeed, its luminosity correlates with M$_{*}$ due to its stellar nature, indicating that they most probably present a characteristic EW(H$\alpha$) \citep[e.g.][]{mori16}, lower than 3\AA, as predicted by \citet{sta08} and \citet{cid11}. \citet{rosa17} already showed that when the SFR is not derived from the H$\alpha$ ionized gas, the linear shape disappears, and the RGs are distributed in a cloud shape within the SFR-M$_{*}$ diagram (as appreciated in Fig. \ref{fig:SFMS}, right panel).

To estimate the shape of the SFMS it is necessary to select those galaxies that are currently star-forming, excluding the RGs. To do so, a cut in the specific SFR (sSFR) is usually adopted, selecting those galaxies with sSFR above a given value, e.g., sSFR$_{\rm lim}\approx 10^{-11}$ yr$^{-1}$ \citep[e.g.,][]{Pozzetti+2010}. For  sSFR$_{\rm lim}= 10^{-10.8}$ yr$^{-1}$,  %$log(sSFR$_{\rm lim}$/yr$^{-1}$)$>$-10.8 dex \citep[e.g., []{Pozzetti+2010}. 
this cut is totally equivalent to a cut in the average EW(H$\alpha$), such as the one used in \citet{sanchez17b}, since both parameters present a tight correlation \citep[e.g.][]{sanchez13,Belfiore17a}. { As a sanity check we explore this correlation using the current analyzed data, confirming the previous results for both the SFR derived using the H$\alpha$ luminosity and the one derived by the SSP analysis. The corresponding sSFRs both present very tight correlations with the EW(H$\alpha$), with correlations coefficients of $r=0.99$ and $r=0.98$ respectively, following the form:

$$\rm log (sSFR_{\rm H\alpha}) = -12.03_{\pm 0.52} + 1.24_{\pm 0.56} log |EW(H\alpha)|, \sigma=0.13$$
$$\rm log (sSFR_{\rm ssp}) = -11.79_{\pm 0.51} + 1.05_{\pm 0.55} log |EW(H\alpha)|, \sigma=0.09,$$
Therefore, we confirm that interchanging any of the three parameters would indeed produce similar results.}

Ideally for this analysis we would like to select only those regions in galaxies where ionization is compatible with star formation, using classical diagnostic diagrams \citep[e.g., following][and references therein]{sanchez17a,sanchez17b,laura17}. However, as pointed out by several authors, the contamination by other sources of ionization in the SFR integrated across the entire optical extent of galaxies is rather low in general, due to either the intrinsically low EW(H$\alpha$) of these non-star-forming ionization regions (e.g., post-AGBs), or the limited extent compared to that of the star-forming regions (e.g., type-II AGNs). Even in the case of type-I AGNs the contamination is rather low, as shown by \citet{catalan15,catalan17}. We would like to note that in any case removing AGNs from our analyzed sample does not affect the results, since the number of AGNs is rather low compared with the bulk population of galaxies \citep[$\sim$3-4\%,  e.g.][]{sanchez17b,remb17}. { As a sanity check we have repeated all the analysis presented here using the SFRs derived by selecting only the individual spaxels where ionization is compatible with star-formation, following the criteria outlined in \citet{sanchez17a}, see Appendix \ref{sec:comp_SFR_spax}. We have found no significant differences.  At this point, we prefer to continue with the original procedure and use the integrated quantities since it is more compatible with what it would be done in a cosmological survey based on single aperture spectroscopic data.}

%%%%%%%%%%%%%%%%%%%%%%%%%%%%%%%%%%%%%%%%%%%%%%%%%%%
% SFMS
\begin{figure}
 \minipage{0.99\textwidth}
    \includegraphics[width=9.5cm]{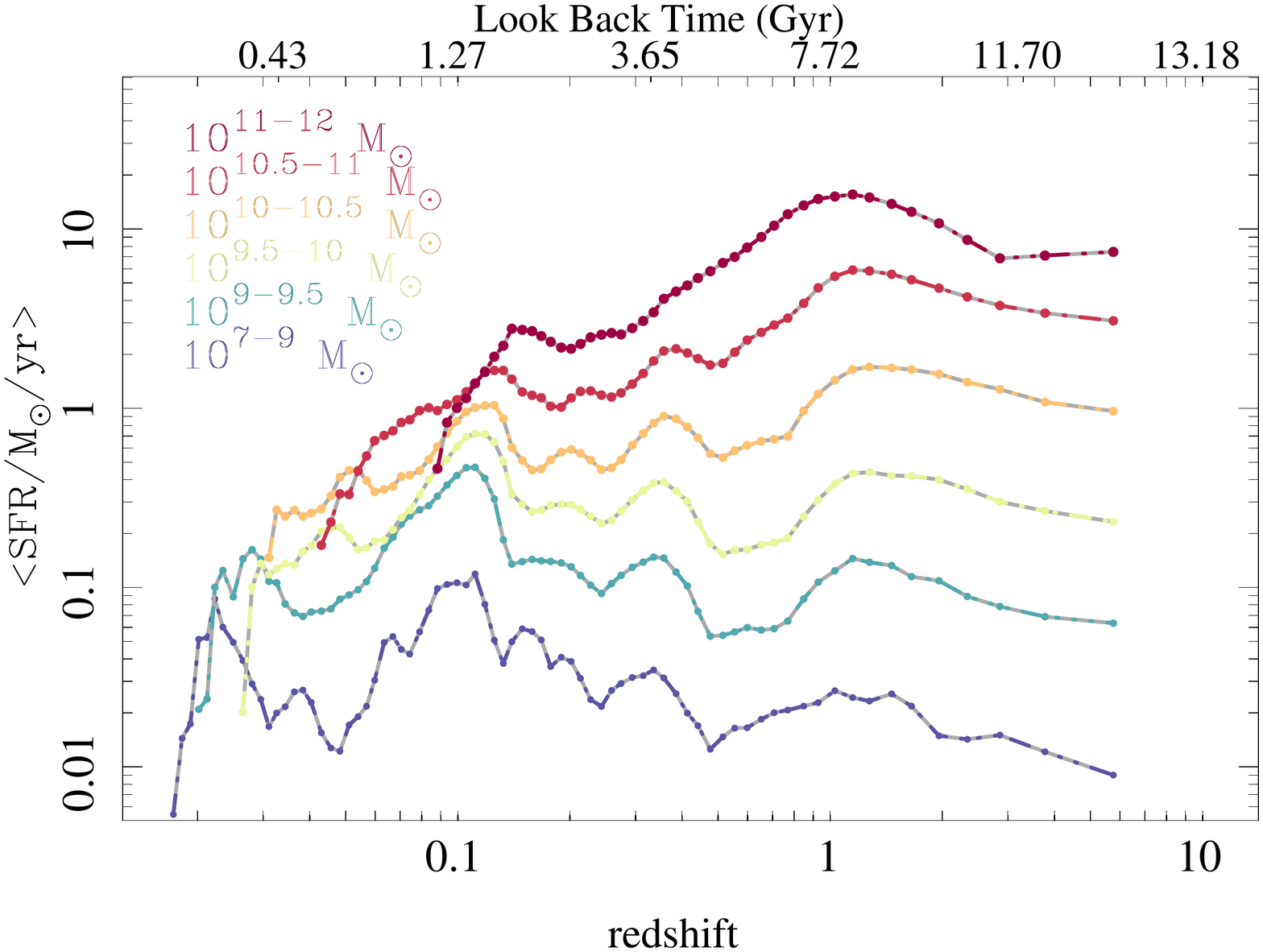}
 \endminipage
\caption{ Typical SFHs derived from our analysis of the stellar populations for galaxies in different mass bins. Each line, color-coded by stellar masss, shows the  median SFR at each look-back-time for all galaxies within the indicated mass range, interpolated to a common look-back-time and re-sampled in a logarithm scale. Solid circles indicate the re-sampled look-back-times. Due to the strong relation between the redshift and the mass in the MaNGA sample the sampled time range changes with the stellar mass.}
  \label{fig:SFH}
\end{figure}
%%%%%%%%%%%%%%%%%%%%%%%%%%%%%%%%%%%%%%%%%%%%%%%%%%%

% Sanchez et al. Cut in 6AA
Different cuts in EW(H$\alpha$) have been proposed to select star-forming regions in galaxies. \citet{sanchez14} proposed a minimum EW(H$\alpha$)$>$6\AA\ to select HII %H$_{\rm ii}$
regions in galaxies, while more recently \citet{Lacerda18} increased that threshold to  $\sim$10\AA\ for star-forming regions in general. However, when integrating across the optical extent of a galaxy this limit could be relaxed toward lower values. If we consider that an EW(H$\alpha$)$<$3\AA\ is the limit for ionization due to post-AGB, HOLMES, i.e., evolved stars \citep[e.g.][]{sta08}, a galaxy that forms stars somewhere within its optical extension, but not everywhere, would have an EW(H$\alpha$)$>$3\AA. Based on this basic assumption \citet{sanchez17b} classified the galaxies with 3\AA$<$EW(H$\alpha$)$<$6\AA\ as green-valley galaxies, somehow  between pure star-forming and totally retired galaxies. These galaxies are less than a 10\%\ of the total sample, and their inclusion or exclusion as SFGs do not modify significantly our results. For simplicity, we will include all these galaxies within the sub-sample of star-forming ones, since indeed they present star formation somewhere within their optical extent. 

In summary, we classify as SFGs at $z\sim$0 (SFGs$_0$) those galaxies with an average EW(H$\alpha$)$>$3\AA. These galaxies are clearly located at the expected location of the SFMS in Fig. \ref{fig:SFMS}. { Those galaxies comprise 41\% of all the objects in the current sample, with 59\% of them being classified as retired galaxies. The fraction of SFGs is slightly lower than that expected in the Local Universe, due to the sample selection of MaNGA aiming to provide  a flat distribution of galaxies in  stellar mass.} Using this sample we perform a log-linear regression between both derivations of the SFR and M$_*$, characterizing the local SFMS:

$$\rm log (SFR_{\rm H\alpha}) = -8.96_{\pm 0.23} + 0.87_{\pm 0.02} log(M_*), \sigma=0.32$$
$$\rm log (SFR_{\rm ssp}) = -6.32_{\pm 0.17} + 0.61_{\pm 0.02} log(M_*), \sigma=0.28,$$
where the SFR is in units of M$_{\odot}$/yr and M$_*$ is in units of M$_{\odot}$ 

Both linear regressions are shown in Fig. \ref{fig:SFMS}. The differences in the slope reflects the differences found between both estimate of the SFR, described in the previous section. The trend is shallower for SFR$_{ssp}$ than for SFR$_{\rm H\alpha}$. The differences at a fixed stellar mass are even smaller. For example, at the characteristic mass M$_*\sim$10$^{10.75}$ M$_{\odot}$, the difference is $\sim$0.12 dex. Thus, we consider that both estimates of the SFR yield similar characterizations of the SFMS at the redshift of the objects once the differences between them are understood. { The slopes derived for both esimates of the SFMS
 are within the range of values reported in the literature for low redshift samples: %the same redshift range: 
 0.77 \citep{elbaz07}, 0.65 \citep{Salim07}, 0.35$\pm$0.09 \citep{chen09}, 0.63-0.77 \citep{oliver10}, $\sim$1  \citep{elbaz11}, 0.67 \citep{Whitaker+2012}, 0.71$\pm$0.01 \citep{zahid12}, 0.63 \citep{sanchez13}, 0.76$\pm$0.01 \citep{Renzini15} and 0.81$\pm$0.02 \citep{mariana16}.} 

%%%%%%%%%%%%%%%%%%%%%%%%%%%%%%%%%%%%%%%%%%%%%%%%%%%
% SFMS_SF
\begin{figure}
  \centering
    \includegraphics[width=9.25cm, clip, trim=70 0 0 0]{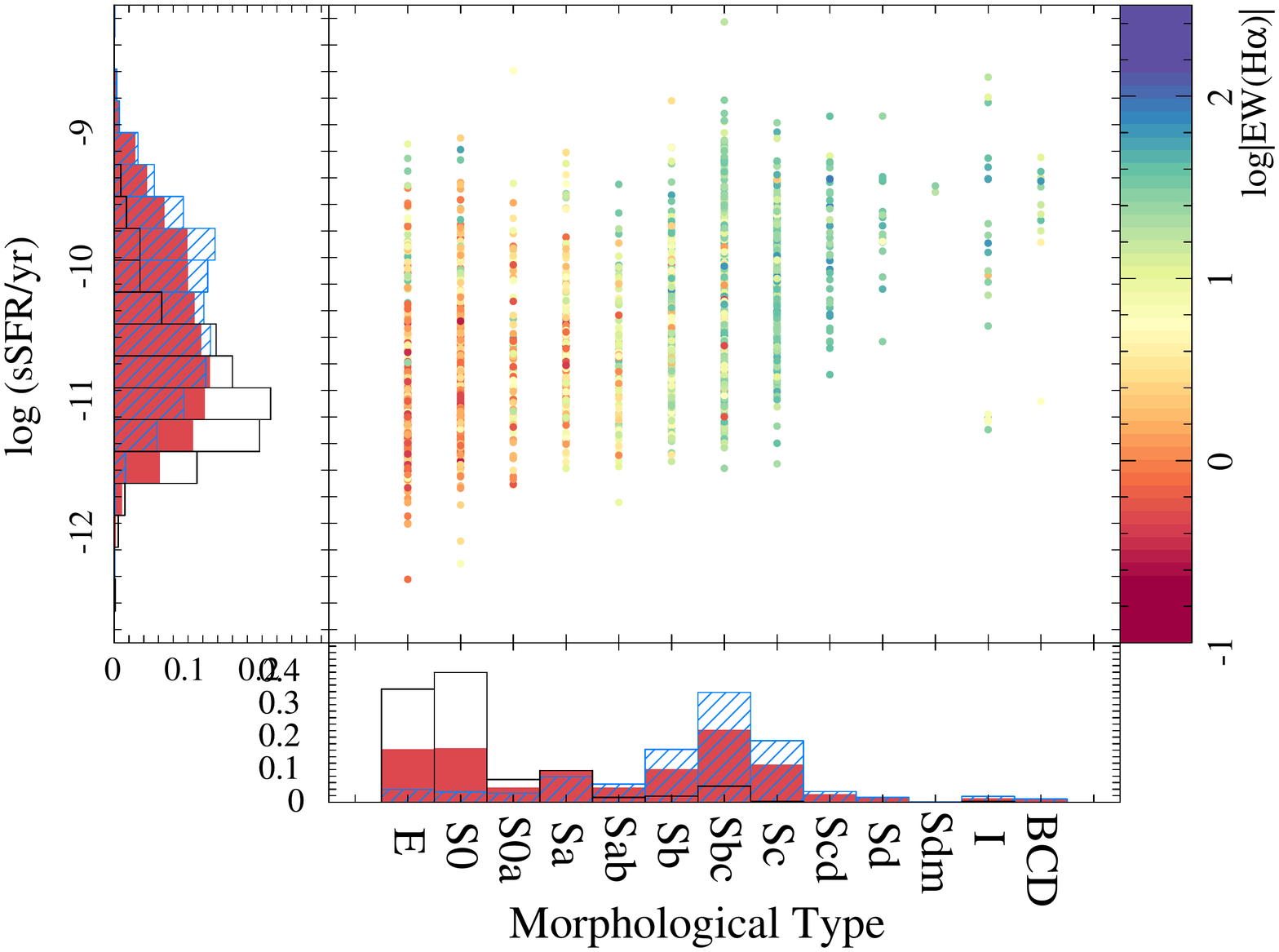}
    \caption{ Distribution of specific star-formation rates versus the morphological type for the full sample of galaxies, color coded by their average EW(H$\alpha$). The normalized histograms of each parameter for the full sample (solid red), star-forming galaxies (hashed light blue), and retired galaxies (open black), are also included.}
  \label{fig:morph}
\end{figure}
%%%%%%%%%%%%%%%%%%%%%%%%%%%%%%%%%%%%%%%%%%%%%%%%%%%

To study the evolution of the SFR on cosmological time-scales we will use not only the local population of star-forming galaxies (i.e., SFGs$_0$), as explained above, but also the population of retired galaxies at z$\sim$0  (RGs$_0$). For doing so, we follow \citet{sta08} and \citet{cid-fernandes10}, and we select those galaxies for which the average EW(H$\alpha$)$<$3\AA: i.e., the remaining sample once removing the SFGs$_0$. This limit recently has been demonstrated to be a good one to select galaxies and regions in galaxies dominated by diffuse, ionized gas not compatible with star formation \citep[e.g.][]{sarzi10,papa13,singh13,Gomes16a,mariana16,Lacerda18}. 

%%%%%%%%%%%%%%%%%%%%%%%%%%%%%%%%%%%%%%%%%%%%%%%%%%%
% SFMS_t
\begin{figure*}
  \centering
    \includegraphics[width=17.5cm]{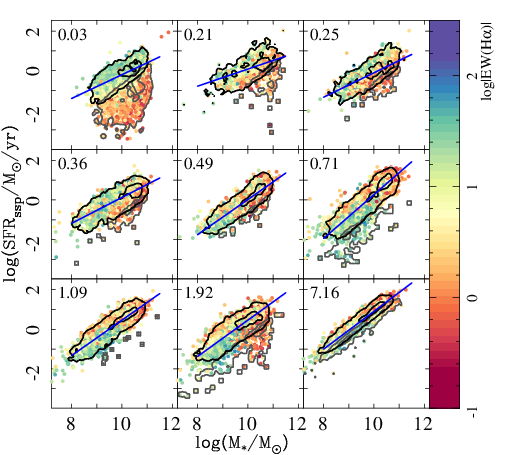}
    \caption{Distribution of the star-formation rate (SFR$_{\rm ssp}$) versus integrated stellar mass (derived from on the analysis of the stellar populations) for each galaxy in the sample at different redshift ranges. For clarity, only those points with an error in the SFR lower than 0.2 dex have been included. The average redshift of each selected range is shown in each panel, and increases from left to right and from top to bottom. Each galaxy in each panel is shown as a solid circle colored according to the EW of H$\alpha$ measured at $z\sim 0$. { Black } contours indicate the density for the SFGs in each redshift bin { while grey contours indicate the density of RGs, both} selected as indicated in the text. Inner and outer contours encircle 25\% and 95\% of the galaxies, respectively, for each type. { Dark-blue solid} lines show the linear fit for the SFMS derived at each redshift range (see text for details). Note that for each particular galaxy within each bin the cut { applied to classify it as SFG or RG } is different, since it depends on its actual redshift, { which explains the overlap between the grey and black contours in the boundary region.} }
  \label{fig:SFMS_t}
\end{figure*}
%%%%%%%%%%%%%%%%%%%%%%%%%%%%%%%%%%%%%%%%%%%%%%%%%%%

%%%%%%%%%%%%%%%%%%%%%%%%%%%%%%%%%%%%%%%%%%%%%%%%%%%
% SFMS_t
\begin{figure*}
  \centering
    \includegraphics[width=17cm, clip, trim=0 100 280 20]{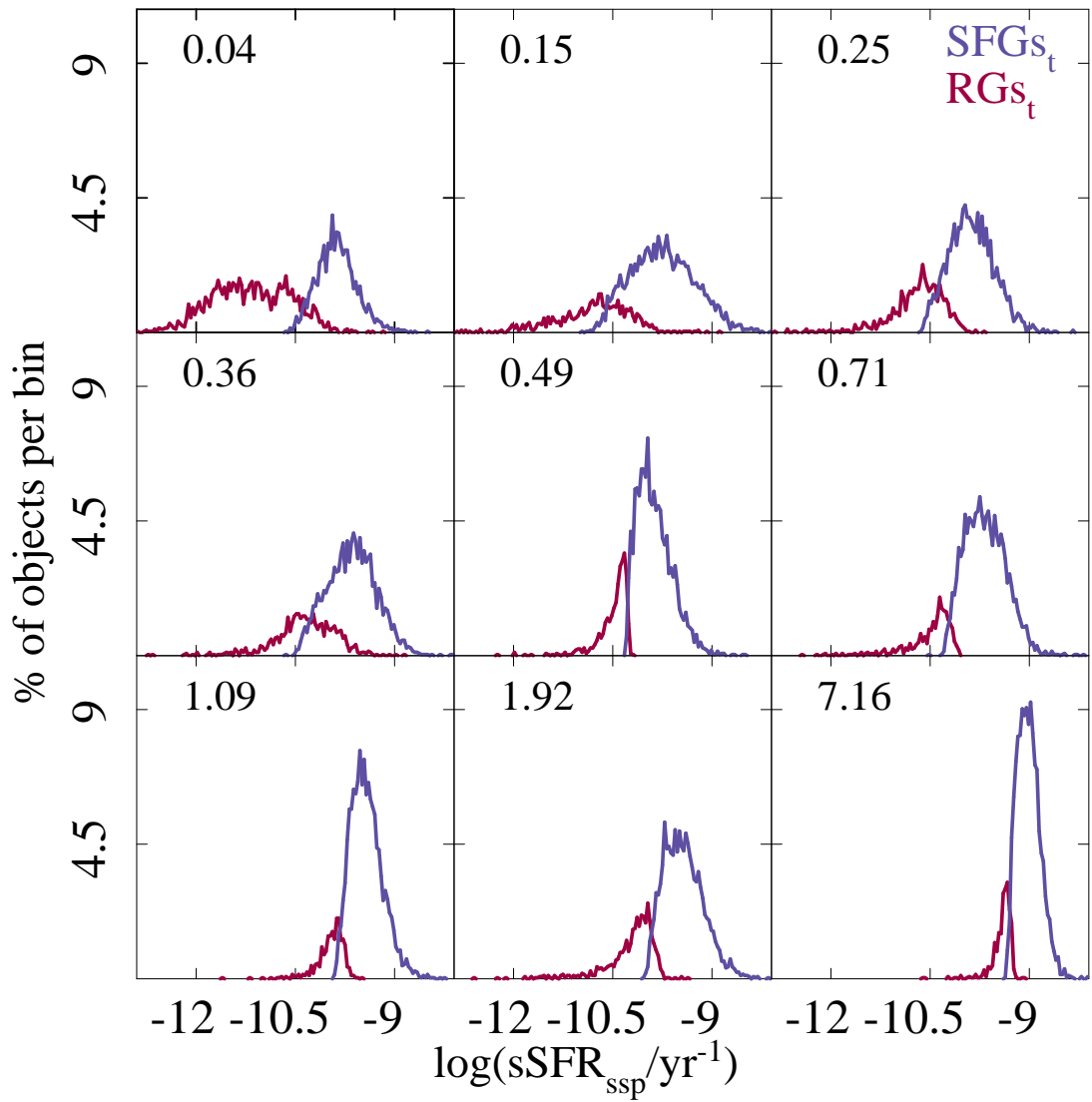}
    \caption{ Histogram of the SFRs for the {\it galaxies} classified as retired (RG$_t$, red line) or star-forming (SFGs$_t$, blue line) in each of the redshift bins shown in Fig. \ref{fig:SFMS_t}, derived from our analysis of the {\it synthetic} catalog described in the text. The mean redshift of each bin is indicated in each panel. The Figure illustrates the evolution of the fraction of retired and star-forming galaxies with cosmic time.}
  \label{fig:SFMS_t_den}
\end{figure*}
%%%%%%%%%%%%%%%%%%%%%%%%%%%%%%%%%%%%%%%%%%%%%%%%%%%

We should indicate that, in essence, this selection criteria between SFGs and RGs is actually selecting galaxies of different morphologies. Figure \ref{fig:morph} shows the distribution of the sSFR with  morphological type, color-coded using the average EW(H$\alpha$), for the subsample of $\sim$2,500 galaxies for which we have a performed a morphological analysis \citep{sanchez17b}. The figure shows clearly that most of the RGs are early-type galaxies, mostly E/S0s, while most of the SFGs are late-type ones (mostly Sb/Sbc/Sc). Thus, in this article when discussing the properties of RGs we should keep in mind that we are basically discussing the properties of today's early-type galaxies, and when discussing the properties of SFGs, we refer mostly to late-type galaxies. The figure also illustrates the correspondence between the EW(H$\alpha$) and the sSFR, described earlier.

%Following a similar argument we select the retired galaxies at 

\subsection{Cosmological evolution of the SFR-M$_*$ diagram}
\label{sec:SFMS_t}

Having characterized the distribution of galaxies in the SFR-M$_*$ diagram (in particular the shape of the SFMS) at the redshift range of the considered sample, and having checked that our estimate of the SFR based on the analysis of the stellar populations is as good as more standard procedures (such as the H$\alpha$ luminosity), we now explore the change of the SFR-M$_*$ diagram accross cosmic time.

To do so we treat our complete sample of 155,838 individual estimates of M$_{*,t}$ and SFR$_{ssp,t}$ as if they comprise a cosmological survey. { As explained in Sec. \ref{sec:SFR} we constructed that synthetic sample by estimating the stellar mass and the SFR at 38 different look-back times for our initial sample of 4101 galaxies based on our multi-SSP fitting procedure. If we consider this synthetic sample as a survey, to explore the evolution with cosmic time we should split the sample in redshift bins that (a) guarantee the required number of objects to do a proper statistical analysis; (b) the resulting sub-sample in each redshift bin should cover well the SFR-M$_{*}$ diagram; and (c) there should be a reasonably good sampling of  redshift  to trace the cosmological evolution. Based on these requirements, and considering that we do not have a homogeneous sampling in redshift  (due to the discrete nature of the distribution of ages within our SSP library), we split the synthetic sample } in nine redshift bins with the following criteria: (i) Each bin must contain at least the same number of galaxies as the original sample (4101). { We have shown that this number is large enough to obtain a good description of the distribution along the SFR-M$_{*}$ diagram, and to characterize well the SFMS}. (ii) The upper-limit of each redshift bin should be at least $\Delta$z=0.07 away from that of the previous bin. The first criterion is restrictive for high redshift bins, where our sampling of the temporal domain is more discrete. On the other hand, the second criteria is particularly restrictive at low redshift, where we have a more continuous sampling in the time domain. { The selected redshift range corresponds roughly to half  the  range covered by the original MaNGA sample. Adopting this range  ensures we are not dominated by galaxy repetition in each analyzed bin, at least at low redshift.}
%With these two criteria we would like to maximize how representative are the sub-samples in each redshift bin and sampling in the best possible way the explored redshift range.
We note that neither criteria are particularly restrictive and they do not affect the results so long as enough {\it galaxies} are included in each redshift bin to sample parameter space well. 
%In particular, they can be relaxed if you are interested in sampling just one of the considered parameters (M$_*$ or SFR).
{ Indeed, repeating the same experiment with different numbers of objects we found that with $\sim$500 {\it galaxies} in each redshift bin we can reproduce all the current results if they are distributed along the SFR-M$_{*}$ in a similar way as the original sample (i.e., if they are a representative subsample in this parameter space).} However, to ensure a good characterization of the distribution along the SFR-M$_*$ diagram, and in particular to study the SFMS across  cosmic time, we have been rather conservative in the selection of these redshift bins.

Figure \ref{fig:SFMS_t} shows the results of this analysis. For each redshift bin, we present  SFR$_{ssp,t}$ versus M$_{*,t}$, labeling each galaxy by its EW(H$\alpha$) at $z\sim$0 (the values directly observed, shown in Fig. \ref{fig:SFMS}). We use this parameter as a direct observational proxy of their {\it final} evolutionary stage (star-forming or retired galaxies) in the local Universe (as shown in Sec. \ref{sec:SFMS_0}): galaxies with EW(H$\alpha$)$<$3\AA\ are considered as retired at $z\sim 0$ (the RGs$_0$ sample; reddish in Fig. \ref{fig:SFMS_t}) while galaxies with EW(H$\alpha$)$>$3\AA\ are considered star-forming at $z\sim 0$  (the SFGs$_0$ sample; blueish in Fig. \ref{fig:SFMS_t}), as explained above. We remind the reader that this cut corresponds well to a cut in sSFR$_0$ at 10$^{-10.8}$ yr$^{-1}$, for our currently adopted IMF, based on the correlation between both parameters \citep[e.g.][]{sanchez13,Belfiore17a}.

We should note that the first two redshift bins sample times within the redshift range of the original observed sample. Therefore, they do not comprise exactly the same galaxies as the remaining bins: i.e., 'low' redshift galaxies ($z<z_{max,bin}$) could be sampled several times in both bins, while 'high' redshift galaxies ($z>z_{max,bin}$) are absent by construction. However, this effect is not particularly strong, since most of MaNGA galaxies ($\sim$80\%) are located at $z<0.06$ (the upper-redshift of the lower redshift bin), and basically all absent galaxies are RGs$_0$ (due to the MaNGA sample selection). { Another caveat to be taken into account is that the diagrams corresponding to the first two redshift bins cannot be easily compared with the distribution shown in Fig. \ref{fig:SFMS}, since the number of galaxies is far larger and therefore visual inspection may lead to wrong conclusions.}

%%%%%%%%%%%%%%%%%%%%%%%%%%%%%%%%%%%%%%%%%%%%%%%%%%%
% SFMS_SF
\begin{figure}
  \centering
    \includegraphics[width=10cm]{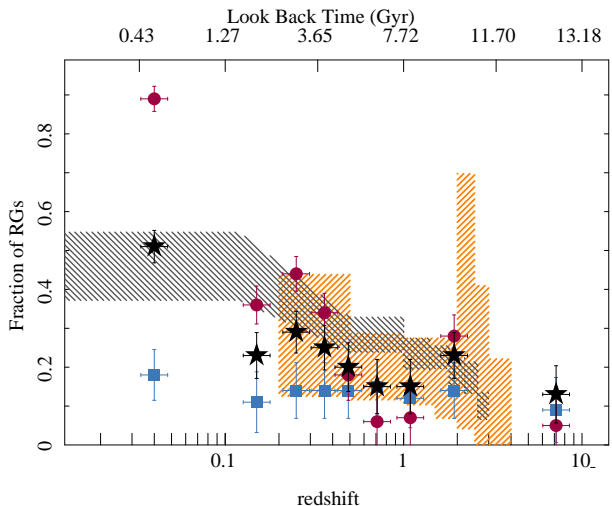}
    \caption{Fraction of RGs as a function of redshift, for all the analyzed galaxies (black solid stars), the currently star-forming (blue solid squares) and retired (red solid squares) galaxies, with their corresponding errors (1$\sigma$ range). The shaded { orange} regions correspond to estimates of the RG fractions from \citet[][see text for details]{Muzzin+2013}, { while the shaded gray regions correspond to the fraction of retired galaxies estimated from \citet{pand17}}.}
  \label{fig:frac}
\end{figure}
%%%%%%%%%%%%%%%%%%%%%%%%%%%%%%%%%%%%%%%%%%%%%%%%%%%

Despite this caveat we can see a clear evolution in the distribution of galaxies within the SFR-M$_*$ diagram with cosmic time. Two main trends arise: (i) The population of retired galaxies at a particular time (RGs$_t$) becomes less common at higher redshifts. { This is shown in Fig. \ref{fig:SFMS_t}, but it requires quantification, as we will discuss later}. (ii) For the star-forming galaxies at a particular time (SFGs$_t$), the SFR at a fixed mass increases with redshift. Both trends were already known, discovered in studies based on cosmological surveys \citep[e.g.][]{Karim+2011,Whitaker+2012,Speagle14}. They reflect that galaxies were more actively forming stars individually in the past and that, in general, the cosmic SFR density was larger in the overall \citep[e.g.][]{lilly96,madau96,Hopkins2006ApJS..163....1H,fardal07,Madau14}. The former is particularly true for high rather than low mass galaxies, resulting in a steeper SFR--M$_*$ relationships at high redshifts. Note that these trends were studied using archaeological methods only once before by \citet{lopfer18}, and using a much reduced sample of galaxies. 

\subsubsection{Fractions of retired galaxies at different redshifts}

{ We have applied an initial redshift-dependent cut in the sSFR following \citet{Pacifici16} to quantify the fraction of RGs at any redshift }:
\begin{equation}
sSFR_{\rm lim}(z) = 0.2/t_H(z),
\label{ssfr-lim}
\end{equation}
where $t_H(z)$ is the cosmic time at redshift $z$. For the cosmology assumed here, $sSFR_{\rm lim}=10^{-10.84}$ yr$^{-1}$ at $z=0$, in excellent agreement with the criterion based on the observed EW(H$\alpha$) used above to divide the \textit{local} population into star-forming and retired galaxies (the samples SFGs$_0$ and RGs$_0$, respectively). { We recall here that both quantities are equivalent, even if the SFR$_{ssp}$ is adopted, as we indicated in Sec. \ref{sec:SFMS_0}.} Thus, for sSFR values at a given time above (below) the limit given in Equation (\ref{ssfr-lim}), we consider that these galaxies are { the first candidates to be classified as } star-forming (retired) at that time and we refer to them as SFGs$_t$ (RGs$_t$). For { this initial selection of} SFGs$_t$, we derive the SFMS at different cosmic times (blue lines in Fig. \ref{fig:SFMS_t}). Once the SFMS is established at any time, we adopted a second cut to classify SFGs$_t$ (RGs$_t$), excluding in an iterative way those galaxies that deviate more than 2$\sigma$ from the derived SFMS relation, or including them if the sSFR is within 2$\sigma$ of this relation. The latter happens only for the highest redshift bins ($z\sim 7.16$), for which we consider that the fraction of star-forming and retired galaxies is not very well defined. { The second cut -- more relevant in the higher redshift bins -- does indeed depend on the stellar mass. Therefore, the final selection of RGs$_t$ comprises two criteria including those galaxies that either (a) lie below the sSFR cut indicated before or (b) lie below 2$\sigma$ of the SFMS, whichever is more restrictive. Their density is represented as gray contours in Fig. \ref{fig:SFMS_t}. The remaining galaxies are classified as SFGs$_t$, and their density is represented as black contours in Fig. \ref{fig:SFMS_t}. The selection criteria described above are close to those adopted in studies based on cosmological surveys \citep[e.g.][]{pand17}, as we will see later.}

{ Figure \ref{fig:SFMS_t_den} illustrates the result of this analysis. It shows the histograms of sSFR of the galaxies classified as retired and star-forming at each epoch (red and blue lines, respectively) following the two cuts described above. Each panel corresponds to the different redshift bins described before (i.e., those shown in Fig. \ref{fig:SFMS_t}). According to the first cut (Eq.\ref{ssfr-lim}), both groups should be separated only by a limiting sSFR at each $z$. The second cut introduces a mass dependence, and this is why the sSFR histograms of SFGs$_t$ and RGs$_t$ have some overlap in each redshfit bin. 
%It shows the histograms of the sSFR of the galaxies classified as retired and star-forming following the previous criteria for each of the subsamples corresponding to the different redshift bins described before (i.e., those shown in Fig. \ref{fig:SFMS_t}). The distributions show the result of the two selection cuts applied, one based on a sSFR threshold dependent on the redshift of the target, and a second one based on the distance to the location of the SFMS. 
The histograms clearly illustrate the evolution of the fraction of RGs$_t$ (F$_{RG}$) to SFGs$_t$ ($1-$F$_{RG}$) accross cosmic time for the complete sample of galaxies. This analysis can be further performed for the subcategories of galaxies classified as star-forming and retired at $z\sim$0 (SFGs$_0$ and RGs$_0$, respectively), to explore the differences in their evolution.}
%The histograms clearly illustrate the evolution of the fraction of SFGs$_t$ and RGs$_t$ accross cosmic times for the complete sample of galaxies, that can be quantified by the fraction of galaxies in each subcategory. The analysis was performed for the galaxies classified as star-forming and retired at $z\sim$0 (SFGs$_0$ and RGs$_0$, respectively), to explore the differences in their evolution.}

{ Figure \ref{fig:frac} shows the derived distribution } of the fraction of RGs as a function of look-back time, t$_{\rm lb}$, for the full sample of galaxies (black stars), and for the two subsamples defined as star-forming (blue squares) and retired (red circles) galaxies at $z\sim 0$.  These values, { extracted from the analysis illustrated in Fig. \ref{fig:SFMS_t_den},} are reported in Table \ref{tab:SFMS_t}, { together with the values found at $z\sim$0 reported in the previous sections for comparison purposes. The reported evolution is now clearly shown in a quantitative way.} In general, we find that the fraction of RGs$_t$ decreases with redshift , although not in a steady way (i.e., there are fluctuations along the main trend). The evolution is consistent with the fraction of RGs$_0$ described in previous sections ($\sim$59\%), { showing a clear increase in the last 4 Gyr ($z<$0.5).} 

Our results are { consistent} with { estimates of the fraction of quiescent/retired galaxies from various cosmological surveys \citep[e.g.,][]{Muzzin+2013,Tomczak+2014,Martis+2016,pand17}. In particular, in Fig. \ref{fig:frac} we compare our results with} those obtained from the COSMOS/UltraVISTA field survey \citep[][orange shaded region]{Muzzin+2013} { and the values reported based on the analysis of the GAMA and CANDELS surveys \citep[grey shaded region]{pand17}}. \citet{Muzzin+2013} report the fraction of RGs$_t$ as a function of mass and $z$. To compare with them we weight masses at a given $z$ with the galaxy stellar mass function at that $z$ to obtain the global fractions as a function of $z$ only. { We propagate the errors; the shaded region in the Figure corresponds to the regime between $\pm\sigma$ around the mean value within each redshift bin (shown as the size of each box) explored by those authors. For the values reported by \citet{pand17}, we adopt the global fractions listed in their Table 4. They classified galaxies in three groups based on their distance to the SFMS at a particular redshift. Galaxies with a SFR larger than 1.5$\sigma$ below the location of the SFMS were classified as star-forming, while galaxies with a SFR lower than 3.5$\sigma$ were classified as retired. Galaxies between the two regimes were classified as transition objects. This classification is not exactly the same as the one presented here, where we use, in practice, a 2$\sigma$ threshold below the location of the SFMS to classify galaxies as star-forming or retired. To compare with our results we consider that only 50\% of their galaxies classified as transition objects would be classified as RGs, while a 50\% would be classified as SFGs. Accordingly, we have combined their two fraction of objects to generate the values shown in Fig. \ref{fig:frac}, with the lower limit corresponding to their reported fraction of RGs minus 1$\sigma$ of their reported error, and the upper limit being their reported fraction of RGs plus a 50\% of transition objects, in addition to 1$\sigma$ of their reported error. In general, our results are consistent with those reported by \citet{Muzzin+2013} and \citet{pand17}. In the first case the fractions of RGs are always within the regime covered by their fractions, showing even a similar increase at $z\sim$2. In the second case our values agree within the errors for the lowest redshift range and at $z\sim$2. Our fraction of retired galaxies follows a similar decreasing shape with redshift, but in general they are $\sim$1$\sigma$ below their reported values for most of the explored redshift regimes.}

{ Going beyond what it was explored by cosmological surveys, we} find that the trend of the RG fraction with $z$ is not universal and is not steady. First, the progenitors of local RGs$_0$ seem to follow the general trend, with a larger fraction of them being SFGs in the past, at least below $z<$1.5 (t$_{\rm lb}<$8 Gyr). However, for the progenitors of local SFGs$_0$, the fraction of RGs$_t$ seems to be rather constant with redshift, reflecting that indeed some of them were retired or less active at high redshift. This fraction is $\sim$15\% at any redshift bin. By construction, it is zero at $z\sim$0. Beyond the nominal error estimated by our procedure, we would be cautious about the significance of any fraction below a 10\%, in particular when the defined boundary between SFGs/RGs is near to the location of the SFMS at the considered redshift. Thus, in general we conclude that the fraction of local SFGs$_0$ that fluctuate in an out of the SFMS$_t$ is of the order of a 10\%\ at all times. { This result cannot be contrasted with cosmological surveys, since by construction, they cannot sample the same galaxies, and trace whether they will become retired or star-forming in the local Universe. However, they can be compared with the expectations from semi-analytical models. Indeed, \citet{pand17} explored that possibility and reported that 13\% of their galaxies classified as SFGs$_0$ and 31\% of the ones classified as RGs$_0$ have experienced some kind of rejuvenation in their SFHs since z$\sim$3. Actually, these fractions would be $\sim$25\% for SFGs$_0$ and $\sim$44\% for RGs$_0$,
if we consider the fifty-fifty sharing of galaxies classified as transition objects by \citet{pand17} between the two groups.}

The second result to highlight is that the fraction of RGs$_t$ seems to increase slightly at $z\sim$2 for both the full sample and for that of local retired galaxies only (RGs$_0$). Interestingly, the results from {\citet{Muzzin+2013}} also shows this feature. We should be cautious about this result for two reasons: (i) the last two redshift bins are in a regime
where the real redshift or cosmological distance sampled by our archaeological method have significant uncertainties, as we will discuss below; and (ii) the boundary between SFGs$_t$ from RGs$_t$ at this redshift is closer to the location of the SFMS$_t$ than at any other redshift ranges, and therefore some of the galaxies classified as retired could well be SFG ones, particularly at the lower end of the SFMS distribution. However, a visual inspection of Fig. \ref{fig:SFMS_t} indicates that indeed the fraction of RGs in this redshift bin seems to be enhanced compared to the adjacent redshift bins. { In any case, this result is limited by our adopted procedure and the reliability of our fractions at that high redshift, and should be explored in detail prior to drawing a firm conclusion.} Thus, the only case in which we { could} claim that we detect a significant population of RGs at very high redshift is at $z\sim$2. { This result is also shown in Fig. \ref{fig:SFH}, where a drop in the SFR is present at about the same redshift.} Further, some of the retired galaxies at this redshift become star-forming again at lower redshifts, { indicated as well by the trend in the mean SFH}. { While we refrain from making a firm conclusion, if these trends in SFH and RG fraction are confirmed, this may indicate} that quenching is not a one-way process at this epoch,{ consistent with the results presented by \citet{pand17} based on their semi-analytical models.} Indeed, exploring the individual behavior of each galaxy in terms of their location within the SFR-M$_*$ diagram we see that it is not uncommon for galaxies to become retired and and then return to star-forming several times. We will explore in more detail these individual tracks in a companion article (Ibarra-Medel et al. in prep.). So far, we should keep in mind that the average fraction of RGs$_t$ declines at high redshift, and that decline is dominated by those galaxies defined as retired at z$\sim$0 (RGs$_0$). { How this decline depends on other properties of the galaxies, such as stellar mass (or morphology), is an  important topic addressed in previous studies \citep[e.g.][]{rgb17,rosa17}. Clear differences in the average SFHs with redshift shown in Fig. \ref{fig:SFH} indicate that these properties (e.g., mass) strongly correlate with their evolution. We will explore these correlations in future analysis, now that we have established the validity of our method to explore the evolution of the SFR activity over a wide range of redshift. The possible rise of retired galaxies at $z\sim$2 should be considered a tentative result at this time.} 
{ We stress out that although we have
detected a possible rise of the fraction of RGs in our analysis, and that this
rise seem to be present in at least one cosmological survey (Muzzin et
al. 2013), it is possible that this result is a spurious effect of
the uncertainties associated with the current adopted method at high redshift. }

%---------------------------------------------------
%---Table----
\begin{table}
\caption{Results of the analysis of the SFMS for each redshift bin and galaxy subsample, including: (1) the average redshift of each bin; (2) the zero-point and slope of the best-fit log-linear regression between the SFR and the M$_*$ for the SFGs at each redshift, with their corresponding errors; (3) the SFR at a mass of 10$^{10.75}$ M$_\odot$, with errors corresponding to half the dispersion in the SFMS; and (4) the fraction of retired galaxies. { For comparison purposes we include in the two first rows the same parameters derived for galaxies at $z\sim$0 extracted from the distributions shown in Fig. \ref{fig:SFMS}, for both the SFRs derived using the H$\alpha$ flux and the SSP-analysis. }}
\label{tab:SFMS_t}
%\footnotesize
%\begin{center}
\begin{tabular} {l c c c c}
\hline
Selection          &  SFR$_0$     & slope    & SFR$_{10.75}$      & F$_{RG}$ \\
Parameter                &  log(M$_\odot$/yr)    &     & log(M$_\odot$/yr)      &  \\
\hline
SFR& \multicolumn{4}{c}{Original sample at z$\sim$0}\\
\hline
%###### SFMS_t ALL
H$_{\alpha}$     &  -8.96$\pm$0.23 & 0.87$\pm$0.02 & 0.32$\pm$0.21 & 0.59 \\
SSP            &  -6.32$\pm$0.17 & 0.61$\pm$0.02 & 0.28$\pm$0.21 & 0.58 \\
\hline
z & \multicolumn{4}{c}{All galaxies at different redshifts}\\
\hline
%###### SFMS_t ALL
0.04 &  -6.47$\pm$0.32 & 0.66$\pm$0.03 & 0.57$\pm$0.21 & 0.51 \\
0.15 &  -5.55$\pm$0.30 & 0.55$\pm$0.03 & 0.38$\pm$0.21 & 0.22 \\
0.25 &  -6.97$\pm$0.43 & 0.69$\pm$0.04 & 0.42$\pm$0.16 & 0.27 \\
0.36 &  -5.78$\pm$0.32 & 0.50$\pm$0.03 & 0.62$\pm$0.16 & 0.27 \\
0.49 &  -9.78$\pm$0.22 & 0.98$\pm$0.02 & 0.72$\pm$0.12 & 0.21 \\
0.71 & -10.82$\pm$0.32 & 1.11$\pm$0.03 & 1.10$\pm$0.21 & 0.15 \\
1.09 &  -8.39$\pm$0.29 & 0.88$\pm$0.03 & 1.11$\pm$0.13 & 0.15 \\
1.92 &  -8.51$\pm$0.27 & 0.90$\pm$0.03 & 1.14$\pm$0.19 & 0.25 \\
7.16 &  -8.67$\pm$0.17 & 0.96$\pm$0.02 & 1.62$\pm$0.12 & 0.13 \\
%###### SFMS_t SF
\hline
z & \multicolumn{4}{c}{SFGs$_0$ at different redshifts}\\
\hline
0.04 &  -6.99$\pm$0.40 & 0.71$\pm$0.04 & 0.67$\pm$0.20 & 0.18\\
0.15 &  -6.73$\pm$0.42 & 0.69$\pm$0.04 & 0.72$\pm$0.19 & 0.11\\
0.25 &  -8.67$\pm$0.27 & 0.87$\pm$0.03 & 0.70$\pm$0.19 & 0.13\\
0.36 &  -6.54$\pm$0.22 & 0.68$\pm$0.02 & 0.75$\pm$0.13 & 0.16\\
0.48 & -10.67$\pm$0.32 & 1.07$\pm$0.03 & 0.83$\pm$0.18 & 0.15\\
0.71 & -11.65$\pm$0.24 & 1.19$\pm$0.02 & 1.09$\pm$0.12 & 0.15\\
1.09 &  -8.74$\pm$0.26 & 0.92$\pm$0.03 & 1.13$\pm$0.12 & 0.13\\
1.92 &  -8.85$\pm$0.19 & 0.92$\pm$0.02 & 1.08$\pm$0.17 & 0.16\\
7.16 &  -8.70$\pm$0.21 & 0.96$\pm$0.02 & 1.58$\pm$0.10 & 0.09\\
%###### SFMS_t RG
\hline
z & \multicolumn{4}{c}{RGs$_0$ at different redshifts}\\
\hline
0.05 &  -7.78$\pm$0.57 & 0.76$\pm$0.06 & 0.37$\pm$0.27 &  0.89 \\
0.15 &  -6.17$\pm$0.35 & 0.60$\pm$0.04 & 0.23$\pm$0.17 &  0.36 \\
0.25 &  -7.02$\pm$0.23 & 0.68$\pm$0.03 & 0.26$\pm$0.22 &  0.40 \\
0.36 &  -5.08$\pm$0.18 & 0.51$\pm$0.02 & 0.43$\pm$0.18 &  0.34 \\
0.49 &  -8.16$\pm$0.15 & 0.82$\pm$0.02 & 0.64$\pm$0.12 &  0.19 \\
0.71 &  -8.55$\pm$0.27 & 0.90$\pm$0.03 & 1.06$\pm$0.16 &  0.06 \\
1.09 &  -6.30$\pm$0.19 & 0.68$\pm$0.02 & 1.04$\pm$0.12 &  0.08 \\
1.92 &  -4.77$\pm$0.26 & 0.52$\pm$0.03 & 0.86$\pm$0.15 &  0.28 \\
7.16 &  -7.56$\pm$0.20 & 0.85$\pm$0.02 & 1.59$\pm$0.11 &  0.05 \\
%#~ & \multicolumn{2}{c}{\hline} & & \multicolumn{2}{c}{\hline} & \\
%\cmidrule{2-3}
%\cmidrule{5-6}
%O3N2-M13 & 0.077 & 8.53 $\pm$ 0.04 & 0.003 $\pm$ 0.037 & 0.060 & -0.02 $\pm$ 0.01 & -0.007 $\pm$ 0.005 & 0.061 \\
\hline
%\bottomrule
\end{tabular}
%\end{center}
\end{table}
%---------------------------------------------------

%%%%%%%%%%%%%%%%%%%%%%%%%%%%%%%%%%%%%%%%%%%%%%%%%%%
% SFMS_t
\begin{figure}
  \centering
    \includegraphics[width=10cm]{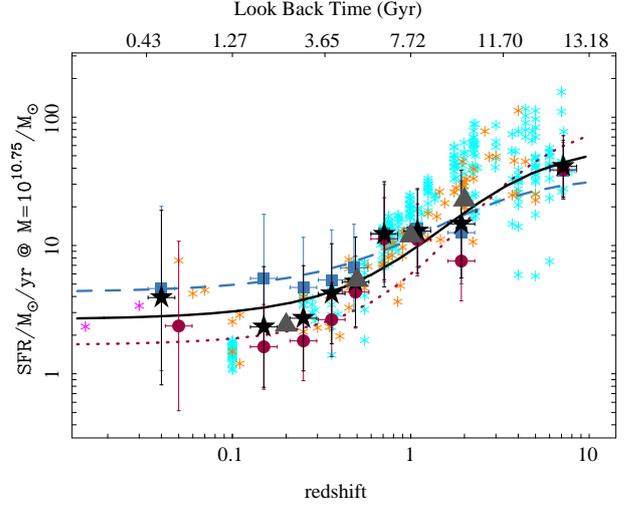}
    \caption{SFR at log(M$_{*}$)=10.75 dex versus redshift for each redshift bin shown in Fig. \ref{fig:SFMS_t}, \ref{fig:SF_SFMS_t} and \ref{fig:RG_SFMS_t} for all the galaxies in the sample (black solid stars), SFGs at $z=0$ (SFGs$_0$; blue solid squares) and RGs at $z=0$ (RGs$_0$; red solid circles).  Error bars indicate the dispersion around each derived SFR. Each line represents the best linear fit of SFRs versus look back time for the complete sample (black solid line), the SFGs$_0$ (blue dashed line) and the RGs$_0$ (red dotted line). { Stars represent the values compiled from the literature by \citet[][orange]{Speagle14} and \citet[][cyan]{Rodriguez-Puebla+2017}.
%, and magenta from \citet{Renzini15} and \citet{mariana16}. Corregi varias cositas aqui, espero que bien.
They are based on different direct estimates of the SFMS from cosmological surveys, and therefore they were shifted when necessary to account for the different adopted calibration of the SFR and IMFs}. { We include the values derived corresponding to the local SFMS by \citet{Renzini15} and \citet{mariana16} as magenta stars. Grey solid triangles correspond to values reported by \citet{lopfer18} based on a parametric SFH derivation of the galaxies from the CALIFA survey. We also shift their SFR$_{10.75}$ to correct for the differences in the IMF, when required.}}
  \label{fig:SFR10}
\end{figure}
%%%%%%%%%%%%%%%%%%%%%%%%%%%%%%%%%%%%%%%%%%%%%%%%%%%

\subsection{Quantifying the evolution of the SFMS}
\label{sec:SFMS_q}

In the previous { section} we have described the qualitative evolution of galaxies in the SFR-M$_*$ diagram, showing that most of the galaxies are more actively forming stars at higher redshift. In this section we describe the actual evolution of the SFMS in a more quantitative way.

To characterize the SFMS at each redshift bin we determine the location of the peak of the density distribution of SFGs in the SFR-M$_*$ diagram, shown in Fig. \ref{fig:SFMS_t}, at different masses. This peak corresponds to the mode of the distribution, by definition. This mode was estimated for mass bins of $\Delta log(M)=0.075$ dex, within a range of masses between 10$^{8.75}$ and 10$^{11}$ M$_\odot$. Only those mass bins comprising at least a 5\%\ of the galaxies were taken into account. The standard deviation along the mode was adopted as the characteristic width of the SFMS for any redshift and mass bin, and used as errors in the subsequent analysis. Then, for each redshift, we perform a linear regression of the set of points defined by these mass bins and the SFRs defined by the peaks/modes of the distribution. The result of this analysis is presented in Fig. \ref{fig:SFMS_t}, and in Table \ref{tab:SFMS_t}, showing for each redshift bin the zero-point and slope of the linear regression, with their corresponding errors derived based on a Monte-Carlo iteration. The analysis was repeated for the two local subsamples of galaxies described before, the SFGs$_0$ and RGs$_0$, to explore if they present different evolution with redshfit. The corresponding SFMS for those subsamples have been included in Appendix \ref{sec:SF_RG}, Figures \ref{fig:SF_SFMS_t} and \ref{fig:RG_SFMS_t}, respectively.

%%%%%%%%%%%%%%%%%%%%%%%%%%%%%%%%%%%%%%%%%%%%%%%%%%%
% SFMS_t
\begin{figure}
  \centering
    \includegraphics[width=10cm]{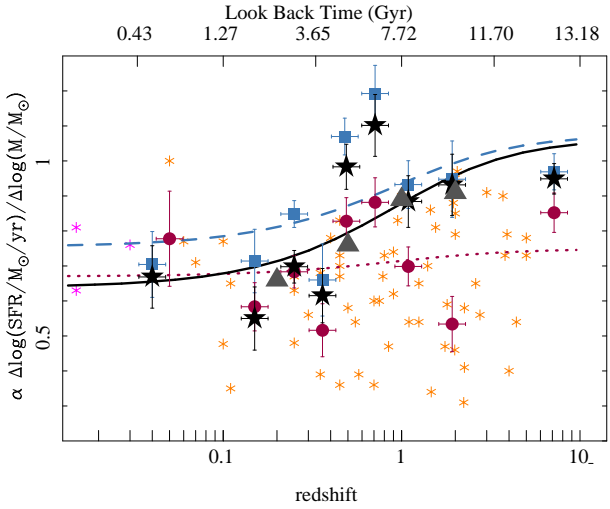}
    \caption{Slope of the SFMS for each redshift bin shown in Figs. \ref{fig:SFMS_t}, \ref{fig:SF_SFMS_t}, and \ref{fig:RG_SFMS_t} for all the galaxies in the sample (black solid stars), the local SFGs (blue solid squares) and the local RGs (red solid circles). The error bars indicate the error in the derived slope. Each line represents the best linear fit of SFR versus look0-back time for the complete sample (black solid line), the SFGs$_0$ (blue dashed line) and the RGs$_0$ (red dotted line). { The orange stars represent values compiled by \citet{Speagle14} based on different direct derivations of the SFMS from various cosmological surveys, and magenta stars correspond to the values reported by \citet{sanchez13}, \citet{Renzini15} and \citet{mariana16}, respectively. Finally, the grey triangles correspond to values reported by \citet{lopfer18} based on a parametric SFH derivation for the galaxies from the CALIFA survey.}}
    \label{fig:alpha}
\end{figure}
%%%%%%%%%%%%%%%%%%%%%%%%%%%%%%%%%%%%%%%%%%%%%%%%%%%

As already noticed by \citet{Speagle14}, the zero-point is not a representative value for exploring the evolution of the SFMS: Since the zeropoint is an extrapolation far beyond the range of sampled masses it is strongly affected by small inaccuracies in the slope. Therefore, in addition to the zero-point, we report the predicted SFR at a characteristic stellar mass of 10$^{10.75}$M$_\odot$, based on the best-fit log-linear regression, to trace the evolution of the average SFR along the SFMS. This stellar mass is similar to the one adopted by \citet{Speagle14}, taking into account the differences in the adopted IMF. The values of this parameter (SFR$_{10.75}$, hereafter), estimated for each redshift bin and each subsample, has been included in Tab. \ref{tab:SFMS_t}. We adopt errors to be half of the dispersion around the best-fit SFMS for each bin. Figure \ref{fig:SFR10} shows the evolution of SFR$_{10.75}$ with cosmic time for the different analyzed subsamples together with the range of values reported by \citet{Speagle14}. As expected, there is a clear increase in the SFR at earlier epochs. We characterize this dependence with a log-linear relation, with these regressions for the different analyzed subsamples: 
$$log(SFR_{10.75,all}) = -0.39_{\pm 0.19}  + 0.05_{\pm 0.03} \ {\rm t_{\rm lb}}$$
$$log(SFR_{10.75,SF})  = -0.22_{\pm 0.14}  + 0.03_{\pm 0.02} \ {\rm t_{\rm lb}}$$
$$log(SFR_{10.75,RG})  = -0.54_{\pm 0.22}  + 0.06_{\pm 0.02} \ {\rm t_{\rm lb}}.$$

The best-fitting curves are shown in Fig. \ref{fig:SFR10}. They illustrate clearly that RGs$_0$ (red dotted line), which were mostly star-forming in the past, show a stronger evolution in SFR than the SFGs$_0$ (blue dashed line). Thus, not only the fraction of RGs$_t$ increases at lower redshfits, but their global SFR strongly declines too. The average evolution of the full population of galaxies (black solid line) was already noticed by \citet{Speagle14} { and more recently by \citet{Rodriguez-Puebla+2017} and \citet{lopfer18}}. We find a good qualitative and quantitative agreement between both results { as can be seen in Fig. \ref{fig:SFR10}. A comparison between the distribution of characteristic SFR derived from cosmological surveys and best-fit regression to our data yields a $\chi^2/\nu =$0.93 \citep[for][]{Speagle14} and 0.95 \citep[for][]{Rodriguez-Puebla+2017}, indicating that they are compatible with a significance level higher than $p>0.05$. If anything, we find that our estimated SFRs are slightly lower at high redshift ($z>$2) than the ones estimated by cosmological surveys, but in very good agreement with the values reported by \citet{lopfer18} using a different archaeological method.} 

As demonstrated, the SFMS is a well defined and tight relation at any redshift bin and for any galaxy subsample, with a standard deviation ranging between $\sigma\sim$0.08 and 0.32 dex, similar to the one reported by different studies \citep[e.g.]{Brinchmann04,noeske07,daddi07,elbaz11,wuyts11,Whitaker+2012,sanchez13,Renzini15,catalan15,mariana16,rosa16a}. Its slope ranges between $\alpha\sim$0.45 and 1.19, and is $<$1 in most cases. Figure \ref{fig:alpha} shows the slope values at different redshifts. Contrary to the results reported by \citet{Speagle14}, we find a clear evolution in the slope of the SFMS that increases with redshift. { This is clearly seen in Fig. \ref{fig:alpha}, where the cloud of points extracted from these authors match the reported ones only at low redshift. Similar results are found by \citet{lopfer18}, using a different procedure to recover the SFHs of galaxies from the CALIFA sample. In this particular case the agreement with our results is particularly good.} We  characterize this trend with t$_{\rm lb}$ again using a log-linear relation, and find the following values:
$$\alpha_{all} = 0.62_{\pm 0.03}  + 0.033_{\pm 0.002} \ {\rm t_{\rm lb}}$$
$$\alpha_{SF_0}  = 0.75_{\pm 0.02}  + 0.034_{\pm 0.002} \ {\rm t_{\rm lb}}$$
$$\alpha_{RG_0}  = 0.69_{\pm 0.02}  + 0.003_{\pm 0.003} \ {\rm t_{\rm lb}},$$
where $\alpha$ is the slope of the SFMS as a function of t$_{\rm lb}$. Based on our results, the slope of the SFMS tend to be closer to one at high redshift, corresponding to an exponential-$\tau$ SFH at high redshift, and sub-linear at low redshifts, corresponding to a SFH following a negative power of time (shallower than an exponential-$\tau$), as discussed by \citet{Speagle14}. The evolution is clearly shallower for the RGs$_0$ than for the SFGs$_0$, that, together with the different evolution in the characteristic SFR$_{10.75}$ indicate RGs$_0$ evolve faster at earlier epochs.

%%%%%%%%%%%%%%%%%%%%%%%%%%%%%%%%%%%%%%%%%%%%%%%%%%%
% SFMS_t
\begin{figure*}
  \centering
    \includegraphics[width=17.5cm]{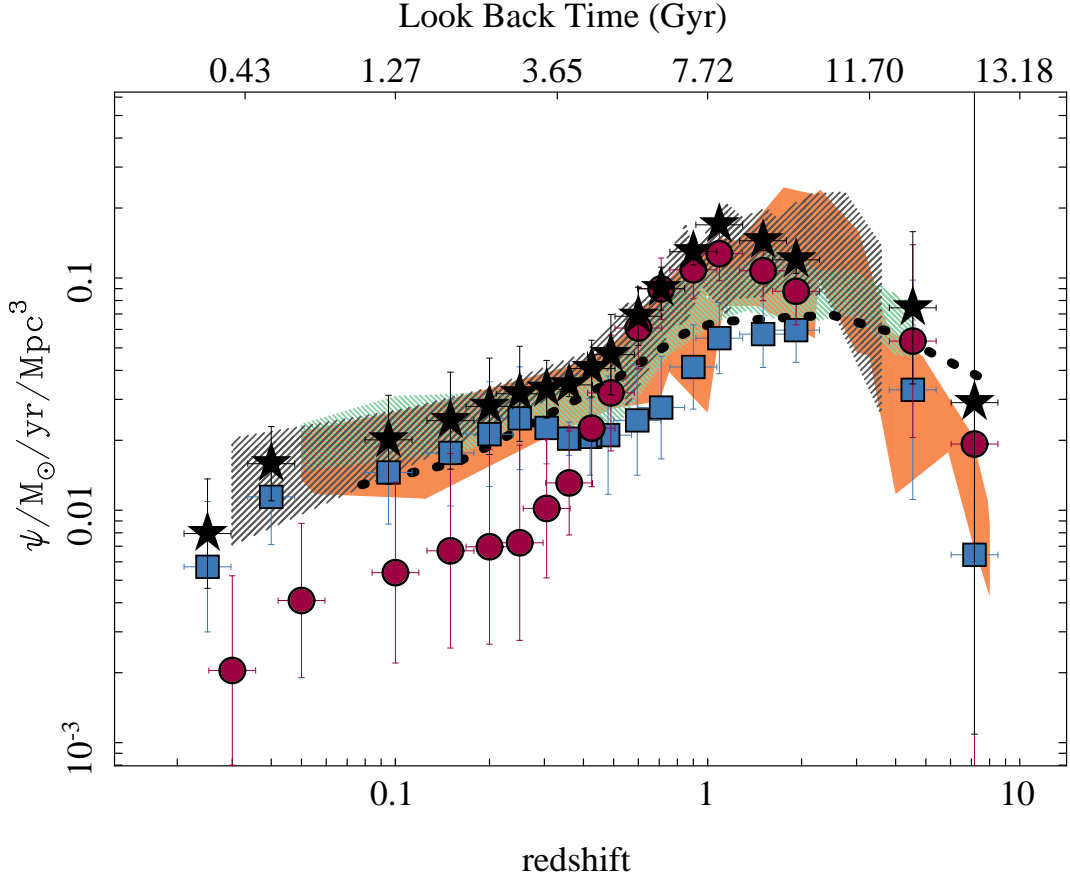}
    \caption{Cosmic evolution of the SFR density derived for all analyzed galaxies (black solid stars), SFGs$_0$ (blue solid squares) and RGs$_0$ (red solid squares), with corresponding errors (1$\sigma$ range). The shadowed regions correspond to the star-formation rate densities derived from direct observations based on cosmological surveys compiled by \citet{Madau14} (the orange solid region corresponds to FIR-derived values while the black hashed region corresponds to FUV-derived values), and \citet{Driver17} (green hashed region). The black dotted points correspond to the derivation based on archaeological methods presented by \citet{lopfer18} based on CALIFA data. When required, literature data has been shifted to account for the different adopted IMFs. The two higher redshift points for our subsamples should be viewed as uncertaint, for reasons described in the text. The average redshift of each subsample in each bin may vary depending on the actual galaxies sampled. This is particularly evident for SFGs and RGs in the lowest redshift bins.}
  \label{fig:madau}
\end{figure*}
%%%%%%%%%%%%%%%%%%%%%%%%%%%%%%%%%%%%%%%%%%%%%%%%%%%

\subsection{Cosmic Star Formation History}
\label{sec:madau}

In previous Sections we have explored how galaxies evolve within the SFR-M$_*$ diagram and
how the SFMS changes quantitatively with time based on the archaeological analysis of the stellar population of a sample of galaxies at low-redshift. Both results are consistent with those reported by cosmological surveys. The basic picture emerging from this analysis is that galaxies were more actively forming stars at earlier times, and that most of the local retired galaxies, RGs$_0$, were once star-forming at higher redshifts. In general, this picture is in agreement with the well-known
evolution of the SFR density (per co-moving volume) of the universe ($\Psi_{SFR}$), characterized by the so-called Madau curve \citep{madau96,madau98,lilly96}. This curve shows a rising of $\Psi_{SFR}$ with redshift (and LBT) up to $z\sim$2-3, and then a possible decline at very high redshift \citep[e.g.][]{Madau14,Driver17}. There remains a large debate about $\Psi_{SFR}$ at high redshift because of the discrepancy between different surveys and the large uncertainties due to dust corrections. Indeed, at high redshift most of the SFR density of the universe is inferred via UV light which, to be converted into SFRs, requires quantifying the amount of light obscured by dust. While there is considerable progress in developing empirical constraints on the amount of dust obsuration, the extact shape of the cosmic SFH remains uncertain \citep[see e.g.,][]{Reddy_Steidel2009,Reddy+2012,Bouwens+2012,Bouwens+2016}.

In principle, if our derived individual SFHs based on archaeological methods are a good representation of the real ones, it is possible to recover the $\Psi_{SFR}$ at any past cosmic time sampled by our SSP-library. It is clear that a detailed comparison between simulated and recovered SFHs is beyond the scope of this article, and will be presented elsewhere (Ibarra-Medel in prep.). We should note that the method has limitations to recover reliable SFHs beyond ages $t\ga 10$ Gyr, since the SSP templates present small differences at those ages. However, the analysis in previous sections indicate that our estimation of the SFHs are in general compatible with the known evolution of the SFR and mass based on cosmological surveys, at least up to z$\sim$2. 

To determine the cosmic SFR density at a given time, $\Psi_{SFR,t}$, we redo the analysis shown in previous sections for a larger number of bins. In this case we relax the condition for the minimum number of galaxies to be present in each redshift bin to half of the original sample ($\sim$2000 galaxies). { We should keep in mind that our previous experiments indicate that we can releact that minimum number to values as low as $\sim$500 galaxies, if they are a representative subsample of the original sample.} Then, for each redshift bin, we co-add the SFR derived for each of the star-forming galaxies at the considered redshift (i.e., those galaxies from the SFGs$_t$ subsample). Each SFR is weighted by the volume sampled by each individual galaxy, derived as describe in Appendix \ref{sec:Vmax}, based on the V$_{max}$ method. In principle, if we have sampled the full MaNGA survey in a random way this would yield $\Psi_{SFR,t}$ at the corresponding redshift bin. However, since we are sampling a different sub-sample of galaxies at different redshifts, and in some cases we are sampling the same galaxy several times in the same redshift bin, it is necessary to apply two corrections: (i) to account for the fraction of volume sampled by the corresponding subsample of galaxies (this would be just the number of sampled galaxies, $\sim$4,000, divided by the total number of the full sample, $\sim$10,000 if all galaxies had the same V$_{max}$, which is not the case); and (ii) to account for the number of repetitions ($n_i$) of the same galaxy in the current subsample. Both corrections are rather simple, and finally the cosmic SFR density can be described by the following formula:
\begin{equation}
\Psi_{SFR}(z) \equiv \Psi_{SFR,t} = f\ \Sigma_{i=0}^{n_{gal,t}} \frac{1}{n_i}\frac{SFR_{i,t}}{V_{max,i}} 
\label{eq:1}
\end{equation}
where SFR$_{i,t}$ is the current SFR of galaxy $i$ at look-back time $t$, V$_{max,i}$ is the volume of this galaxy, and $n_i$ is the number of repetitions of this galaxy in the considered redshift range. { As indicated in Sec. \ref{sec:SFMS_t}, the number of repetitions is different for each redshift bin since the time-sampling of the SSP library is more fine at lower redshifts than at higher ones. For z$<$0.1, galaxies can be repeated up to 10 times in the given redshift range of a bin, while between 0.1$<z<$2 the number of repetitions is modest to negligible in general. For the higher redshift bins, because they are wider, the repetitions range between 1 and 3. We stress that this parameter is only important because we allowed galaxies to be repeated within the same bin. }

The correction fraction $f$ is expressed as:
\begin{equation}
f= \frac{n_{tot}}{n_{gal}} \Sigma_{i=0}^{n_{gal,t}} \frac{n_i V_{max,i}}{V_{tot}},
\label{eqn:correction}
\end{equation}
where 
$$ V_{tot} = \Sigma_{i=0}^{n_{gal}} V_{max,i} $$
is the total volume sampled by the n$_{gal}$ galaxies currently observed by the survey and n$_{tot}$ is the expected number of galaxies of the survey when completed (assumed to be 10,000 by this calculation). A different approach would be to ignore the origin of the considered SFRs, and treat the data as a {\it synthetic} catalog. In this case the solution would not require any correction factor, and the cosmic SFR history would be derived by the simple formula:
$$ \Psi_{SFR}(z) \equiv \Psi_{SFR,t} =   \frac{n_{bin,t}}{n_{tot}} \Sigma_{i=0}^{n_{bin,t}} \frac{SFR_{i,t}}{V_{max,i}} $$
where $n_{bin,t}$ would be the number of galaxies of the {\it synthetic} catalog in the considered redshift range. We repeated our calculations using this formula without any significant change.
 
%%%%%%%%%%%%%%%%%%%%%%%%%%%%%%%%%%%%%%%%%%%%%%%%%%%%%%%%%%%%%%%%%%%%%%%%%%%%%
\begin{landscape}
\begin{table}
\caption{Star-formation and stellar mass density of the universe at different redshift bins, together with the average sSFR, derived from our analysis of the stellar populations of the full sample and the two subsamples of SFGs$_0$ and RGs$_0$. For each redshift we provide the number of galaxies used to derive the parameters, together with the estimated values and the corresponding errors. The listed redshift indicates the average value for the different bins corresponding to each subsample. Those values may change between each subsample, due to the different sampled galaxies.}
\label{tab:madau}
\centering
\begin{tabular} {c r r r r r r r r r r r r}
\hline
z     & N$_{gal}$  &  log($\Psi$)     &  log($\rho$)     & log(sSFR) & N$_{gal}$  &  log($\Psi$)     &  log($\rho$)     & log(sSFR)  & N$_{gal}$  &  log($\Psi$)     &  log($\rho$)     & log(sSFR) \\
      &     &  M$_\odot$/yr/Mpc$^3$ & M$_\odot$/Mpc$^3$& yr$^{-1}$ &     &  M$_\odot$/yr/Mpc$^3$ & M$_\odot$/Mpc$^3$& yr$^{-1}$  &     &  M$_\odot$/yr/Mpc$^3$ & M$_\odot$/Mpc$^3$& yr$^{-1}$\\
\hline
& \multicolumn{4}{c}{All galaxies} & \multicolumn{4}{c}{SF galaxies} & \multicolumn{4}{c}{Retired galaxies}\\
\cline{3-5}
\cline{7-9}
\cline{11-13}
 0.03 & 1980 & -2.10$\pm$0.29 & 8.79$\pm$0.17 & -10.89$\pm$0.59 &  1480 & -2.24$\pm$0.28 & 8.49$\pm$0.15 & -10.73$\pm$0.57 &  721 & -2.69$\pm$0.27 & 8.47$\pm$0.14 & -11.16$\pm$0.53\\
 0.04 & 3694 & -1.80$\pm$0.29 & 9.09$\pm$0.17 & -10.89$\pm$0.59 &  2719 & -1.94$\pm$0.28 & 8.79$\pm$0.15 & -10.73$\pm$0.57 &  1396 & -2.39$\pm$0.27 & 8.77$\pm$0.14 & -11.16$\pm$0.53\\
 0.10 & 3541 & -1.70$\pm$0.23 & 8.99$\pm$0.18 & -10.69$\pm$0.48 &  2641 & -1.84$\pm$0.24 & 8.68$\pm$0.16 & -10.51$\pm$0.48 &  1319 & -2.27$\pm$0.18 & 8.69$\pm$0.14 & -10.95$\pm$0.36\\
 0.15 & 3389 & -1.61$\pm$0.20 & 8.86$\pm$0.18 & -10.47$\pm$0.41 &  2564 & -1.75$\pm$0.21 & 8.53$\pm$0.16 & -10.28$\pm$0.43 &  1242 & -2.17$\pm$0.13 & 8.58$\pm$0.14 & -10.76$\pm$0.25\\
 0.20 & 3065 & -1.55$\pm$0.18 & 8.71$\pm$0.18 & -10.26$\pm$0.38 &  2414 & -1.67$\pm$0.19 & 8.43$\pm$0.16 & -10.10$\pm$0.39 &  1013 & -2.16$\pm$0.12 & 8.37$\pm$0.14 & -10.53$\pm$0.24\\
 0.25 & 2741 & -1.50$\pm$0.17 & 8.48$\pm$0.18 & -9.97$\pm$0.36 &  2264 & -1.60$\pm$0.18 & 8.31$\pm$0.16 & -9.91$\pm$0.37 &  784 & -2.14$\pm$0.12 & 7.96$\pm$0.14 & -10.10$\pm$0.23\\
 0.30 & 2843 & -1.48$\pm$0.25 & 8.54$\pm$0.20 & -10.02$\pm$0.52 &  2251 & -1.65$\pm$0.26 & 8.32$\pm$0.16 & -9.97$\pm$0.53 &  919 & -1.99$\pm$0.20 & 8.18$\pm$0.16 & -10.18$\pm$0.41\\
 0.36 & 2946 & -1.46$\pm$0.32 & 8.60$\pm$0.22 & -10.06$\pm$0.67 &  2238 & -1.69$\pm$0.35 & 8.34$\pm$0.17 & -10.03$\pm$0.72 &  1054 & -1.88$\pm$0.25 & 8.33$\pm$0.16 & -10.21$\pm$0.51\\
 0.42 & 2993 & -1.39$\pm$0.23 & 8.60$\pm$0.20 & -10.00$\pm$0.48 &  2234 & -1.68$\pm$0.26 & 8.30$\pm$0.16 & -9.98$\pm$0.53 &  1126 & -1.65$\pm$0.16 & 8.38$\pm$0.16 & -10.03$\pm$0.33\\
 0.49 & 3041 & -1.33$\pm$0.16 & 8.61$\pm$0.18 & -9.94$\pm$0.34 &  2230 & -1.68$\pm$0.17 & 8.25$\pm$0.15 & -9.93$\pm$0.34 &  1199 & -1.49$\pm$0.13 & 8.43$\pm$0.16 & -9.92$\pm$0.26\\
 0.60 & 3080 & -1.16$\pm$0.17 & 8.63$\pm$0.20 & -9.80$\pm$0.36 &  2209 & -1.61$\pm$0.17 & 8.22$\pm$0.15 & -9.83$\pm$0.35 &  1271 & -1.22$\pm$0.13 & 8.52$\pm$0.17 & -9.74$\pm$0.27\\
 0.71 & 3119 & -1.05$\pm$0.17 & 8.66$\pm$0.21 & -9.71$\pm$0.38 &  2188 & -1.56$\pm$0.17 & 8.19$\pm$0.16 & -9.75$\pm$0.36 &  1343 & -1.05$\pm$0.13 & 8.60$\pm$0.18 & -9.65$\pm$0.28\\
 0.90 & 3332 & -0.89$\pm$0.17 & 8.69$\pm$0.26 & -9.58$\pm$0.39 &  2389 & -1.38$\pm$0.17 & 8.24$\pm$0.17 & -9.62$\pm$0.36 &  1350 & -0.96$\pm$0.12 & 8.56$\pm$0.20 & -9.53$\pm$0.27\\
 1.09 & 3545 & -0.77$\pm$0.17 & 8.72$\pm$0.30 & -9.49$\pm$0.42 &  2591 & -1.26$\pm$0.17 & 8.28$\pm$0.18 & -9.54$\pm$0.37 &  1358 & -0.90$\pm$0.12 & 8.52$\pm$0.22 & -9.42$\pm$0.26\\
 1.50 & 3584 & -0.84$\pm$0.19 & 8.67$\pm$0.30 & -9.51$\pm$0.45 &  2641 & -1.24$\pm$0.19 & 8.32$\pm$0.19 & -9.57$\pm$0.41 &  1361 & -0.97$\pm$0.13 & 8.45$\pm$0.22 & -9.42$\pm$0.30\\
 1.92 & 3623 & -0.92$\pm$0.21 & 8.62$\pm$0.29 & -9.54$\pm$0.50 &  2692 & -1.22$\pm$0.21 & 8.36$\pm$0.21 & -9.59$\pm$0.47 &  1365 & -1.06$\pm$0.15 & 8.37$\pm$0.22 & -9.43$\pm$0.35\\
 4.54 & 2703 & -1.13$\pm$0.30 & 8.35$\pm$0.53 & -9.47$\pm$0.84 &  1899 & -1.48$\pm$0.32 & 8.08$\pm$0.39 & -9.56$\pm$0.83 &  1155 & -1.27$\pm$0.26 & 8.11$\pm$0.41 & -9.38$\pm$0.68\\
 7.16 & 1783 & -1.54$\pm$0.26 & 7.43$\pm$0.64 & -8.97$\pm$0.74 &  1106 & -2.19$\pm$0.27 & 6.81$\pm$0.60 & -9.00$\pm$0.76 &  945 & -1.71$\pm$0.52 & 7.31$\pm$0.66 & -9.03$\pm$1.46\\
\hline
\end{tabular}
\end{table}
\end{landscape}
%%%%%%%%%%%%%%%%%%%%%%%%%%%%%%%%%%%%%%%%%%%%%%%%%%%%%%%%%%%%%%%%%%%%%%%%%%%%%%%%

Figure \ref{fig:madau} shows the cosmic SFR density of the universe as a function of look-back time (and redshift), { for the full sample of galaxies analyzed and the two subsamples of star-forming and retired galaxies at $z\sim$0, i.e., the SFGs$_0$ and RGs$_0$. The result of all this analysis is included in Table \ref{tab:madau}. In addition we include in the figure } similar derivations extracted from the literature, based on compilations of different cosmological surveys \citep[e.g.][]{Madau14,Driver17} or recent archaeological studies \citep[e.g.][]{lopfer18}. The estimated $\Psi_{SFR}(z)$ presents the well know trend, rising from a value of $\sim$0.01 M$_\odot$/yr/Mpc$^3$ in the nearby universe toward a value $\sim$10 times larger at $z\sim$1-2 ($t_{\rm lb}\sim$8 Gyr), and then declining afterwards. Our results agree both qualitatively and quantitatively withing the errors with previously reported values extracted from the literature. This agreement is particularly good for the last 8 Gyr. The main difference is found in the peak of the cosmic SFH { , which in our case seems to be located at $z\sim$1}, while most cosmological surveys find a peak at $z\sim$1.5-2.5, particularly for derivations based on far-infrared observations \citep[e.g.][]{Madau14}. { We should be cautious about this result, due to the poorly constrained shape of $\Psi_{SFR}(z)$ beyond $z>3$ for our method, as we will discuss in Sec. \ref{sec:caveat}, Fig. \ref{fig:madau_NC} and Sec. \ref{sec:madau_cuts}.} The bump in the $\Psi_{SFR}$ is in general broader than the one reported by \citet{Madau14}, and more similar to the one described by \citet{Driver17} and \citet{lopfer18}. Despite these differences, the agreement is remarkable good { , particularly in the regime where our results are most reliable, i.e, $<$8 Gyrs.}

A substantial difference concerning inferences between our archaeological approach and cosmological surveys is that we can trace the SFHs of the same sub-sample of galaxies without any assumption of how a particular sub-sample evolves. This was already illustrated in Sec. \ref{sec:SFMS_t}, where we explored the evolution of the SFR-M$_*$ diagram for the sub-samples of SFGs$_0$ and RGs$_0$. Following the same approach, we include in Fig. \ref{fig:madau} the contribution to the SFR density history of both families of galaxies, finding strong differences. The galaxies that are retired in the local Universe (RGs$_0$), with a halted/quenched SFR in the near universe, present a sharper trend in their cosmic SFH, being on average passive only in the last $\sim$3.5 Gyr ($z<0.3$). On the contrary, at higher redshifts (particularly between z$\sim$0.5 and z$\sim$2; t$_{\rm lb}\sim$5-10 Gyr) they are the galaxies that contributed most to the total SFR density of the Universe. The transition between when RGs$_0$'s and SFGs$_0$ dominate the comoving SFR seems to be a rather fast process that took place in a few Gyr at maximum, in a redshift range between $z\sim$0.3-0.5. However, if we consider the results presented in Sec. \ref{sec:SFMS_t}, not all the RGs$_0$ halt at once, and in some cases individual RGs$_0$ become active and inactive in star-formation several times. Therefore, although on average the star formation rapidly diminishes below { $z~0.7$} for RGs$_0$, their individual histories may depart considerably from this average behavior.

The cosmic SFH of the SFGs$_0$ is considerably different. They present a shallower rise of $\Psi_{SFR}$, with an increase of a factor $\sim$5 from the nearvy universe to their peak. The peak is less pronounced, broader, and shifted to slightly higher redshifts ($z\sim 2$). In contrast, the drop at higher redshift ($z>3$) seems to be sharper, indicating that the star-formation activity started later in these galaxies than in the retired ones (RGs$_0$). { However we should be cautios about this result due to the large uncertainites at this redshift range.} Contrary to the RGs$_0$, SFGs$_0$ do not have a strong drop in SFR at lower redshifts, being in general the galaxies that contribute most to the $\Psi_{SFR}$ in the last $\sim$4 Gyr ($z<0.4$).

{ We dicuss now two aspects of our method of determining $\Psi_{SFR}(z)$. On one hand, if we had selected only one SFR measurement per galaxy and redshift bin, then the correction fraction $f$ given by Equation (\ref{eqn:correction}) should not have been applied. In this case we should have selected instead either the average SFR along the redshift interval, interpolating our data as a sequence of SFHs, or picked a particular SFR, if we adopt the current approach (i.e., consider our collection of SFR and M$_*$ as a cosmological survey). The latter case is illustrated in Fig. \ref{fig:madau_no_rep} in the Appendix \ref{sec:madau_no_rep}. As seen, the differences with Fig. \ref{fig:madau} are small.

On the other hand, as indicated before we used only the SFGs$_t$ sub-sample for the calculation of $\Psi_{SFR}(z)$. First, these would be the galaxies for which the SFR could be estimated in a cosmological survey. Second, below the considered cut in sSFR, the SFR$_{ssp}$ becomes very uncertain since it is derived from a very small fraction of young stars (e.g., Bitsakis et al., in prep). To reflect the effect of this selection we have included in the error budget the differences between using the full sample, without any cut in the sSFR, and using the SFGs$_t$ to derive the cosmic SFR. We confirm that this selection only affects the derived $\Psi_{SFR,t}$ at very high redshift (z$>2$), as reflected in the very large error bars in this regime. { This is shown in Fig. \ref{fig:madau_NC}, and discussed in the Appendix \ref{sec:madau_cuts}.} }

%%%%%%%%%%%%%%%%%%%%%%%%%%%%%%%%%%%%%%%%%%%%%%%%%%%
% SFMS_t
\begin{figure*}
  \centering
    \includegraphics[width=10cm,clip, trim=35 0 55 0]{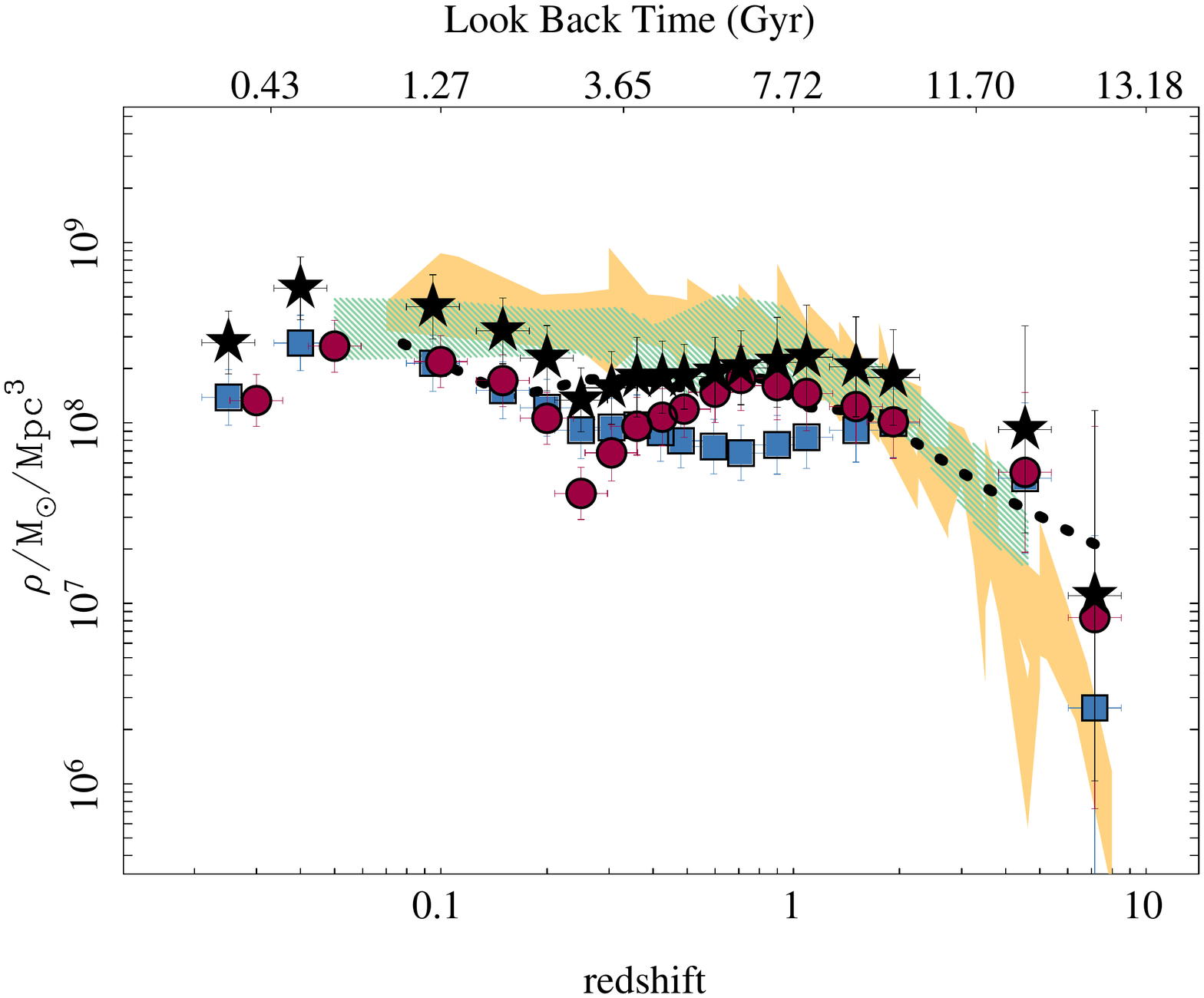}\includegraphics[width=10cm, clip, trim=35 0 55 0]{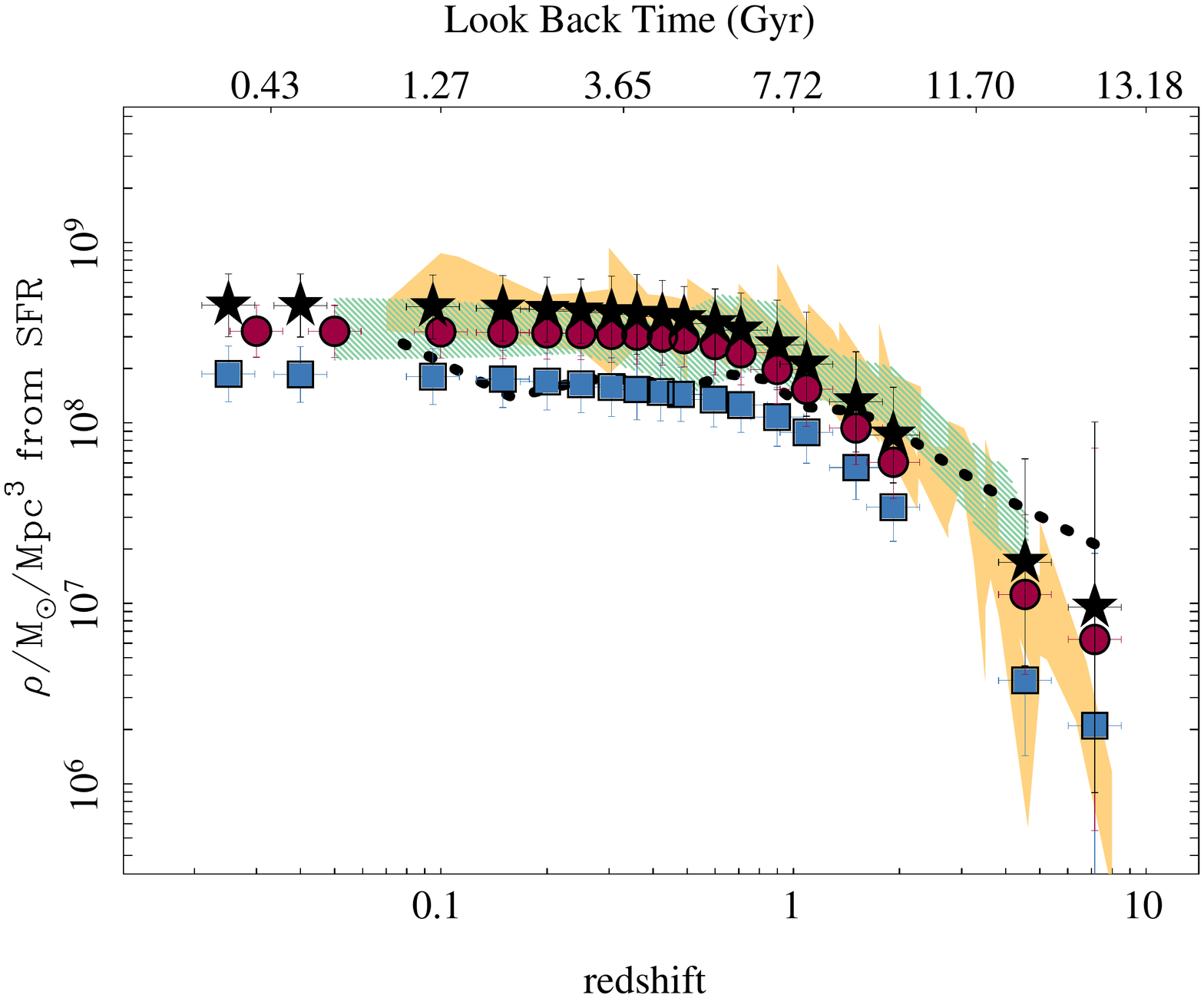} 
    \caption{{\it Left panel}: Cosmic evolution of the stellar mass density derived by co-adding the individual stellar masses of each galaxy in each redshift bin corrected by their volume. The density is presented for all the analyzed galaxies (black solid stars), the SFGs$_0$ (blue solid squares) and RGs$_0$ (red solid circles), with their corresponding errors (1$\sigma$ range). The shadowed regions corresponds to the stellar mass densities derived by different compilations presented by \citet{Madau14} (yellow hashed region), and \citet{Driver17} (green hashed region). The black dotted points corresponds to the derivation presented by \citet{lopfer18}. As in Fig. \ref{fig:madau} shifts are applied in the literature data to take into account the different adopted IMFs, as required. {\it Right Panel}: Similar plot, but for the stellar mass densities derived from the integration of the star-formation density shown in Fig.\ref{fig:madau}.}
    \label{fig:Mass}
\end{figure*}
%%%%%%%%%%%%%%%%%%%%%%%%%%%%%%%%%%%%%%%%%%%%%%%%%%%

\subsection{Evolution of the cosmic stellar-mass density}
\label{sec:m_den_t}

Our methodology allow us to recover not only the SFR of galaxies at different redshifts, but also their stellar masses, as shown in Sec. \ref{sec:SFMS_t}. By using theses masses and repeating the calculations described in Sec. \ref{sec:madau}, it is possible to estimate the cosmic stellar-mass density at different epochs, $\rho_{*,t}$, by using the formula:
$$ \rho_{*}(z) \equiv \rho_{*,t} = f\ \Sigma_{i=0}^{n_{gal,t}} \frac{1}{n_i}\frac{M_{*,i,t}}{V_{max,i}} $$
where $f$, n$_{gal,t}$, n$_i$ and V$_{max,i}$ have the same meaning as described in the previous subsection. The results of this calculation for the different analyzed subsamples and the different redshift bins have been included in Tab. \ref{tab:madau}. 

Figure \ref{fig:Mass}, left-panel, shows the estimated cosmic stellar-mass density history, $\rho_*(z)$, together with similar derivations extracted from the literature. As in the case of $\Psi_{SFR}(z)$, we find a trend similar to those reported in previous studies based mostly on cosmological surveys. We find a fast rise in $\rho_*$ at early epochs, with an increase in the first $\sim$3-4 Gyr from 10$^{7}$ M$_{\odot}$ Mpc$^{-3}$ to near 10$^{8.3}$ M$_{\odot}$ Mpc$^{-3}$, followed by a shallower increase in the last $\sim$9-10 Gyr towards the current value near to 10$^{9}$ M$_{\odot}$  Mpc$^{-3}$ (at $z\sim$0). This characteristic shape is well described by inferences from different cosmological surveys \citep[e.g.][]{perezgonzalez08,Pozzetti+2010,ilbert13,Muzzin+2013,Madau14,Driver17}.

However, the agreement is not as good as it was for the cosmic SFH. Our measurements exhibit a less smooth trend, with a decrease at $z\sim$0.3 (t$_{\rm lb}\sim$3.5 Gyr) dominated by the contribution of the RGs$_0$. Unless the stellar mass loss dominates the evolution of the stellar-mass density, which is not the case, this decrease is unphysical. Near this redshift bin we find a considerable decrease in the number of RGs; this may be affecting our statistics (see Tab. \ref{tab:madau}). The fact that this drop is not present in the contribution of the SFGs$_0$ supports this suspicion that that the down-turn is an artifact of low-number statistics in this redshfit bin. However, we should note that \citet{lopfer18} shows a similar drop based on their archaeological analysis of a different sample of galaxies. 
% Therefore, we cannot exclude that this drop is an effect of the methodology.

The cosmic stellar-mass density histories of the SFGs$_0$ and RGs$_0$ show differences, but not as distinct as the ones shown in the cosmic SFHs, largely because the stellar-mass density is an integral quantity. The RGs$_0$ contribute more to $\rho_*$ in the first $\sim$9 Gyr of the cosmic history of the universe, while in the last $\sim$4 Gyr both subsamples of SFGs$_0$ and RGs$_0$ have similar stellar-mass densities. The strongest difference happens in the highestredshift bin, at $z\sim$7, where the contribution of the RGs$_0$ is ten times larger than that of the SFGs$_0$. However, the errors in this very last redshift bin are large, and therefore the result is not robust. 

The above estimates of $\rho_*$ ares derived considering that our different estimates of the stellar masses are independent; we treat occurences of the same galaxy in different redshift bins like individual galaxies {\it observed} at different redshifts, i.e., as if we had a cosmological survey covering a very large redshift range. This is why we observe fluctuations, such as the drop in the stellar-mass density at t$_{\rm lb} \sim 3$ Gyr. However, we can derive $\rho_*(z)$ also by integrating the SFR density, $\Psi_{SFR}(z)$, shown in Fig. \ref{fig:madau}, obtaining an alternative and truly integral estimate of this parameter. For doing so, we adopt the formula:
$$ \rho_{*,\Psi}(z) \equiv \rho_{*t,\Psi} = \Sigma_{Th}^{t} (1-R_{Ti}) \Psi_{Ti,SFR} \Delta Ti$$
where (i) $\Psi_{Ti,SFR}$ is the SFR density at the look-back time $Ti$, with $Ti$ running through the times sampled by our fossil record method;  (ii) $\Delta Ti$ is the time period between two adjacent redshift bins; and (iii) R$_{Ti}$ is the average stellar mass loss at this considered time. The result of this calculation is shown in Fig. \ref{fig:Mass}, right-panel. In general we recover a similar trend as before, with a sharp rising in the early epochs of the universe, up to $\sim$1, and a much shallower increase in the last $\sim$5 Gyr. However, there are also clear differences. By construction, there is no drop in the stellar-mass density, since in this derivation this parameter can only grow with time, reinforcing the suggestion that the decrease at $z \sim 0.3$ described before is a statistical fluctuation. In our second estimate, the cosmic stellar-mass density is dominated by the RGs$_0$ at all redshifts, while in the initial estimate SFGs$_0$ have a similar contribution to the stellar-mass density as the RGs$_0$ over the last $\sim$3 Gyr. Finally, note that our estimated $\rho_{*,\Psi}(z)$ agrees reasonably well with inferences based on cosmological surveys, within the scatter and uncertainties.
%although there is a good qualitative agreement with literature results, quantitatively our estimated $\rho_{*,\Psi}(z)$ is below any previous estimation at high redshift ($z>1$). 
%In particular, it is worth noticing that our result is very similar to the one presented by \citet{lopfer18}, based also in an archaeological procedure, up to $z\sim$3.

It seems that our first estimate of the cosmic stellar-mass density is more prone to statistical fluctuations in the sampled number of galaxies. However, by construction it should be most comparable to a cosmological survey, and indeed matches fairly well with the literature results  (if we exclude the redshift range between $z\sim$0.2-0.3). Our second estimate is less affected by statistical fluctuations and agrees well at all redshfits with derivations based on cosmological surveys.

%%%%%%%%%%%%%%%%%%%%%%%%%%%%%%%%%%%%%%%%%%%%%%%%%%%
% SFMS_t
\begin{figure*}
  \centering
  \includegraphics[width=10cm,clip, trim=35 0 55 0]{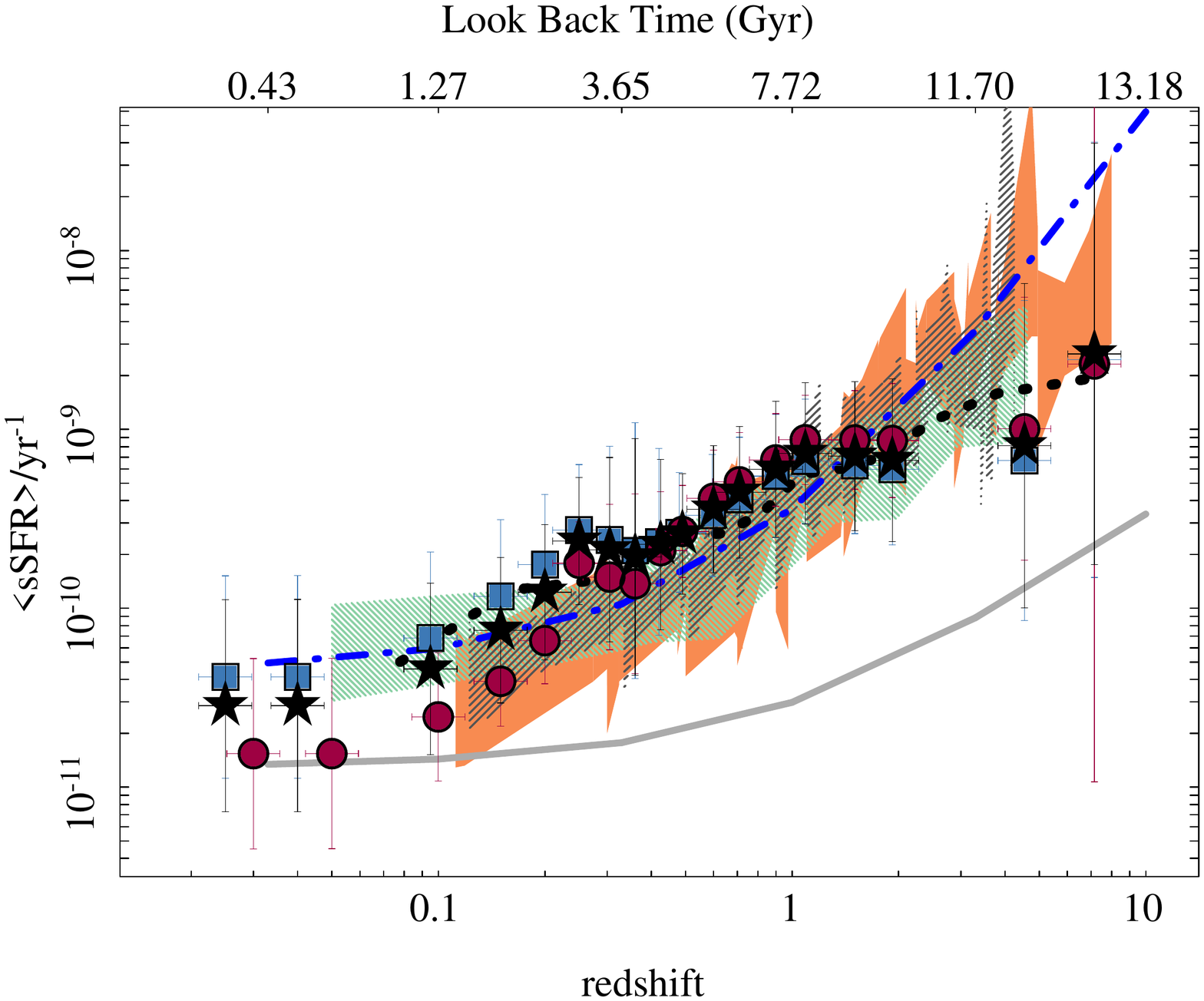}\includegraphics[width=10cm, clip, trim=35 0 55 0]{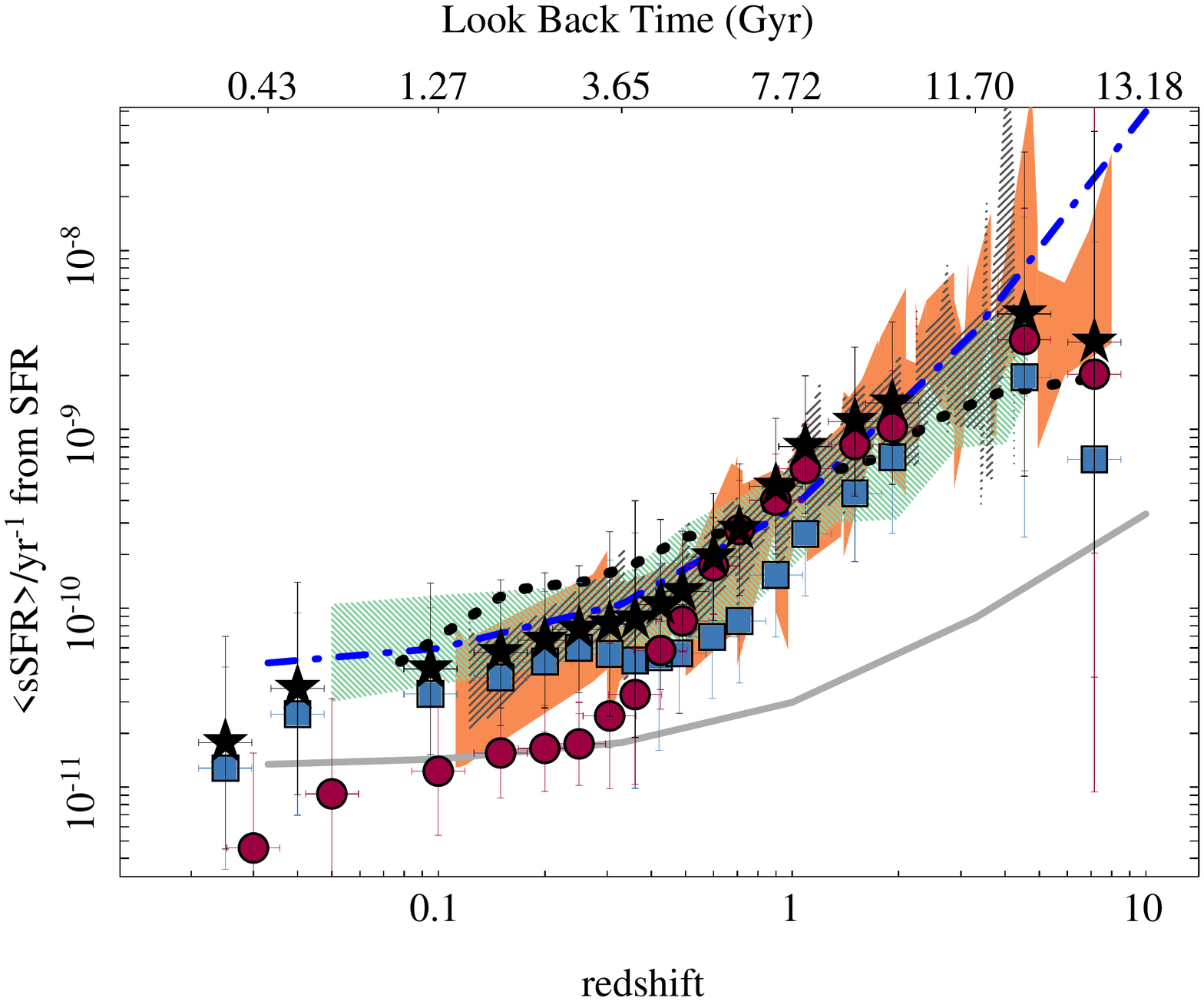} 
    \caption{{\it Left panel}: Cosmic evolution of the average sSFR derived by co-adding the individual sSFR's of each galaxy corrected by their volume. The $<$sSFR$>$ is presented for all  analyzed galaxies (black solid stars), SFGs$_0$ (blue solid squares) and RGs$_0$ (red solid squares), with their corresponding errors (1$\sigma$ range). Shadowed regions and the black dotted line correspond to the same literature results shown in Fig. \ref{fig:madau}. The blue dashed-dot line shows the best fit to the results by \citet{elbaz11}. The grey-solid line shows the lower limit to the sSFR adopted to select SFGs at each redshift. {\it Right panel}: Similar figure when the $<$sSFR$>$ is derived by dividing the star-formation density shown in Fig. \ref{fig:madau} by the stellar-mass density shown in right panel of Fig. \ref{fig:Mass}. }
  \label{fig:sSFR}
\end{figure*}
%%%%%%%%%%%%%%%%%%%%%%%%%%%%%%%%%%%%%%%%%%%%%%%%%%%

\subsection{Evolution of the cosmic specific star-formation rate}
\label{sec:sSFR}

Having derived the cosmic SFR and stellar-mass densities, it is possible to estimate the average sSFR of the universe by dividing these two quantities. This parameter was estimated at each redshift bin and galaxy subsample, and it is listed in Tab. \ref{tab:madau}. Figure \ref{fig:sSFR}, left-panel, shows the distribution of $<$sSFR$>$ versus cosmic time compared with similar derivations extracted from the literature. In this case we adopt our first estimate of the stellar-mass density, as listed in Tab. \ref{tab:madau} and shown in the left panel of Fig. \ref{fig:Mass}. As in the case of $\Psi_{SFR}(z)$ and $\rho_*(z)$, we find a similar trend between our reported $<$sSFR$>$ and the values reported in the literature. There is a general and steady decrease of $<$sSFR$>$ from the early universe, with values near 10$^{-9}$ yr$^{-1}$, to the current time, with values near 10$^{-11}$ yr$^{-1}$. There is an offset between our reported $<$sSFR$>$ and those found in the literature from cosmological redshift surveys, while our results agree well at all redshifts with  \citet{lopfer18}. Our values are higher than those using FIR surveys reported by \citet{Madau14} by about $\sim$0.1-0.2 dex from $0.1<z<1$, and lower at $z>2$. The difference is stronger at high redshift ($z>2$), where ours clearly depart from previously reported results.

%When comparing this distribution with the ones reported for $\Psi_{SFR}$ and $\rho_{*}$ 
We see that these differences are mainly induced by the miss-match between the stellar-mass densities reported from cosmological redshift surveys and those estimated from our analysis. The larger values of $\rho_{*}$ reported at high redshift implies a lower values of $<$sSFR$>$ in the same redshift range. On the other hand, the artificial drop in $\rho_{*}$ at $z\sim$0.2-0.3 produces an increase of the $<$sSFR$>$. Beyond this particular feature, there is a similar trend for both the full sample and the subsamples of SFGs$_0$ and RGs$_0$, with a general decrease of $<$sSFR$>$ from early epochs to the present-day. However, the trend is sharper for the RGs$_0$ which contribute more to the $<$sSFR$>$ at $z>$0.5 and then strongly decrease at lower redshifts. We can characterize this general trend by fitting a log-linear relation with $1+z$, following \citet{elbaz11}. The results of this analysis for the different subsamples are:
$$\rm log (sSFR_{all}) = -10.09_{\pm 0.06}  + 1.82_{\pm 0.24} \ log(1+z)$$
$$\rm log (sSFR_{SF})  =  -9.01_{\pm 0.05}  + 1.57_{\pm 0.24} \ log(1+z)$$
$$\rm log (sSFR_{RG})  = -10.30_{\pm 0.08}  + 2.29_{\pm 0.40} \ log(1+z),$$
where the errors in the derived parameters were estimated using a Monte-Carlo iteration accounting for uncertainties in individual data points. The trend with redshift is shallower than the one reported by \citet{elbaz11}, who found a slope of $\sim$3 for this trend. The main reason for this discrepancy arise from the values of $<$sSFR$>$ in the last two redshift bins that depart considerably from previously reported estimates, as seen in Fig. \ref{fig:sSFR}. 

Following the methodology reported in the previous subsection regarding the analysis of the stellar-mass density, we repeated our estimate of the average sSFR by dividing the SFR density, $\Psi_{SFR}$, shown in Fig. \ref{fig:madau}, by the stellar mass density derived from the integration of this very same quantity, $\rho_{*,\Psi}(z)$, shown in the right panel of Fig. \ref{fig:Mass}. This second estimate of $<$sSFR$>$ is shown in the right panel of Fig. \ref{fig:sSFR}. As for the case of $\rho_{*,\Psi}(z)$, the second estimate has much smoother behavior with time. There is a steady decrease in $<$sSFR$>$ from the early universe, with the only exception in the highest redshift bin, where there is a drop similar to to our initial estimate (Fig. \ref{fig:sSFR}, left panel). Agreement with estimates from cosmological surveys is very good except below $z<0.1$ and above $z\sim$5. In particular, the reported values match  well with the best fit reported by \citet{elbaz11}, with only two values departing from this trend (again, those at the highest and lowest redshift bins). Qualitatively, the trends found for SFGs$_0$ and RGs$_0$ are very similar with this second estimate of $<$sSFR$>$, although the differences between both subsamples are more clearly highlighted now. In general, the contribution of RGs$_0$ dominates the $<$sSFR$>$ at redshifts beyond $z>$0.5, with a sharp decrease below this redshift (sharper with this second estimate than for the first one). When compared to the contribution of the SFGs$_0$, the trend for the progenitors of local RGs appears delayed at early epochs even though the progenitors of local SFGs have lower values of sSFR at these times. SFS have a shallower evolution below $z<1$, and hence dominate the local average sSFR (as expected by construction of this subsample).

The dependence of this second derivation of the average sSFR seems to be sharper, with a stronger evolution with redshift. This is reflected in the slopes derived when fitting against the redshift, following the same procedure as described before:
$$\rm log (sSFR_{all,\Psi}) = -10.71_{\pm 0.06}  + 3.55_{\pm 0.27} \ log(1+z)$$
$$\rm log (sSFR_{SF,\Psi})  = -10.90_{\pm 0.05}  + 3.06_{\pm 0.24} \ log(1+z)$$
$$\rm log (sSFR_{RG,\Psi})  = -11.40_{\pm 0.07}  + 4.59_{\pm 0.42} \ log(1+z)$$
The slopes of this log-linear relations are closer to the value of $\sim$3 reported by \citet{elbaz11}, in particular for the SFGs$_0$. Together with the results shown in the previous section, it seems that estimates based on this second method fit better with estimates from cosmological surveys.

The adopted limit in sSFR to pre-select SFGs at a given redshift (Eq. \ref{ssfr-lim}) has been included in both panels of Fig. \ref{fig:sSFR} (gray line). We see that the average $<$sSFR$>$ is above this limit for most of the considered redshift ranges, in particular for the direct estimate shown in the left-panel. As expected, the progenitors of local RGs  approach this limit at low redshift, and those that are still forming stars present a lower SFR than the progenitors of local SFGs at the same look-back time.

\section{Discussion}
\label{sec:disc}

We have shown how archaeological methods can be used (i) to trace the evolution of galaxies in the SFR-M$_*$ diagram, and (ii) to estimate the cosmic SFR and stellar-mass density histories in the Universe. In this section we discuss the caveats and implications of these results.

\subsection{Caveats on the methodology}
\label{sec:caveat}

Prior to any discussion we should mention the main caveats of the current adopted methodology. As extensively explained in \citet{Pipe3D_I} and \citet{Pipe3D_II}, the SFHs of galaxies can be recovered only up to a certain limit of reliability, and the derived results are strongly attached to the adopted methodology. The results depend particularly on the adopted SSP template library. In the current analysis we adopted the GSD156 library(Sec. \ref{sec:ana}). This particular library samples time in 39 non-continuous, non-regular ages with almost a logarithmic sampling. The main caveat for this library is that the age of the oldest SSP, 14.13 Gyr, is indeed older than the universe under the assumed cosmology (13.72 Gyr). The reason to include this old SSP is a technical one, since in this way we avoid the oldest stellar population being over represented due to saturation of the sampling and the effects of random noise. Since, due to the adopted procedure, we sample the cosmological times at the look-back times between sampled ages (to derive the SFR one needs to obtain the $\Delta$M$_*$ between to consecutive SSPs, separated by $\Delta$t), we in fact do \textbf{not} sample ages older than the Universe. However, the measurements in the very last redshift bin shown in Fig. \ref{fig:SFMS_t}, and the two last bins discussed in Sec. \ref{sec:madau} and subsequent subsections could be affected in two different ways: (i) the actual sampled look-back time could have large uncertainties; (ii) since the time variable may contain uncertainties, the derivation of the SFR would indeed contain these uncertainties as well. We have tried to determine the uncertainties by repeating the analysis assuming the different possible time ranges $\Delta$t, making a random sampling of the time, and associating them to a particular SSP based on their vicinity to the nominal age. Then, we recalculate the time and the SFR considering this new time range. The standard deviation of the distribution of values found using this method with respect to the original value of the SFR has been included in the error budget of the values shown in Fig. \ref{fig:madau}, \ref{fig:Mass} and \ref{fig:sSFR}, and listed in Table \ref{tab:madau}. The effect is strong in the very last redshift bins ($z>$3) for the $\Psi_{SFR}$ and $<$sSFR$>$ derivations.

The second major caveat regarding our results is the selection of SFGs at different redshifts. We have adopted a cut in the sSFR to perform a first selection of those galaxies that actively form stars at given time. Then, based on the parametrization of the SFMS, we perform a second cut, as described in Sec. \ref{sec:SFMS_t}. These two selections are adopted to identify which galaxies contribute to the $\Psi_{SFR}$ at each redshift bin. The first cut is broadly used, and it is the one that affects more the scale of the distributions up to redshift $z\sim$2-3. However, this first cut does not affect considerably its shape; if no cut in the sSFR is performed we would find the same rising shape in the SFR density, but at slightly larger values than the predictions by cosmological surveys. Beyond this redshift, it is the second selection that affects more the results, shaping the drop in the SFR density at $z>$3. Thus, not performing this second cut, the derived values for the $\Psi_{SFR,t}$ would remain almost constant in the last two redshift bins (although they would be clearly within the estimated errors). { This is well illustrated in Fig. \ref{fig:madau_NC}, and discussed in Appendix \ref{sec:madau_cuts}}. The fact that by doing these two selection cuts we reproduce very well the well known shape of the $\Psi_{SFR}(z)$, $\rho_{*}(z)$ and $<$sSFR$>(z)$ is reassuring. However, due to the uncertainties on both cuts, we should remain cautious about the shape of these distributions at high redshifts.

Finally, the third major caveat concerns the very nature of the procedure adopted here. The stellar-population analysis that we perform based on FIT3D/Pipe3D provides us with the fraction of light that comes from a particular set of stars of a particular age. These light fractions are then transformed to mass-fractions via the M/L ratio of each particular SSP, and finally to masses, considering the luminosity of the galaxy. Therefore, we have assumed that all mass increase in galaxies from one age to another is due to star formation in situ. This argument neglects the mass growth by galaxy mergers, either major or minor. We know that wet mergers had a strong influence at early epochs, in particular beyond t$_{\rm lb}>$8 Gyr \citep[e.g.][]{papo05}, where these mergers induce strong star-formation events. However, these events happen in the primary galaxy and can be considered as in-situ star formation. At later epochs, the merger rate is much lower \citep[e.g.][]{jogee09}, although for massive early-type galaxies it is expected that they have suffered at least one major dry-merger since $z\sim$0.7 to the present time \citep[e.g.][]{bell06}. 

%%%%%%%%%%%%%%%%%%%%%%%%%%%%%%%%%%%%%%%%%%%%%%%%%%%
% SFMS_t
\begin{figure}
  \centering
  \includegraphics[width=10cm,clip, trim=35 0 55 0]{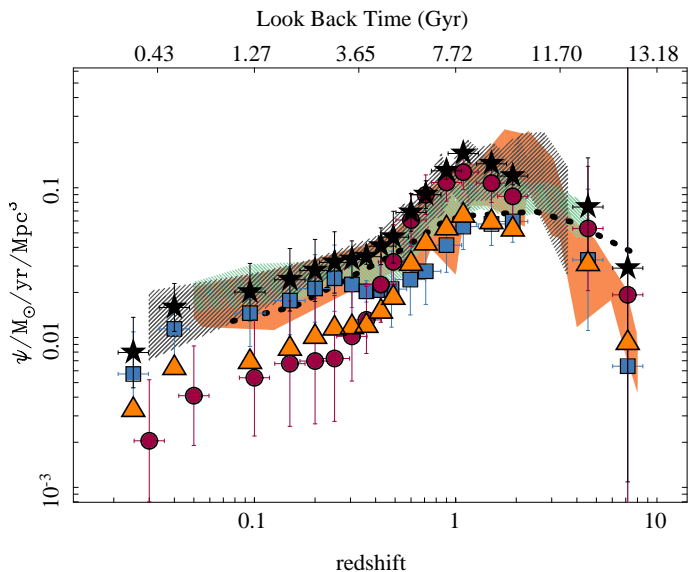}
    \caption{Cosmic evolution of the SFR density derived for a simulated set of galaxies that result from 1-1 mergers between pairs of star-forming or retired galaxies that maintain the SFH of their progenitors (orange triangles). All the other symbols included in the figure are the same as the one shown in Fig. \ref{fig:madau}}
  \label{fig:mergers}
\end{figure}
%%%%%%%%%%%%%%%%%%%%%%%%%%%%%%%%%%%%%%%%%%%%%%%%%%%

However, mergers do not alter the shape of $\Psi_{SFR}(z)$, $\rho_{*}(z)$ and $<$sSFR$>$ as a function of $z$ based on our procedure. The reason for that is the very nature of the calculation described above. What we determine with the archaeological method is the mass fraction of stars of a certain age. In a dry merger, the global stellar mass is increased. However, if both galaxies had a similar SFH, i.e., they have a similar distribution of stars per age, then the combined SFH would not be affected. On the other hand, if they have different SFHs, the combined SFH would be a mix of both of them. If all  galaxies suffer similar merging histories, then the general shape of the explored parameters would not change significantly. This is illustrated in Figure \ref{fig:mergers} where the cosmic SFH is shown for a set of simulated galaxies that are the result of a 1:1 merger between galaxies following the evolution found for SFGs$_0$ and for RGs$_0$ respectively. The resulting $\Psi_{SFR}(z)$ follows a shape in between the one found for both subsamples of galaxies, having the same qualitative properties as the one found for the full sample: a smooth rising from the current time to $z\sim$1-2, with a broad peak, and sharp decline at $z>3$ { (again noting that derivations beyond that redshift are uncertaint due to the limitations of the adopted methodology)}. Thus, dry mergers (or any merger that does not result in an increase in the SFR) by themselves would not affect our calculations. 

In addition to these three major caveats we should keep in mind that our results maybe strongly affected by the methodology of the archaeological procedure adopted here, by its implementation in Pipe3D, and by the selection of the SSP library. We aim to narrow-down their effects in future work by performing a detailed analysis of N-body + Hydrodynamical simulations (Ibarra-Medel et al. in prep.).

\subsection{Merging rate effects on $\rho_*(z)$}
\label{sec:merging}

In the previous subsection we discussed the effects of mergers in our calculations. We found that they should not affect significantly the derivation of $\Psi_{SFR}(z)$, however, they should affect somehow the integrated stellar mass density. As shown in Sec. \ref{sec:m_den_t}, we found differences between the direct estimates of $\rho_{*}(z)$, based on co-adding the volume-corrected stellar masses of galaxies sampled at different redshifts (Fig.\ref{fig:Mass}, left panel), and the estimates based on the integration of $\Psi_{SFR}(z)$ over time (Fig.\ref{fig:Mass}, right panel). Some of these differences are indeed due to sampling problems at particular redshifts. However, another possible difference could be due to the contribution of mergers that would not affect the fraction of stars at different ages but could affect the total amount of mass. Actually, we found a clear deviation between our derived stellar-mass density integrating the SFR at high redshift when compared to cosmological surveys (Fig. \ref{fig:Mass}). The deviations in the two highest redshift bins are uncertaint, however the trend starts at lower redshift ($z>1$). Thus, although we are cautious about deriving a quantitative estimates from our current analysis, we consider that a feasible explanation for these differences could be the contribution of mergers at high redshift.  

\subsection{The SFHs of star-forming and retired galaxies}
\label{sec:dSF}

Despite the caveats indicated before, there is a clear result that emerges from our analysis: the galaxies that dominate the SFR budget in the local universe are not the ones that dominated at higher redshifts. Most of these later galaxies are now RGs. They dominated $\Psi_{SFR}$ at $z>$0.5 and then had a sharp drop/quenching of their star-formation activity below $z<$0.3. This drop was general, although as we have seen in Sec. \ref{fig:SFMS_t}, a fraction of these galaxies retired at earlier epochs ($z\sim$2) and then re-started their star-formation activity more recently ($z\sim 1-2$). This switch-off in the star-formation activity is clearly connected with the present-day morphology of these galaxies, since, as we indicated in Sec. \ref{sec:SFMS_0}, most of them are ellipticals and S0, which is a well established result \citep[e.g.][]{Blanton+2005}. This change in the SFH with morphology has been recently described by \citet{rgb17,rosa17,lopfer18} based on fossil record procedures applied to IFS data. On the other hand, there is also a dependence on stellar mass that we have not explored fully in the current study. In earlier studies by \cite{heavens04} and \cite{panter07}, they showed that more massive galaxies had sharper, more peaked SFHs, and as mass decreases, the SFHs becomes smoother. This result is similar to the one recently presented by \citet{rgb17}. Therefore, galaxies of different masses present also different contributions to the global cosmic SFH, $\Psi_{SFR}(z)$. Our current result does not discriminate between stellar mass and or morphology, at least directly. Therefore, we cannot yet determine which is the dominant factor. However, it is clear that local RGs suffered a sharp downturn in their SFR $\sim$4 Gyr ago.

More recently, \citet{lopfer18}, making use of the spatial information provided by the IFS data of the CALIFA survey showed that the central regions of galaxies contribute to the global $\Psi_{SFR}(z)$ with a sharp, peaky, shape similar to that of the early-type or massive galaxies. In contrast, the contribution of the outer regions is shallower, less peaky, similar to that of the late-type or less massive galaxies. Compared to our results, the central regions of galaxies behave like the progenitors of local RGs (RGs$_0$), while the outer regions behave like the progenitors of local SFGs (SFGs$_0$). This fits with our current understanding of galaxy evolution, since we consider that in those galaxies were star formation was starting to halt, this quenching happens from the inside out, being stronger in more massive and earlier-type galaxies. There is different evidence for this process, such as the direct observation of a drop in the sSFR in the center of these galaxies \citep[e.g.][]{rosa16a,Belfiore17a,sanchez17b}, connected with a drop in the gas content \citep[e.g.][]{sanchez17b}, and even the flattening or drop of the abundance gradient in the central regions of galaxies \citep[e.g.][]{sanchez14,laura16,belf16a,laura17}. The inside-out quenching process has also been seen in cosmological simulations of galaxy evolution \citep[][]{Tacchella+2016,Avila-Reese+2018} { and observed at high-redshift \citep[e.g.][]{Tacchella+2015} .} Therefore, the picture that emerges is that SFR was quenched from the inside out, being connected with the growth of the bulge. For early-type spirals, like Sa and Sb, this quenching is still underway. However, for more early-type galaxies, like S0 and Ellipticals, the quenching happens earlier such that these galaxies dominate the budget of RGs in the local Universe. { In summary, there is a mass-grow downsizing and a quenching downsizing, and both seems to present an inside-out pattern and a dependence with morphology.}

{ 

\subsection{The shape of the Cosmic SFR density at high redshift}
\label{sec:highz}

Different cosmological surveys report slightly different shapes for $\Psi_{SFR,t}$ beyond z$\sim$1-2, as it can be appreciated in Fig. \ref{fig:madau}. In general, surveys based on FIR observations estimate larger SFRs, and a less sharp drop in the cosmic star-formation rate than observations involving rest-frame FUV observations \citep[e.g.][]{Madau14,Driver17}. In some cases there is no clear drop, with $\Psi_{SFR,t}$ showing a flat distribution or values more consistent with a plateau at this redshift range \citep[e.g.,][]{mclure18}.

Our results in this regards should be taken with care, for the reasons described in this article, and discussed in Sec. \ref{sec:caveat}. In particular, the definition of star-forming and retired galaxies in the highest redshift bins is critical. As  shown in Appendix \ref{sec:madau_cuts} the inclusion in the computation of the cosmic SFR of the RGs$_t$ changes dramatically its shape at high redshift, flattening it.

The galaxies that we classify as retired at high redshifts are only slightly below the SFMS (see Fig. 4). We could speculate that these are galaxies at the end of their star-burst phase, emitting then mainly in the FIR (sub/millimetric wave-lenghts at $z\sim 0$) due to the large amount of accumulated dust. Since the redshift determination of submillimetric/dusty star-forming galaxies is not very accurate, the cosmological surveys from which cosmic SFR densities are calculated, might be missing the contribution of this population at $z\ga 4$. The fossil record method is not affected by this issue. However, as discussed in this paper, our uncertainties at this cosmological times are too large to make any clear claim in this regard.

%If indeed the galaxies that we classify as retired at high redshift do indeed make a significant contribution to the cosmic SFR, then it is possible that cosmological surveys suffer from a significant bias in this regime. However, our uncertainties are too large to make any clear claim in this regards.

}

\subsection{The evolution of the SFMS}
\label{sec:dSFMS}

Our results confirm previous findings indicating that the SFMS is a tight relation, with a very low scatter over a  wide range of redshifts \citep[e.g.][]{Brinchmann04,Salim07,sanchez13,catalan15,mariana16,rosa16a}. We also confirm that this relation evolves strongly with redshift, both in scale (zero-point) and shape (slope), in agreement with \citet{Speagle14} and \citet{chiosi17}. The evolution of the SFMS has been reported in several studies \citep[e.g.][]{noeske07,elbaz07,Salim07,Schiminovich+2007,pannella09,Whitaker+2012,lopfer18}. However, the nature of that evolution, either physical or related to observational biases, was questioned due to (i) the differences in the selection of different samples at different redshifts, and (ii) the different observational techniques applied to derive the relevant parameters. Therefore, in many cases the homogenization of the dataset involved different corrections that may affect the results \citep[see for a discusion][]{Speagle14}. Our study shows this evolution for the same sample of galaxies, where the involved parameters are derived using the same techniques, thus confirming its physical origin.

%The only study of the evolution of the SFMS along cosmological times using the archaeological method to our knowledge is the recently presented by \citet{lopfer18}. Our results agree both quantitatively and qualitatively with those presented by these authors regarding the derivation of the $\Psi_{SFR,t}$, $\rho_t$ and $<sSFR>$. However, regarding the SFMS they report a strong evolution in the zero-point, but a mild or no evolution in the slope

Due to the nature of the analysis, this is one of the few studies where we can compare the evolution of the SFMS between local populations of star-forming and retired galaxies, SFGs$_0$ and RGs$_0$, respectively. Indeed, although we find a similar qualitative trend between both subsamples, the detailed evolution seems to be different. SFGs$_0$ show a shallower evolution in amplitude, while RGs$_0$ show a shallower evolution in slope in the SFMS relation. In other words, SFGs$_0$ have a larger average SFR (by a factor $\sim$2) at low redshift relative to RGs$_0$. However, at high redshift RGs$_0$ have similar amounts of star formation per unit mass. If star-formation happens following the SFMS that is present at high redshift, when both subsamples share this relationship, then those galaxies that are retired in the local universe have become less and less active, in particular in the last $\sim$4 Gyr. { However, SFGs$_0$ also form less stars at the same mass than their progenitors at high redshift.}

If quenching happens inside out, as described in the previous section, a natural explanation would be that certain regions of galaxies that still form stars follow the same basic rules while other regions are already quenched. The fact that there is an almost universal spatial resolved SFMS of star-forming regions \citep[e.g.][]{sanchez13,mariana16,hsieh17}, with a mild dependence on the morphological properties of galaxies \citep[e.g.][Cano-Diaz et al., in prep.]{rosa17}, supports this conclusion. In future studies we will try to explore the spatially-resolved SFMS along cosmological times using the currently adopted procedures to confirm this hypothesis.

\subsection{Turnover mass of the SFMS}
\label{sec:dturn}

Different authors have reported that the SFMS has a turnover mass at which the relation deviates from linearity, dropping towards lower SFRs at high masses. This turnover mass has been reported both in the Local Universe \citep[e.g.][]{catalan15}, and at different redshifts \citep[e.g.][]{lee15,Tomczak+2016}, observed with many different surveys. The mass at which the SFR drops seems to increase with the redshift. Recent cosmological hydrodynamical simulations of galaxy formation, such as {\it Illustris} \citep{nelson15}, reproduce this result that the SFR seems to be weaker at higher redshifts and at larger stellar masses \citep{Sparre15}. This has been recently compared with observations by \citet{lopfer18}.

We cannot reproduce this drop in the SFR at high-masses as shown by previous studies. This could be a consequence of the detailed adopted method to select SFGs at different redshift ranges in our study, involving (i) a cut in the sSFR (that evolves with redshift), and (ii) excluding those galaxies that deviate significantly from the derived SFMS. Indeed, a visual exploration of the distributions of points along the M$_*$-SFR diagram in several redshift bins (Fig. \ref{fig:SFMS_t}) shows that most of the galaxies that deviate from the SFMS are located in the high mass regime { (although it is not a general trend, with a counter example at $z\sim$0.7)}. If we had considered them as actively forming stars they would have traced a deviation or drop as the one described by previous studies. 

\subsection{Fraction of Retired galaxies}
\label{sec:dRGs}

Different galaxy surveys at high redshift have shown that there are already passive galaxies even at z$\sim$2 \citep[][]{cimatti04,glazebrook04,Muzzin+2013,pand17}. The red-sequence is well established at z$\sim$0.7-1 \citep[][]{bell04,wuyts11}, although the fraction of galaxies that populate this sequence is lower than the one found in the local universe. We find indeed RGs$_t$ at any redshift bin from our analysis (Fig. \ref{fig:SFMS_t}), even at z$\sim$2. { However, its } fraction decreases very quickly with redshift (Fig. \ref{fig:frac}) { with a clear increase at $z<0.5$}. { Our results are consistent and in some cases agree within the errors with the trends and values reported from cosmological surveys, as we have shown in Sec. \ref{sec:SFMS_t}. We also have found that the transition between star-forming and retired galaxies does not seem to be a single event or a one-way process. A fraction of galaxies retired at high-$z$ become active at lower redshifts, to be quenched later on. This result is not totally new, since \citet{pand17} already predicted, based on semi-analytical simulations that between a 13\% and a 31\% of galaxies have experienced rejuvenation events since $z\sim$3. We will explore that possibility in future analysis, trying to trace the evolution of single objects along the SFR-M$_*$ diagram.}

{ Finally, our result} disagrees { somewhat} with those recently presented by \citet{lopfer18}. { Since this is the most recent result based on an archaeological analysis of similar data to the one presented here it is worthy to compare results in more detail}. They found almost no RGs$_t$ beyond $z>0.5$. Those authors speculate that the progenitors of current early-type, mostly retired galaxies, are not those retired galaxies found by cosmological surveys at high redshift, { since most of the early-type galaxies at $z\sim$0 were actively forming stars at high redshift.} Our results agree somewhat with this interpretation, since we find that the population of RGs$_0$ were mostly actively forming stars at high redshift ($\sim$80\% of them at $z\sim$2). However, a fraction of them were already retired galaxies at high redshift ($\sim$10-20\%). Thus, a fraction of the retired galaxies discovered by cosmological surveys at high redshift are indeed the progenitors of retired galaxies observed in the local Universe, but not of all of them. { The main discrepancy with the results by \citet{lopfer18} is that they found basically no retired galaxies beyond $z>$0.5, a result that indeed also disagrees with the results by direct observations based on cosmological surveys \citep[e.g.][]{Muzzin+2013,pand17}.}

There are three main reasons that could explain the differences found between our results and the ones presented by \citet{lopfer18}. First, they used a different sample with fewer galaxies, which may limit the fraction of detected RGs at high redshift, in particular in the regime of very massive galaxies. Some authors \citep[e.g.][]{toft12,belli17} have shown that RGs$_t$ at high redshift are already very massive objects, with stellar masses between 10$^{10.5-11}$ M$_\odot$. Thus, they would be most probably the progenitors of galaxies at the edge or not sampled by the CALIFA survey, but maybe sampled by larger surveys, as used here. Second, they adopted a parametric SFH to model their data.
{ On one hand, the analytic SFH laws, as the exp-$\tau$, present a smoothly declining SFR after a peak, biasing the inferences in some cases to a smooth and late passage towards the retired regime, in such a way that the fraction of retired galaxies becomes negligible at high  redshifts. Some quenching processes in reality can be much faster than the limits imposed by an exponential decline, thereby producing retired galaxies at high redshift. On the other hand, according to these analytical laws,} once retired, a galaxy cannot become active at a more recent cosmic time.
We have seen that in some cases RGs$_t$ at a certain look-back time could present a rejuvenation and become active in more recent times. { This result is in agreement with several results reported in the literature: (i) blue colors and high star formation rates in a fraction of local isolated ellipticals \citep[][and more references therein]{Lacerna+2016}; (ii) weak star-forming, spiral-like, structures found in {\it bona-fide} elliptical galaxies in the Local Universe \citep[e.g.][]{Gomes16b}; and (iii) star-formation activity reported in these galaxies based on the combination of near-UV, SDSS $r$ and mid-IR color studies \citep[e.g.][]{Kaviraj+2007,gildepaz07,ko14}. Even more, it agrees with the presence of a fraction of retired galaxies at $z\sim$2  found by cosmological surveys \citep[e.g.][]{Muzzin+2013}.} 

Finally, \citet{lopfer18} are sampling galaxies up to larger galactocentric distances, since almost all galaxies are sampled up to 2.5 R$_e$ while only a fraction of them are sampled by the MaNGA observations up to these radii. { In principle, since the MaNGA FoV does not guarantee that we cover the outer regions of galaxies for all the galaxies, it is possible that our results are biased to tracing more the evolution of the inner regions. This would require an aperture correction to estimate that effect. However, a direct comparison between the integrated stellar masses derived within the MaNGA FoV by Pipe3D and those listed in the NSA catalog do not show any significant aperture effect in our data. Therefore, this should not be the source of discrepancy.} 

In summary, we should recall that our results fit better with expectations from cosmological surveys. Of the three reasons, the second is the one that could explain the differences in the more natural way. { However, the only way to be totally sure that this is the case it is to repeat the current analysis over the dataset analyzed by \citet{lopfer18} and compare with the results. This analysis, that it is clearly beyond the current study, will be presented in a forthcoming article (S\'anchez et al., in prep.).}

\section{Conclusions}
\label{sec:con}

In this study we explore { statistically} the evolution of galaxies in the SFR-M$_*$ diagram accross cosmic time adopting the archaeological approach to follow the stellar population of $\sim$4000 galaxies in the local Universe observed by the MaNGA survey. The main conclusions of this analysis are as follows:

\begin{itemize}

\item We confirm that the SFMS holds as a tight relation between the SFR and the stellar masses over a wide range of look-back times, from the local Universe to z$\sim$7. The SFMS evolves strongly in its zero-point, illustrating that galaxies form stars in a more active way in the past, and in slope (from $\approx 0.7$ at $z\sim 0$ to $\approx 1$ at $z>0.5$), reflecting that even star-forming galaxies are becoming less active in the more massive range in recent times. This evolution is different for the local populations of star-forming and retired galaxies, with the latter presenting a stronger evolution in the zero-point, and a weaker evolution in the slope.

\item  The fraction of retired galaxies decreases rapidly with increasing redshift, in particular for the local population of retired galaxies {, rising fast at low-redshift ($z<0.5$)}.  However, we are able to detect retired galaxies over a wide range of redshifts up to $z\sim$2. { A fraction of those retired galaxies become active again at lower redshift, although most of them, in particular the more massive ones, end as retired galaxies in the local universe.}
%Only a small fraction of the retired galaxies at high redshift end as retired in the local universe.  

\item We are able to recover the cosmic SFR and stellar-mass density histories over a wide range of  look-back times, with good precision up to $z\sim$2. Our results agree qualitatively with the distributions found in studies from different cosmological surveys in all of the explored redshift range, and quantitatively at least for the last 8 Gyr. 

\item The contribution of the local star-forming and retired galaxies to star-formation and mass densities is significantly different, with the local RGs being very active in the past, contributing to most of the $\Psi_{SFR}$ and $<$sSFR$>$ of the Universe at $z>0.5$, and to most of the $\rho_*$ at any redshift. On the other hand, local SFGs dominate the SFR density in the universe in recent times, $z<0.2$, while RGs present a general halting/quenching in their star formation between z$\sim$0.2-0.5.

\end{itemize}

Our results all together indicate that the quenching of the star formation in galaxies is strongly related to their morphology and mass, since most of the currently RGs are indeed early-type massive galaxies. If the quenching happens inside-out, as supported by recent results, then bulge growth in combination with other processes (like gas starvation in dense environments) should be strongly connected with this quenching process. We will address this hypothesis in future studies by exploring the evolution of the SFR at different galactocentric regimes, making use of our spatial resolved information, now that we have confirmed that the integrated properties of star formation and stellar mass match reasonably well with the results found from cosmological surveys on the co-moving SFR and stellar mass density evolution.

\section*{Acknowledgements}

We would like to thanks the referee for his/her comments and suggestions.

We thank the CONACyT programs CB-285080 and DGAPA IA101217 grants for their support to this project. The data products presented in this paper  benefited from support and resources from the HPC cluster Atocatl at IA-UNAM.

We thanks R. Gonzalez-Delgado, R. L\'opez Fern\'andez, R. Cid Fernandes and E. Lacerda for their pioneering work on this kind of studies, for the generous discussions and the enlightening ideas.

Funding for the Sloan Digital Sky Survey IV has been provided by
the Alfred P. Sloan Foundation, the U.S. Department of Energy Office of
Science, and the Participating Institutions. SDSS-IV acknowledges
support and resources from the Center for High-Performance Computing at
the University of Utah. The SDSS web site is www.sdss.org.

SDSS-IV is managed by the Astrophysical Research Consortium for the 
Participating Institutions of the SDSS Collaboration including the 
Brazilian Participation Group, the Carnegie Institution for Science, 
Carnegie Mellon University, the Chilean Participation Group, the French Participation Group, Harvard-Smithsonian Center for Astrophysics, 
Instituto de Astrof\'isica de Canarias, The Johns Hopkins University, 
Kavli Institute for the Physics and Mathematics of the Universe (IPMU) / 
University of Tokyo, Lawrence Berkeley National Laboratory, 
Leibniz Institut f\"ur Astrophysik Potsdam (AIP),  
Max-Planck-Institut f\"ur Astronomie (MPIA Heidelberg), 
Max-Planck-Institut f\"ur Astrophysik (MPA Garching), 
Max-Planck-Institut f\"ur Extraterrestrische Physik (MPE), 
National Astronomical Observatories of China, New Mexico State University, 
New York University, University of Notre Dame, 
Observat\'ario Nacional / MCTI, The Ohio State University, 
Pennsylvania State University, Shanghai Astronomical Observatory, 
United Kingdom Participation Group,
Universidad Nacional Aut\'onoma de M\'exico, University of Arizona, 
University of Colorado Boulder, University of Oxford, University of Portsmouth, 
University of Utah, University of Virginia, University of Washington, University of Wisconsin, 
Vanderbilt University, and Yale University.

%\end{}



\appendix

{
\section{Examples of the spatially binning scheme}
\label{sec:bin}

%%%%%%%%%%%%%%%%%%%%%%%%%%%%%%%%%%%%%%%%%%%%%%%%%%
% SFMS_t
\begin{figure*}
  \centering
    \includegraphics[width=18cm, clip, trim=40 397 130 100]{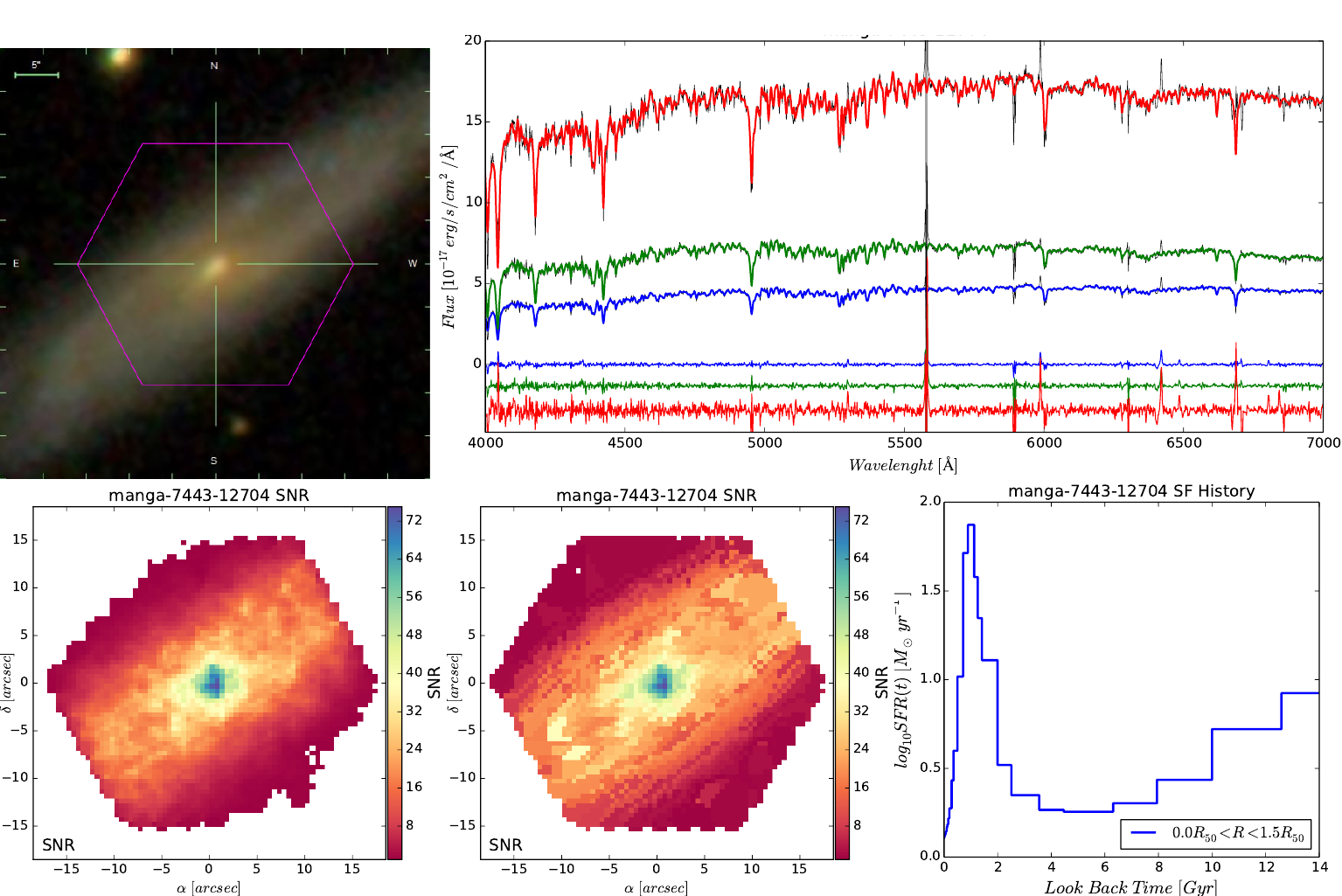}
    \caption{Figure illustrating the effects of the binning process. Top-left panel shows the SDSS image of the galaxy manga-7443-12704, with the FoV of the hexagonal MaNGA IFU over-plotted. Top-right panel shows the average spectra (black-solid line) within three elliptical apertures following the ellipticity and position angle of the galaxy at three galactocentric distances: 0-0.5, 0.5-1.0 and 1.0-1.5 effective radius of the galaxy. In addition it is shown the average of the individual stellar population models fitted for the individual tellessas in each bin as colored solid-lines (red for the inner-most region, green for the intermediate one, and blue for the outer most one), together with the residuals from the comparison between the average original and model spectra (arbitrary offset for a better visualization). {  Bottom-left and bottom-middel panels show the S/N in each spaxel before the spatial binning and in each tessella after the binning, respectively, illustrating the pattern of the binning too.}
 Finally the integrated star-formation history derived for this individual galaxy is shown in the bottom-right panel, as derived by the fitting routine, without considering the different corrections described in the text. }
  \label{fig:bin1}
\end{figure*}
%%%%%%%%%%%%%%%%%%%%%%%%%%%%%%%%%%%%%%%%%%%%%%%%%%%

%%%%%%%%%%%%%%%%%%%%%%%%%%%%%%%%%%%%%%%%%%%%%%%%%%
% SFMS_t
\begin{figure*}
  \centering
    \includegraphics[width=18cm, clip, trim=40 397 130 100]{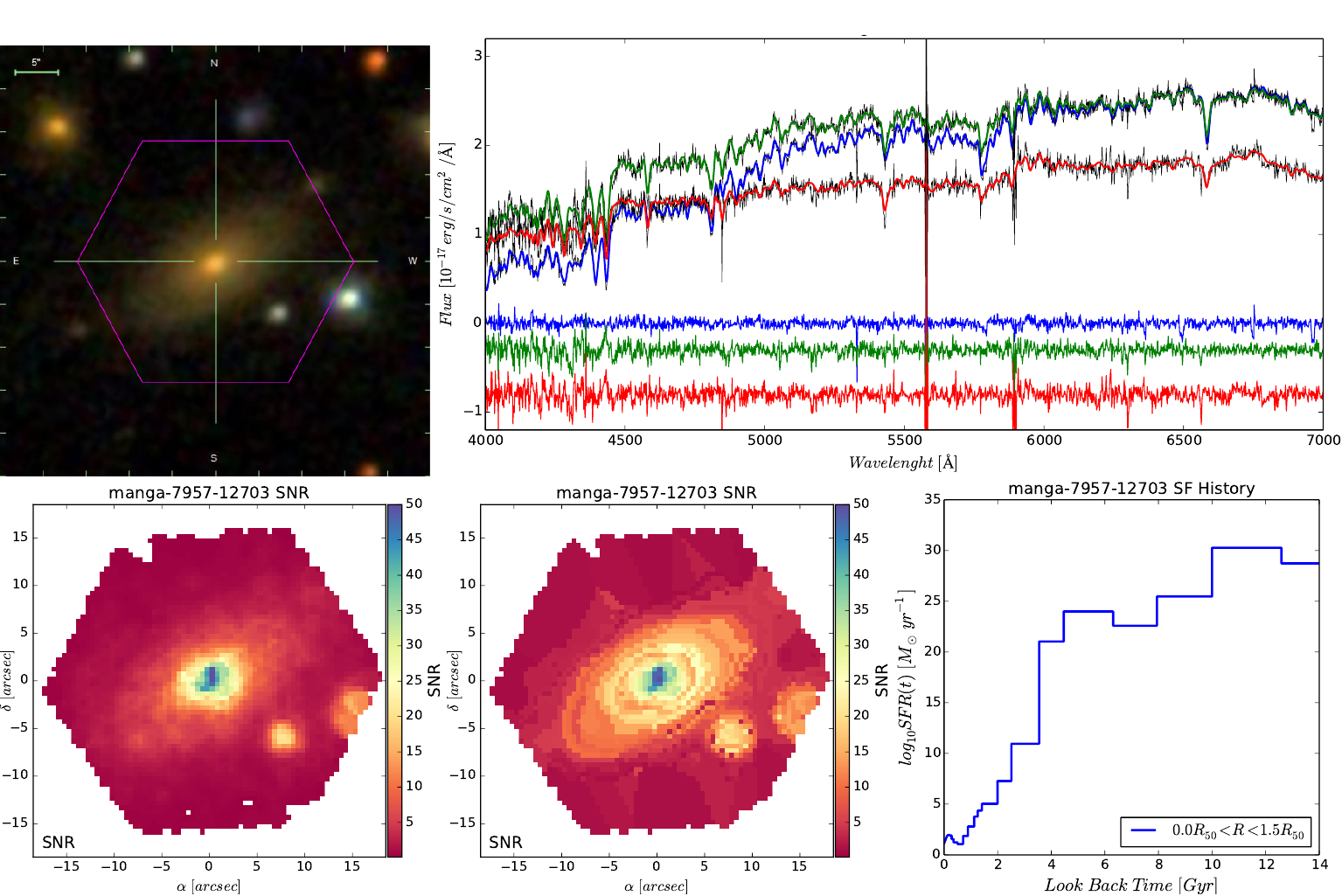}
    \caption{Similar figure as \ref{fig:bin1} for the galaxy manga-7957-12703.}
  \label{fig:bin2}
\end{figure*}
%%%%%%%%%%%%%%%%%%%%%%%%%%%%%%%%%%%%%%%%%%%%%%%%%%%

The binning scheme adopted in {\sc Pipe3D} was described in detail in
\citet{Pipe3D_II}. { It aims to increase the S/N above a
  certain selected value, but to preserve as much as possible the shape
  of the original distribution of the light within the galaxy. For
  this reason,} the procedure adopts two criteria to group adjacent
spaxels. { The first one is just a S/N threshold or limit.}
If the original spaxel already has a S/N above that goal one, the
bin/tessella comprises that original spaxel only. If not, it co-adds
adjacent spaxels { trying to reach the S/N limit}, taking into account the
covariance between adjacent spaxels in the S/N analysis. However,
contrary to other binning schemes {, as a second and more restrictive criterion,} it forces the adjacent spaxels to
have similar flux intensities in the considered wavelength range. In
other words, it does not allow adjacent spaxels of very different flux
intensities to be co-added, within a defined fractional difference
between considered fluxes. When no adjacent spaxels are found
fulfilling this criterion, the aggregation process stops. For this
reason this binning scheme preserves better the original shape of the
observed galaxy. However, it does not guarantee that the foreseen
S/N { goal} is reached, what it is particularly true for the outer regions of
galaxies. Importantly, this method is guaranteed that the areas
clearly dominated by sky-noise are well separated from the rest of the
tessellas.

%%%%%%%%%%%%%%%%%%%%%%%%%%%%%%%%%%%%%%%%%%%%%%%%%%
% SFMS_t
\begin{figure}
  \centering
    \includegraphics[width=8.5cm, clip, trim=0 30 60 30]{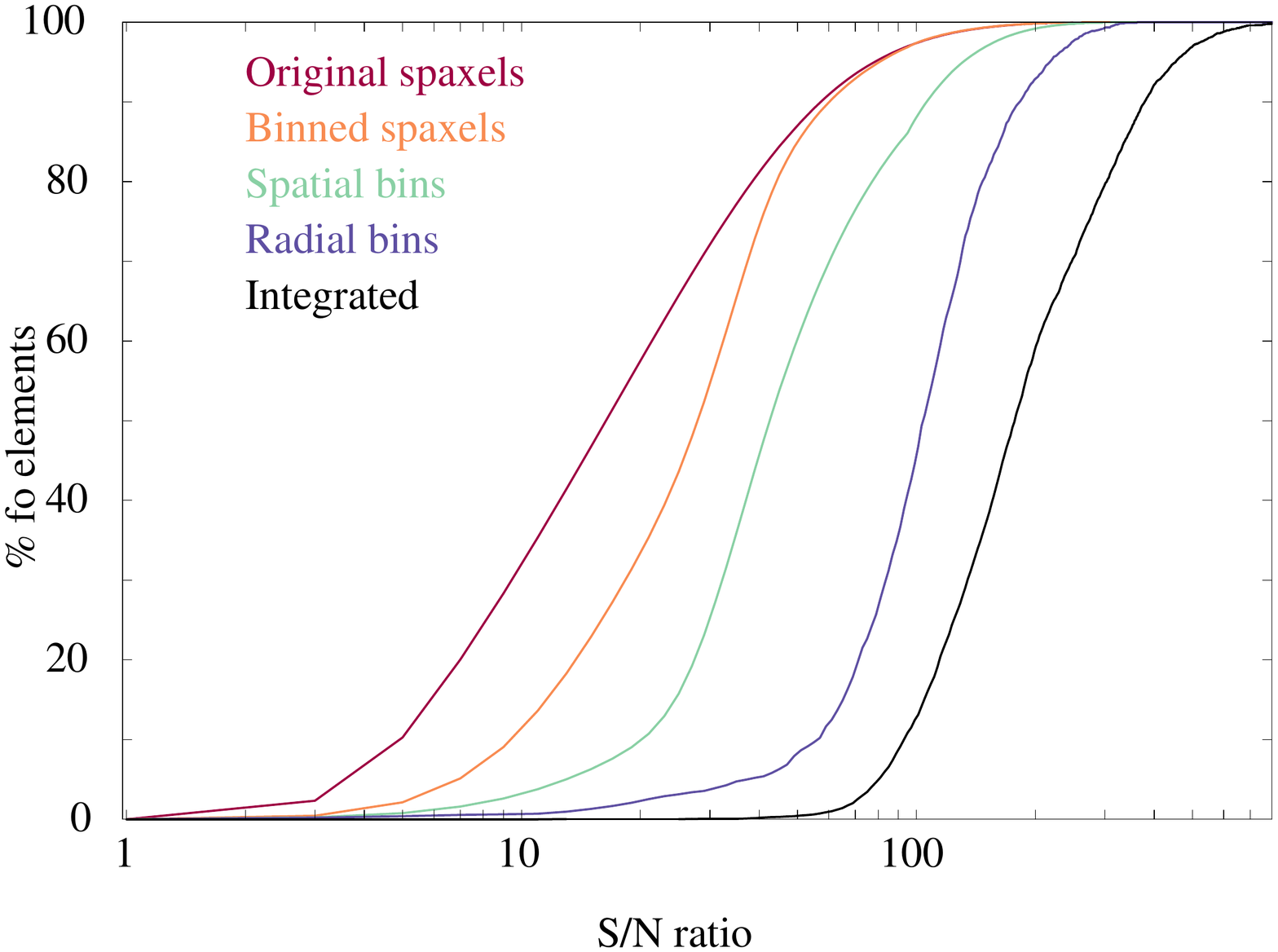}
    \caption{ Cumulative functions of the signa-to-noise ratios for the orginal spaxels (red), the individual spaxels once performed the binning procedure (orange), the individual spatial bins or tessellas (light green), radial binning the spectra in annulae of 0.5 effective radius, like the ones shown in Fig. \ref{fig:bin1} and \ref{fig:bin2} (dark blue), and the integrated spectra across the entire FoV (black).}
  \label{fig:SN}
\end{figure}
%%%%%%%%%%%%%%%%%%%%%%%%%%%%%%%%%%%%%%%%%%%%%%%%%%%

This is well illustrated in Fig. \ref{fig:bin1} and \ref{fig:bin2} where it is shown the S/N { distributions prior and after performing the
segmentation/binning scheme} along the FoV of two galaxies selected randomly from the sample. It is clear the goal S/N (50) is only reached, after binning, for the very central regions ($\sim$15-25\% of the tessellas). In general the S/N is larger than 10-20 for most of the considered tessellas. Those tessellas corresponding to sky-dominated areas are much larger, irregulars, and clearly different than the remaining ones. They were excluded from the analysis by an intensity and S/N cut. { The effect of the binning can be quantified by the comulative functions of the signal-to-noise ratio per spaxel and spatial bin shown in Figure \ref{fig:SN}. As indicated before the goal S/N is not reached in all the tessellas after the binning procedure, however, the S/N is clearly increased, both spaxel-wise and within each tessella. For the original spaxels a 20\% have a S/N below 7, while for the binned spaxels this percentage corresponds to the double of the S/N. On the other hand for the individual tessellas this percentage corresponds to S/N of $\sim$25, and performing a radial bin as the one shown in the upper-right panel of Fig. \ref{fig:bin1} and \ref{fig:bin2}, it corresponds to a S/N of $\sim$75. Finally, for the integrated spectra through the entire FoV, it corresponds to a S/N$\sim$100. Consistently, only a 20\% of the original spaxels have a S/N above 30, while for the binned spaxels this percentage corresponds to a S/N$\sim$40. For the final tessellas, it corresponds to a S/N$\sim$70, and for the radial bins and the integrated spectra to a S/N$\sim$100 and $\sim$200, respectively. }

Due to the peudo-stochastic nature of {\sc FIT3D} analysis \citep{Pipe3D_I}, the tessellas with lower S/N ($\sim$10) present a lower { precission} in the derivation of the stellar population properties, and in particular the star-formation histories. { This corresponds to just a few percentage ($\sim$5\%) of the final tessellas, as appreciated in Fig. \ref{fig:SN}. Even more }, the stacking of the derived models and their corresponding properties are a good representation of the data. This is illustrated in Fig. \ref{fig:bin1} and \ref{fig:bin2}, upper-right panel, where it is shown the average spectral within three different apertures together with the average of the individual models derived for the individual tessellas at the same apertures. We should remark here that we do not show a fitting to the average spectra, but an average of the results of the individual fits. The precision with which the average of the models describe the average spectra is remarkable. Indeed, the residuals of the subtraction of the data to the average of the models is better for the outer-rings than for the inner ones. This reflects that indeed the S/N of the average spectrum in the outer-ring is larger than that of the inner ones, and that using tessellas of lower S/N does not produce an appreciable bias in the average model. Therefore, we are confident of the stellar population model derived using this approach, in particular for the integrated properties.

Finally, as an illustration of the differences derived by the analysis for these two different galaxies considered, Fig. A1 and A2 show their integrated the star-formation histories as they are originally derived by Pipe3D without any of the corrections described in this article (i.e., in terms of sampled time ranges). It is clear that manga-7443-12704 has formed stars more recently than manga-7957-12703, what is it is reflected in the colors exhibited by the two galaxies in the SDSS images included in Fig. \ref{fig:bin1} and Fig. \ref{fig:bin2}.

}

\section{Effects of selecting the ionizing source in the derivation of the SFR$_{\rm H\alpha}$}
\label{sec:comp_SFR_spax}

%%%%%%%%%%%%%%%%%%%%%%%%%%%%%%%%%%%%%%%%%%%%%%%%%%
% SFMS_t
\begin{figure}
  \centering
    \includegraphics[width=10cm, clip, trim=70 0 0 0]{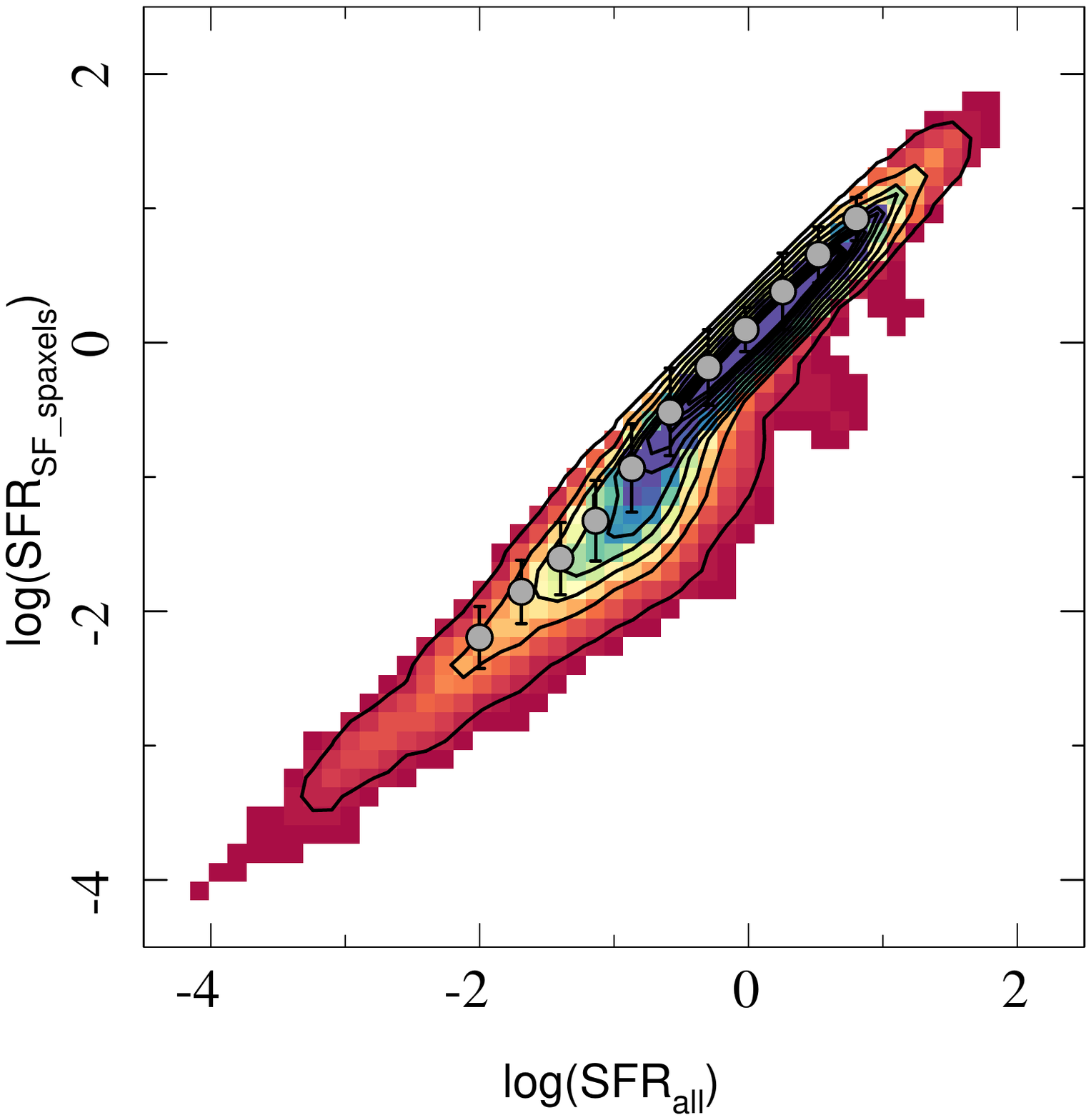}
    \caption{ Density map showing the distribution of the SFRs based on the dust-corrected H$\alpha$ luminosity selecting only those spaxels fully compatible with being ionized by star-formation (SFR$_{SF\_spaxels}$) versus just integrating through all the optical extent of the galaxies (SFR$_{all}$). The contours illustrate the density of points, with the same contour encircling a 95\% of the points, with a decrease of 20\% in the fraction of points between each consecutive contour. The grey circles indicate the mean values in a bins of $\sim$0.3 dex in SFR, which the errorbars indicate the standard deviations around these mean values. }
  \label{fig:comp_SFR_spax}
\end{figure}
%%%%%%%%%%%%%%%%%%%%%%%%%%%%%%%%%%%%%%%%%%%%%%%%%%%

{ We perform a comparison between the derived SFRs using the dust-extinction corrected H$\alpha$ luminosity integrated along the optical extension of the galaxies and the values estimated by selecting only those regions where ionization is classified as star-formating (HII-like) on the kpc scales of the MaNGA data. As extensively discussed in  \citet{sanchez17a} and references therein, the selection of the source of ionization is not simple, even given all the line ratios accessible with the MaNGA data. In particular, for regions with low EW(H$\alpha$) the best we can do is to indicate that it unclear if the gas is ionized by star-formation. However, the contrary cannot be stated. Therefore, excluding those regions from the H$\alpha$ budget to derive the SFR could lead us to wrong results. In any case, we performed the current comparison to constrain the possible effects of this selection.

Figure \ref{fig:comp_SFR_spax} shows the distribution of SFRs derived using the two different procedures. There is very good agreement between both quantities, following an almost one-to-one relation, with a dispersion of $\sim$0.2 dex and a global offset of $\sim$0.01 dex. This difference is even smaller for galaxies dominated by star-formation ionization in their integrated properties ($\sim$50\% of the galaxies), with no offset and a difference lower than a 2\% in the measurements. For those galaxies dominated by LINER-like ionization there is a larger difference, with an offset of $\sim$0.16 dex and a dispersion of 0.29 dex. However, for those galaxies we cannot guarantee that the ionization comes from SF or other sources of ionization, and in most cases they will be classified as retired, as we described in Sec. \ref{sec:SFMS_0}. Therefore, the selection of any of both methods to derive the SFR$_{\rm H\alpha}$ will not affect the results. Finally, for the $\leq$100 AGNs detected in MaNGA \citep{mall18,sanchez17b} we find a significant offset, of the order of $\sim$0.5 dex. However, as in the case of the RGs we cannot guarantee that part of this flux is due to SF, we can only determine that it is not the dominant source of the ionization. Including or removing them from the total number of analyzed galaxies does not produce any significant different. For a comparison with single aperture cosmological surveys, where mixing of ionization does not guarantee that they would be classified as AGNs \citep[e.g.][]{davies16}, we prefer to keep them in the described analysis and stick to the integrated quantity described in Sec. \ref{sec:SFR}.}

\section{Relation between the SFR$_{H\alpha}$ and SFR$_{ssp}$}
\label{sec:comp_SFR}

%%%%%%%%%%%%%%%%%%%%%%%%%%%%%%%%%%%%%%%%%%%%%%%%%%
% SFMS_t
\begin{figure}
  \centering
    \includegraphics[width=10cm, clip, trim=70 0 0 0]{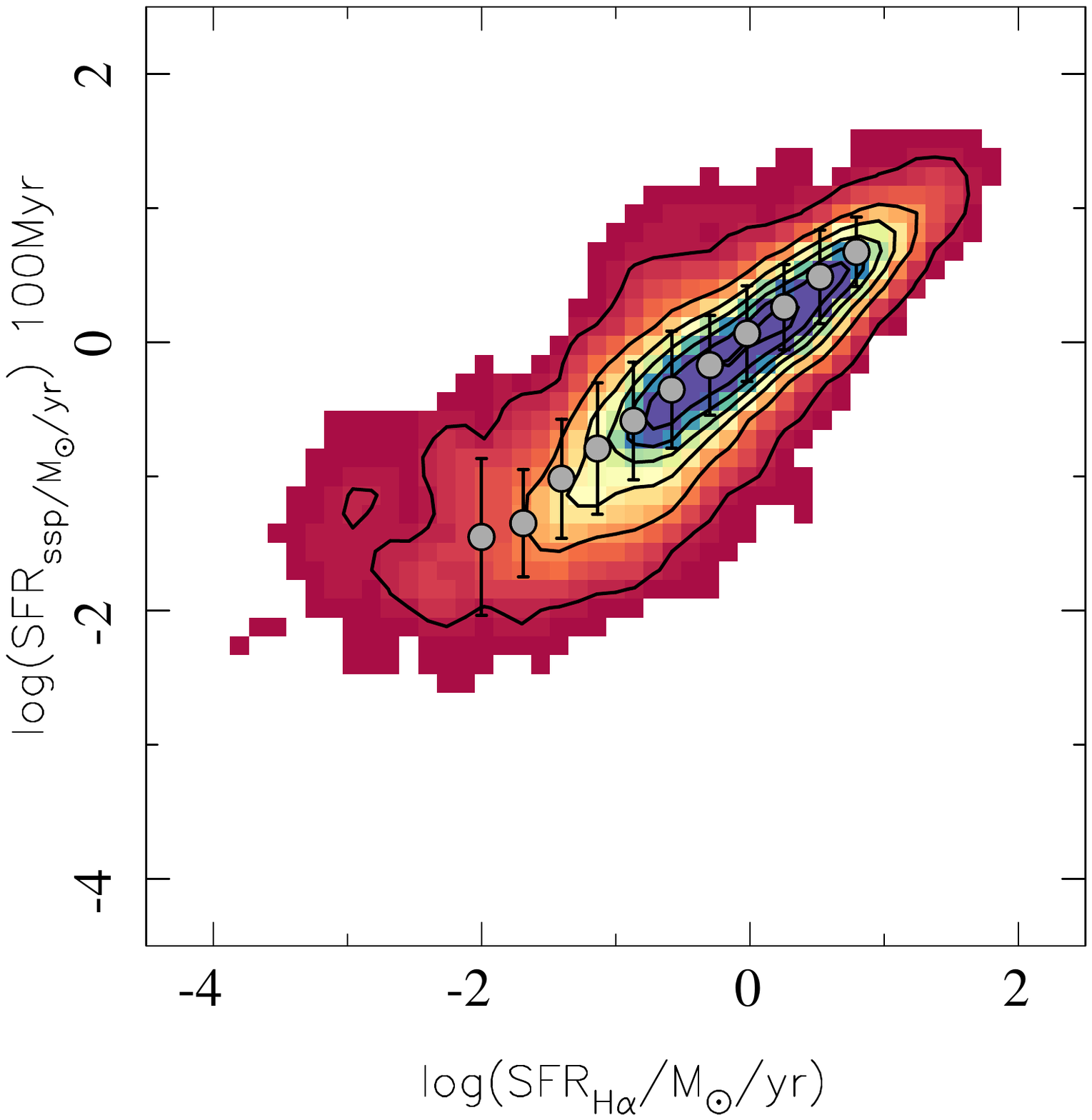}
    \caption{ Density map showing the distribution of the SFRs derived using the SSP analysis (SFR$_{ssp}$) versus the SFRs estimated from dust-corrected H$\alpha$ luminosity (SFR$_{H\alpha}$). The contours illustrate the density of points, with the same contour encircling a 95\% of the points, with a decrease of 20\% in the fraction of points between each consecutive contour. The grey circles indicate the mean values in a bins of $\sim$0.3 dex in SFR;  errorbars indicate the standard deviations around these mean values. }
  \label{fig:comp_SFR}
\end{figure}
%%%%%%%%%%%%%%%%%%%%%%%%%%%%%%%%%%%%%%%%%%%%%%%%%%%

{ In this article we use SFRs derived at each cosmological time for galaxies based on the multi-SSP analysis outlined in Sec. \ref{sec:ana} (i.e., SFR$_{ssp}$). For comparison purposes, when discussing the distribution of galaxies in the SFR-M$_*$ diagram at $z\sim$0, in Sec. \ref{sec:SFMS_0}, we make use of the SFR derived based on the dust-extinction corrected H$\alpha$ luminosities, presented in \ref{sec:SFR} (i.e., SFR$_{\rm H\alpha}$). Both quantities do not present a one-to-one correspondence, as already indicated. We explore here the relation between both quantities and discuss the nature of their differences.}

{ Figure \ref{fig:comp_SFR} shows the distribution of SFR$_{ssp}$ along the SFR$_{\rm H\alpha}$ derived for the $\sim$4000 galaxies analyzed here.} A direct comparison between the two different estimates of SFR  shows they have an offset of $\sim$0.11 dex, with a dispersion of $\sim$0.32 dex one each other. A linear regression between both parameters shows that they depart slightly from a one-to-one relation, although they present a very strong ($r=0.99$) and tight ($\sigma=0.1$ dex) correlation:

$$SFR_{ssp} = 0.1_{\pm 0.3} + 0.8_{\pm 0.3} SFR_{\rm H\alpha}$$

These differences are a consequence of: (i) The different stellar libraries adopted { in the current study compared to the one used for the SFR$_{\rm H\alpha}$ calibrator, since this would change the correspondence between the optical emission (analyzed in the current study) and the UV emission (considered in the SFR$_{\rm H\alpha}$ calibration)}. (ii) The amount of ionizing photons { that depends on assumptions of the emission in the UV from young stars to derive that calibrator too. This depends strongly on the effects of the metallicity, that has not been discussed here. Most of the calibrations of the SFR based on the H$\alpha$ luminosity are anchored to the solar metallicity \citep{kennicutt98}. However, the UV luminosity output by a stellar population depends on its metallicity, which affects stellar temperatures and line blanketing. Generally speaking, less-metal-rich stars produce more UV light \citep[e.g.][]{Madau14}. The amplitude of this effect is not insignificant, and depends on the details of the SFH, but in general, for the same SFR, less-metal rich galaxies would present more ionizing photons and therefore larger H$\alpha$ luminosities, and the contrary for metal rich galaxies. In contrast to the SFR$_{\rm H\alpha}$ estimate, our derivation of the SFR$_{ssp}$ considered metallicity as part of the decomposition of the stellar population using the multi-SSP analysis. This would indeed generate a difference between both estimations of the SFR}. (iii) The time scale used to derive the SFR { (as we have shown before)}, and the differences between the assumed { SFHs adopted for the calibration of the relation between the H$\alpha$ luminosity and the SFR \citep{kennicutt98}, and the reconstructed SFHs that could be derived from our stellar decomposition analysis. Although the conversion factor between SFR and H$\alpha$ luminosity is almost independent of the considered SFHs \citep[e.g., Sec. 2.3 of][]{kennicutt98}, its correspondence with the SFR$_{ssp}$ derived at different time scales depends strongly on the shape of that SFHs. 
We should recall here that we have not assumed any particular shape for the SFH, although those could be reconstructed from the weights of the decomposition based on the multi-SSP analysis. 
%In any case, it is not expected a one-to-one correspondence and the correspondence with strongly depends on the real SFHs of galaxies. 
Finally, (iv) the differences between the assumed conditions of the ionized gas to derive the SFR$_{H\alpha}$ calibration \citep[Case B recombination, T$_e$=10$^4$ K][]{kennicutt98} and the real conditions of the observed ionized gas would produce significant differences. Just changing the geometry of the ionized gas distribution around the young stellar clusters would produce significant differences in the derived emission line fluxes \citep[e.g.][]{mori16}. All together it is not surprising that the two derivations of SFR exhibit differences.} This is quite usual when comparing different calibrations \citep[as clearly illustrated by][]{Speagle14,catalan15,Davies+2016}. Indeed, when comparing with cosmological simulations we found a similar discrepancy between both estimations of the SFR (Ibarra-Medel et al., in prep.). { The derived relation has been included for future reference, but it is not used in the current study.}
 
\section{Evolution of the SFMS segregated between SFGs and RGs}
\label{sec:SF_RG}

\begin{figure*}
  \centering
    \includegraphics[width=17.5cm]{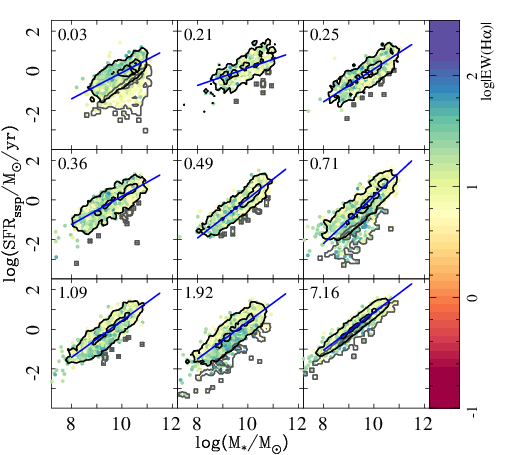}
    \caption{Distribution of the star-formation rate (SFR$_{\rm ssp}$) versus integrated stellar mass, as shown in Fig. \ref{fig:SFMS_t} for the same redshift bins, here only for those galaxies that are star-forming at the observed time ($z<0.04$). Symbols and labels are the same. As in the previous figure, for clarity, only those points with an error in the SFR lower than 0.2 dex have been included.}
  \label{fig:SF_SFMS_t}
\end{figure*}

\begin{figure*}
  \centering
    \includegraphics[width=17.5cm]{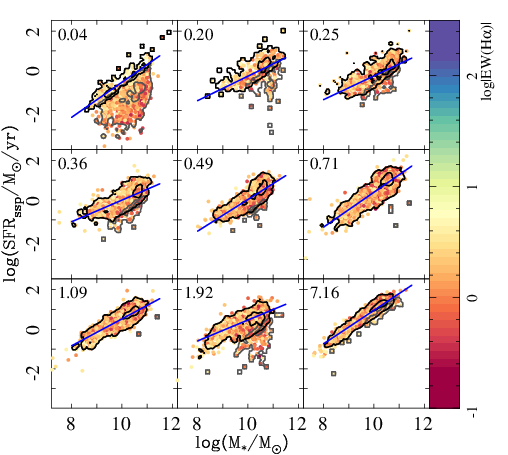}
    \caption{Distribution of the star-formation rate (SFR$_{\rm ssp}$) versus integrated stellar mass, as shown in Fig. \ref{fig:SFMS_t} for the same redshift bins, here only for those galaxies that are retired at the observed time ($z<0.04$). Symbols and labels are the same. As in the previous figure, for clarity only those points with an error in the SFR lower than 0.2 dex have been included.}
  \label{fig:RG_SFMS_t}
\end{figure*}

We explored the evolution of the SFMS with cosmic times in Sec. \ref{sec:SFMS_t}, for both the full sample of galaxies and the two sub-samples of currently (z$\sim$0) star-forming and retired galaxies. For clarity we showed the distribution in the SFR-M$_*$ plane only for the full sample (Fig. \ref{fig:SFMS_t}). The same distribution for the two subsamples of SFGs and RGs are shown in Figures \ref{fig:SF_SFMS_t} and \ref{fig:RG_SFMS_t}. The results of the analysis of the SFMS for those two subsamples of galaxies are listed in Table \ref{tab:SFMS_t}.

%%%%%%%%%%%%%%%%%%%%%%%%%%%%%%%%%%%%%%%%%%%%%%%%%%%
% SFMS_SF
%%%%%%%%%%%%%%%%%%%%%%%%%%%%%%%%%%%%%%%%%%%%%%%%%%%

%%%%%%%%%%%%%%%%%%%%%%%%%%%%%%%%%%%%%%%%%%%%%%%%%%%
% SFMS_RG
%%%%%%%%%%%%%%%%%%%%%%%%%%%%%%%%%%%%%%%%%%%%%%%%%%%

\section{Volume Corrections}
\label{sec:Vmax}

In Section \ref{sec:madau} we derived the cosmic SFR density and the  stellar-mass density as predicted from arhcaelogical methods for $\sim4000$ local galaxies from the MaNGA survey. The MaNGA survey is a sample with approximately equal numbers of galaxies at a fixed stellar mass. In other words, MaNGA has approximately a flat stellar mass distribution. In this Appendix we, briefly, describe the  volume corrections applied in order to project the mass distribution from MaNGA into a  galaxy stellar mass function, GSMF, that would be measured from a volume-limited sample that is complete in stellar mass. Here we use the stellar masses, $M_*$, galaxy colors, $g-r$ and redshifts
$z$ from \citet{Blanton+2017}. We use  K-corrected magnitudes at $z=0$ tabulated in \citet{Blanton+2017}. In addition, we introduce the empirical model for the E-correction at $z=0$  described in \citet{Dragomir+2018}. Finally, we derive the GSMF from the galaxy sample in  \citet{Blanton+2017} denoted herafter as $\phi_*(M_*)$. We will use this model of the GSMF to estimate the expected number of galaxies as a function of redshift and stellar mass. 

In a volume-limited sample that is complete in stellar mass, the GSMF can be simply estimated as $\phi_*(M_*) d\log M_{*}= N_g(M_*) / V$ where $N_g$ is the observed number of galaxies with masses between $\log M_*\pm d\log M_* /2$ and $V$ is the comoving volume of the observed sample. In the case of the MaNGA survey the above is not true and using the above estimator would lead to an incorrect GSMF. Note, however, that is possible to use the above estimator to determine the average fraction of galaxies that are expected for a given $M_*$, $z$ and $g-r$ from the MaNGA sample to the expected number of galaxies that fully samples the GSMF, $f_{m}$. While this fraction depends on the above three factors, in order to simplify our discussion, we consider next the case in which the fraction, $f_{m}$ depends only on $z$ and $M_*$ leaving the general case to the end of this section. 

Based on the above definition of the GSMF, we can estimate the expected number of galaxies with stellar masses between 
$\log M_*\pm d\log M_* /2$ in a redshift range of $z\pm dz/2$ as  $\langle N_{g}(M_*,z)\rangle = \phi_*(M_*) d\log M_*dV(z)$.  If $\langle N_{M}(M_*,z)\rangle$ is the expected  number of galaxies in the MaNGA sample over the same stellar mass and  redshift range then $f_{m} (M_*,z)= \langle N_{M}(M_*,z)\rangle / \langle N_{g}(M_*,z)\rangle$. Equivalently,  $f_{m} (M_*,z)= \langle \phi_{M,*}(M_*)\rangle / \phi_*(M_*) $ with $\langle \phi_{M,*}(M_*)\rangle$ as the average uncorrected GSMF from MaNGA. We calculate $\langle \phi_{M}(M_*)\rangle$ as:
\begin{equation}
\langle \phi_{M}(M_*)\rangle = \sum_i^{N_z} \phi_{M,i}(M_*,z_i) \times \omega_i,
\end{equation}
where $\phi_{M,i} (M_*,z_i) d\log M_* = N_{M,i}(M_*,z_i) / dV(z_i)$ with $N_{M,i}(M_*,z_i)$ as the real observed number of galaxies in MaNGA with stellar mass $\log M_*\pm d\log M_* /2$ at $z_i\pm dz/2$ while the weights $\omega_i$ are defined as: $\omega_i = N_{M,i} / \sum_i N_{M,i}$. The summation over $i$ refers to redshift and   $\sum_i \omega_i = 1$. Thus the expected number of galaxies in a volume-limited sample that is complete  in stellar mass based on MaNGA is: $N_{M,V} (M_*,z_i) =  N_{M} (M_*,z_i) / f_{m} (M_*,z)$. In principle, we can define  stellar mass bins and redshift bins as small as we want in order to have $ N_{M} (M_*,z_i) = 1,0$ objects  per bin. Thus for the $j$th galaxy in the sample with stellar mass $M_*$ and redshift $z$ the volume correction is given by:
\begin{equation}
\frac{1}{V_{j}} =  \frac{\omega_i(M_*,z)}{f_{m} (M_*,z)dV(z)}.
\end{equation}
In practice we do not requiered $ N_{M} (M_*,z_i) = 1,0$ objects per $M_*$ and $z$ bin since the  resulting volume corrections would be dominated by Poisson noise due to the low number of galaxies in MaNGA. Instead, we calculate an average volume correction $\langle 1 / V\rangle$ by creating a grid of 30 bins redshift bins within $0.03<z<0.17$ and 50 stellar mass bins within  $10^{8}<M_{*}/M_{\odot}<10^{12}$. Thus for every galaxy with $M_*$ and redshift $z$  we use cubic spline interpolations to find its corresponding volume correction from a grid of $\langle 1 / V_{i,j}\rangle$ over the $i$th stellar mass bin and  the $j$th redshift bin. 

In the general case in which $f_{m}$ also depends on galaxy colors, the above procedure is still valid. The fraction will be  given by $f_{m} (M_*,z|c)= \langle \phi_{M,*}(M_*|c)\rangle / \phi_*(M_*|c) $ where $c$ indicates that we are estimating our GSMF as a function of color. For our final volume correction we have divided the sample into five color bins between $0<g-r<1.2$. 

{

\section{Cosmic SFR History using original redshift bins}
\label{sec:madau_org}

%%%%%%%%%%%%%%%%%%%%%%%%%%%%%%%%%%%%%%%%%%%%%%%%%%%
% SFMS_t
\begin{figure}
  \centering
    \includegraphics[width=9.5cm]{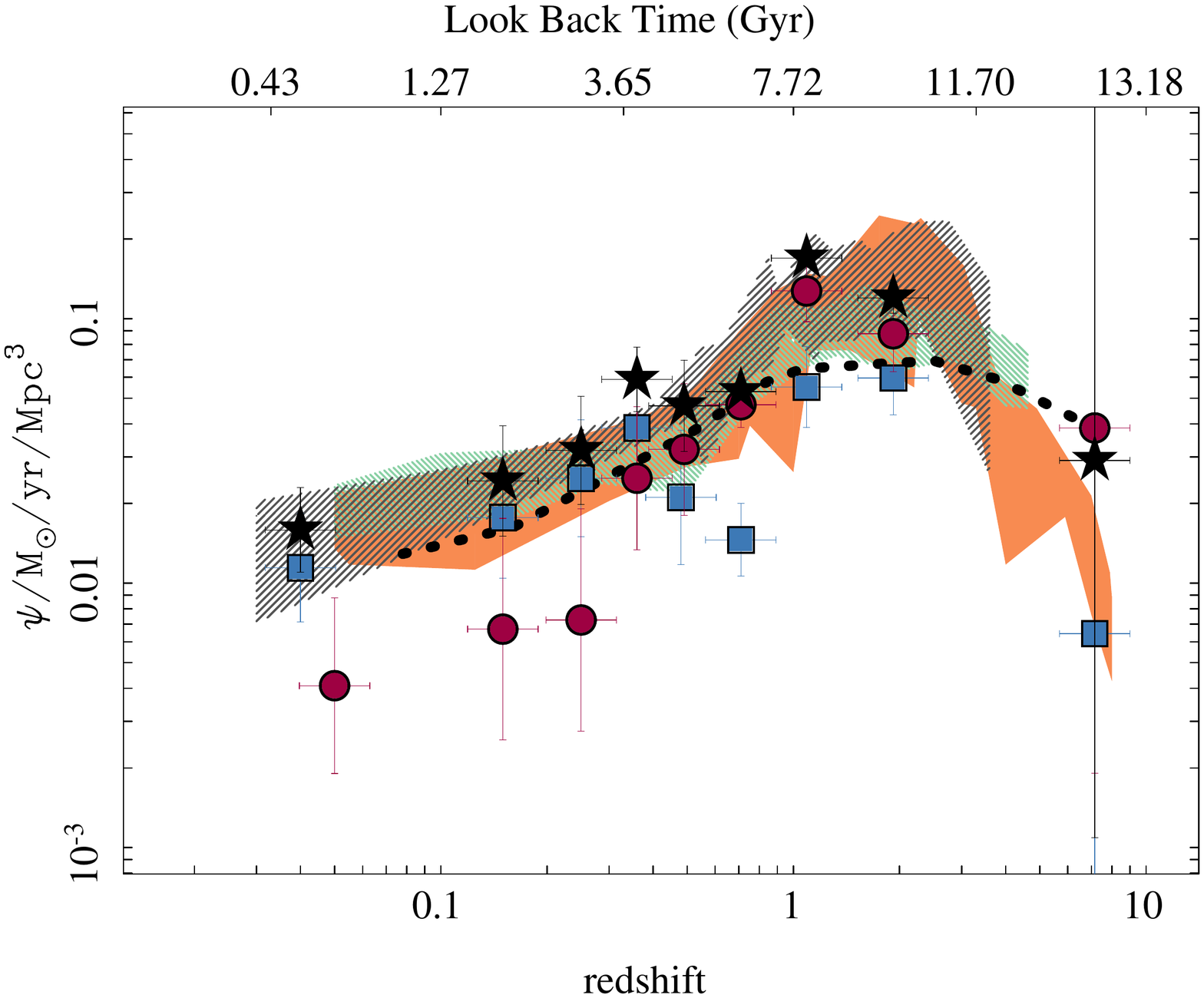}
    \caption{ Similar figure to Fig. \ref{fig:madau}, showing the cosmic evolution of the SFR density and using the same symbols. Here, we use the same redshift bins as the ones used from Sec. \ref{sec:SFMS_t} to Sec. \ref{sec:SFMS_q}. The main trends are similar among the two figures.}
  \label{fig:madau_org}
\end{figure}
%%%%%%%%%%%%%%%%%%%%%%%%%%%%%%%%%%%%%%%%%%%%%%%%%%%

The cosmic SFR history discussed in Sec. \ref{sec:madau} was built using a different set of redshift bins as the ones used to analyze the evolution of galaxies along the SFR-M$_*$ diagram, shown from Sec. \ref{sec:SFMS_t} to Sec. \ref{sec:SFMS_q}. These new redshift bins were introduced to maximize the coverage along cosmological times of the sampled parameter, $\Psi_{SFR,t}$. However, the analysis could be performed using exactly the same redshift bins used before, to the penalty of a poorer time sampling. Figure \ref{fig:madau_org} shows the result of this analysis. As expected the number of bins is reduced by a factor two, and the trends shows in Fig. \ref{fig:madau} are less clearly defined. However, the main results discussed in Sec. \ref{sec:madau} hold. The global rise of  $\Psi_{SFR,t}$ along earlier cosmological times, with a broad peak around $z\sim$1-3, and the differences outlined between the cosmic evolution of RGs$_0$ and SFGs$_0$ remain visible when using the original redshift bins. 
}

{  
\section{Cosmic SFR History without galaxy selection}
\label{sec:madau_cuts}

As indicated in Sec. \ref{sec:madau}, to build the cosmic SFR history from our analysis we selected only the SFGs at each look-back time for reasons already indicated in that section. To evaluate the impact of that selection in the derived $\Psi_{SFR,t}$ we repeated the analysis without performing any galaxy selection, thus, assuming that (1) all the measurements of SFR at any time are reliable and (2) that our {\it survey} is deep enough to measure the SFR of all those galaxies. Then, we added quadratically the difference between the values derived using this estimate with the ones derived by selecting only the SFGs to our error budget. We claimed that the main difference is present in the two higher redshift bins. Figure \ref{fig:madau_NC} illustrates that claim by showing the $\Psi_{SFR,t}$ derived by adopting no sample selection. As already indicated in Sec. \ref{sec:madau}, the main trends described for this distribution for $z<3$ hold, without any significant modification. There are some fluctuations of the values, most of them consistent with the original distribution for both the full sample and the SFGs$_0$ and RGs$_0$ subsamples. As already indicated the main difference is in the two higher redshift bins, in which instead of the decrease described before we see either a rising or a flat distribution. We should note that this is consistent with the SFHs shown in Fig. \ref{fig:SFH}, within the errors. Based on this analysis we claim along the article that out SFHs are not reliable beyond $z>$3.

%%%%%%%%%%%%%%%%%%%%%%%%%%%%%%%%%%%%%%%%%%%%%%%%%%%
% SFMS_t
\begin{figure}
  \centering
    \includegraphics[width=9.5cm]{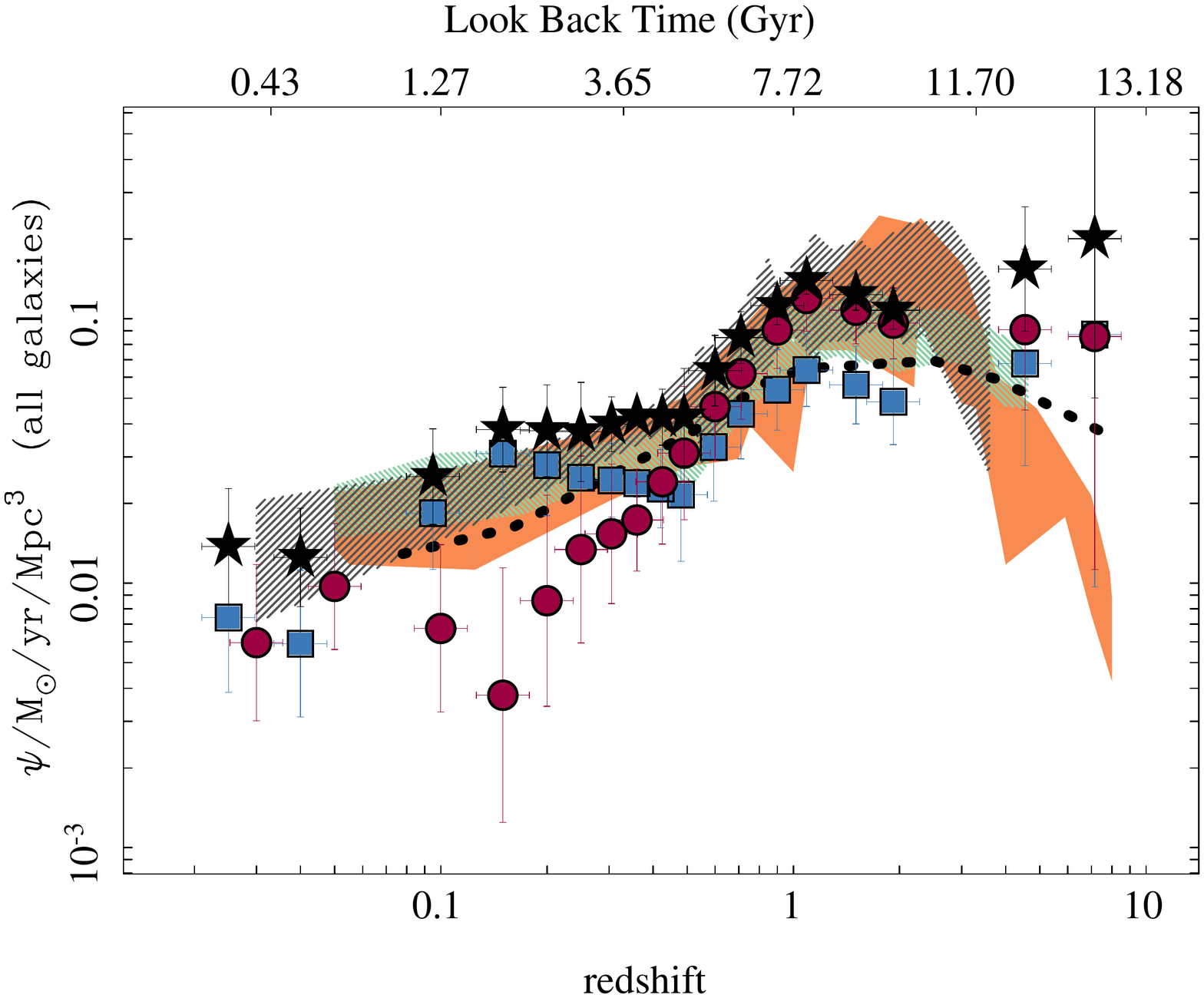}
    \caption{Similar figure to Fig. \ref{fig:madau}, showing the cosmic evolution of the SFR density and using the same symbols. Here, no cut is made in the sSFR$_t$ of the targets. Note that most of the differences between these two figures are found at very high redshifts.}
  \label{fig:madau_NC}
\end{figure}
%%%%%%%%%%%%%%%%%%%%%%%%%%%%%%%%%%%%%%%%%%%%%%%%%%%

\section{Cosmic SFR History without galaxy repetition}
\label{sec:madau_no_rep}

%%%%%%%%%%%%%%%%%%%%%%%%%%%%%%%%%%%%%%%%%%%%%%%%%%%
% SFMS_t
\begin{figure}
  \centering
    \includegraphics[width=9.5cm,clip,trim=0 100 0 0]{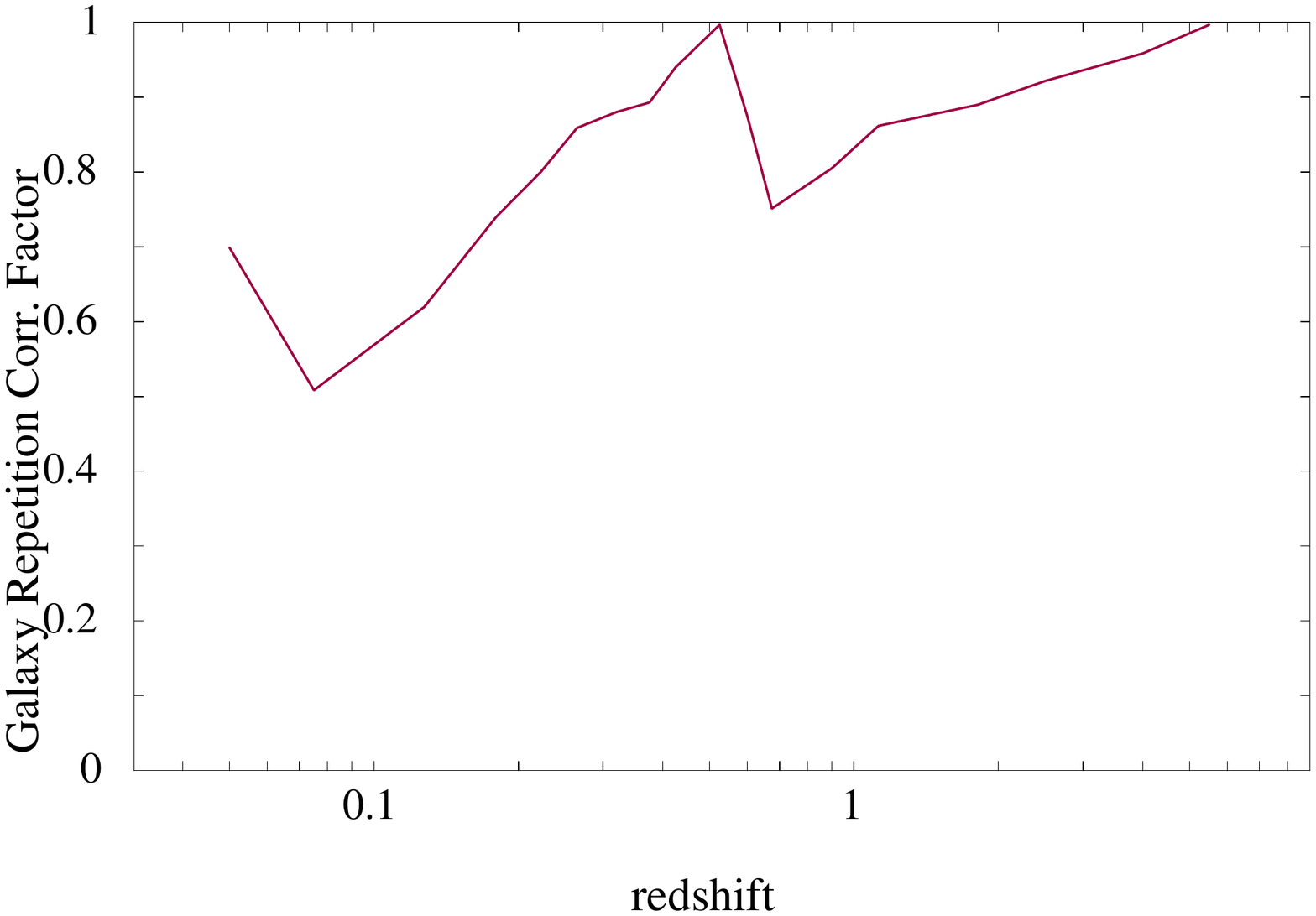}
    \caption{ Galaxy repetition correction factor introduced in Eq. \ref{eq:1} to derive the cosmic star-formation rate density to take into account the repetition of the same galaxies sampled at different look-back times in the same redshift bin.}
  \label{fig:f_cor}
\end{figure}
%%%%%%%%%%%%%%%%%%%%%%%%%%%%%%%%%%%%%%%%%%%%%%%%%%%

We estimated the cosmic SFR history of the universe in this article by treating our collection of SFRs and stellar masses as an observed cosmological survey.  Thus, when we select {\it synthetic galaxies} in a given redshift bin, the values could be due to the same original galaxy that is sampled several times in that given redshift bin. This introduces a correction factor for the number or repetitions of a galaxy within each redshift bin (Sec. \ref{sec:madau}). { Figure \ref{fig:f_cor} shows the distribution of this correction factor, defined in Eq. \ref{eqn:correction} along the redshift. To avoid the introduction of this correction factor a }  different approach would be to average the SFRs coming from the same galaxy within each bin to produce the same mathematical result. However, this would not respect the basis of our approach, i.e., to treat the dataset in a similar way as cosmological surveys are treated.

A different approach would be to allow each galaxy to be sampled only once in each redshift bin. This will introduce a certain degree of uncertainty, since it would require selecting one of the sampled SFRs. This approach is less precise, and can lead to considerable biases depending on how this selection in done. However, it can be tested to illustrate if the current procedure provides better results, and if so, at what redshifts (i.e., matching better to the values reported from cosmological surveys). For this test we repeated our estimates of $\Psi_{SFR,t}$ following the prescriptions shown in Sec. \ref{sec:madau} with the only difference that when a galaxy is sampled once it is not allowed to be sampled again. Thus, we forbid repetition of the same original galaxy in a given redshift bin. Since by construction the dataset is ordered by redshift, those values correspond to smaller look-back times within each bin. Since in general most of the SFHs of galaxies exhibit a decline in SFR with the redshift, the accepted values correspond with lower SFRs than in the original calculation. Figure \ref{fig:madau_no_rep} shows the result of the calculation. In general the more affected bins are between $z<$0.1 and $z>$1.5, as expected since those bins are the ones for which the galaxies are repeated most frequently. In particular, this approach (biased, as indicated before), under-predicts the cosmic SFR in those bins where repetitions are frequent, highlighting the fact that we have selected preferentially lower values of SFR than is representative for a given galaxy within a given redshift bin.

%%%%%%%%%%%%%%%%%%%%%%%%%%%%%%%%%%%%%%%%%%%%%%%%%%%
% SFMS_t
\begin{figure}
  \centering
    \includegraphics[width=9.5cm]{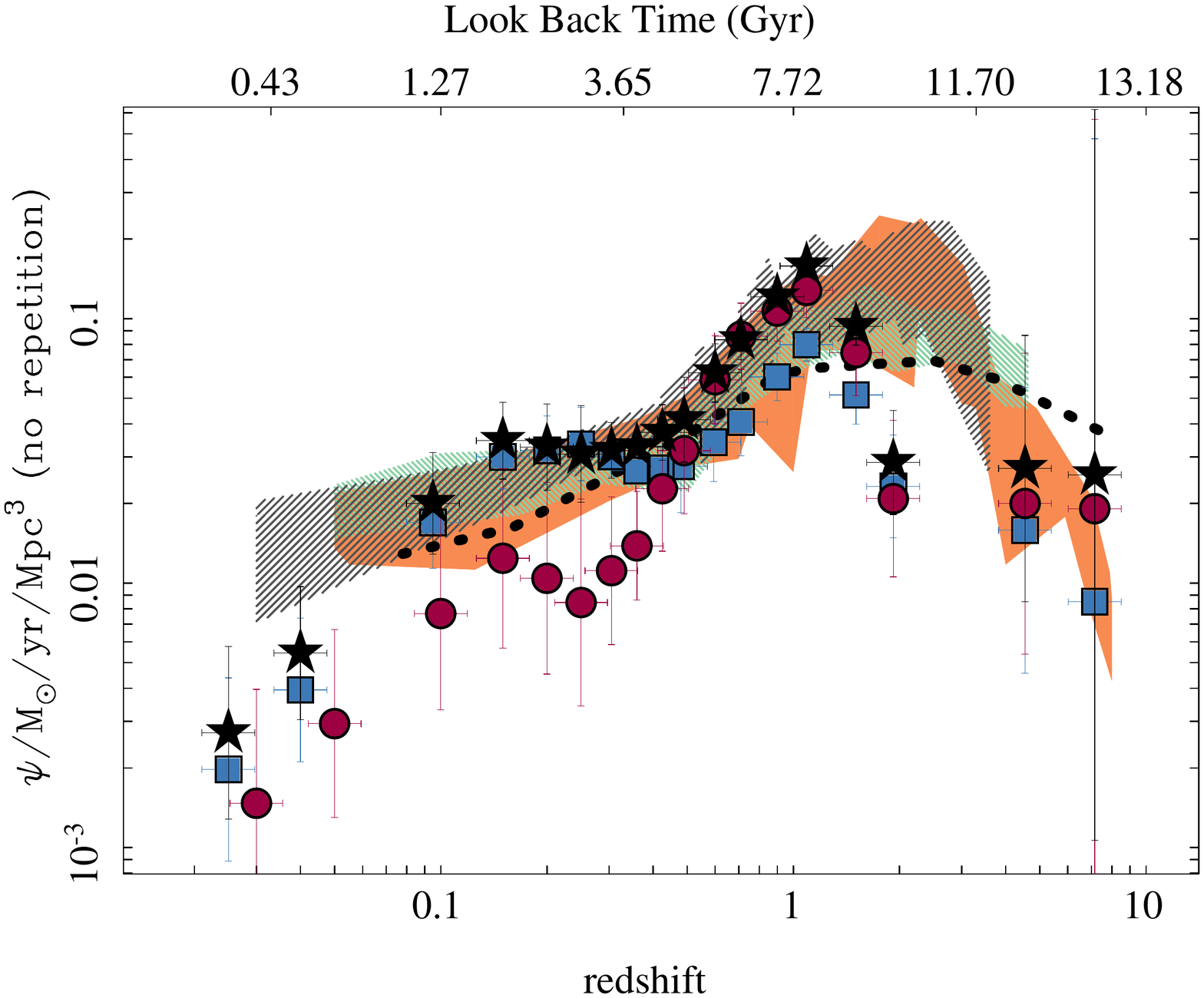}
    \caption{Similar figure as Fig. \ref{fig:madau}, showing the cosmic evolution of the SFR density, using the same symbols, but this time without allowing repetitions of values extracted from the analysis of the stellar populations of the same original galaxy. In this particular approach we adopted the first (lower redshift) value encounter for a particular galaxy as the one that it is representative of the different SFRs sampled within the considered redshift bin.}
  \label{fig:madau_no_rep}
\end{figure}
%%%%%%%%%%%%%%%%%%%%%%%%%%%%%%%%%%%%%%%%%%%%%%%%%%%

In summary, the considered correction factor is required due to the adopted approach in which
we treat our dataset as a synthetic cosmological survey, and therefore each {\it synthetic galaxy} should be weighted individually, irrespectively of the number of times that it is sampled within a particular redshift range. 

Despite of the clear bias introduced by the approach adopted in this new derivation of $\Psi_{SFR,t}$, and the fact that the numerical values depart more from the ones reported by cosmological surveys, it is interesting to note that most of the qualitative properties of the cosmic SFR history reported along this article are still the same: (1) The reported distribution seems to follow at least qualitatively the known shape for the Madau curve; (2) The peak in the cosmic SFR density seems to be at a slightly lower redshift than the one reported in the literature; (3) Retired galaxies and star-forming galaxies in the local universe present different contributions to $\Psi_{SFR,t}$, with the former being more active in the past, and having a sharper evolution.

}

\bibliographystyle{aasjournal}
%\bibliography{ref} 

\bibliography{CALIFAI,references-VAR,Alenka,ref_raom,Califa8_SFH}

\end{document}